\documentclass[traditabstract]{aa} 
\usepackage{graphicx}
\usepackage{lscape}
\usepackage{txfonts}
\usepackage{natbib}
\usepackage{url}
\usepackage[T1]{fontenc}

\newcommand{\fuden}{10$^{-17}$ erg~s$^{-1}$~cm$^{-2}$~arcsec$^{-1}$}

\newcommand{\Funits}{10$^{-16}$ erg~s$^{-1}$~cm$^{-2}$}

\newcommand{\degree}{\ensuremath{^\circ}}

\newcommand{\nodata}{ ~$\cdots$~ }

\DeclareRobustCommand{\ion}[2]{%
\relax\ifmmode
\ifx\testbx\f@series
{\mathbf{#1\,\mathsc{#2}}}\else
{\mathrm{#1\,\mathsc{#2}}}\fi
\else\textup{#1\,{\mdseries\textsc{#2}}}%
\fi}

\begin{document}
   \title{Integral field spectroscopy of a sample of nearby galaxies:}

   \subtitle{II. Properties of the \ion{H}{ii} regions}

   \author{
     S.\,F. S\'anchez\inst{1,2}
     \and
     F.\,F. Rosales-Ortega\inst{2,3}
     \and
     R.\,A. Marino\inst{4}
     \and
     J. Iglesias-P\'aramo\inst{1,2}
     \and
     J.\,M. V\'\i lchez\inst{1}
     \and
     R.\,C. Kennicutt\inst{5}
     \and
     A.\,I. D\'\i az\inst{3}
     \and
     D. Mast\inst{1,2}
      \and
     A. Monreal-Ibero \inst{1}
     \and
     R. Garc\'\i a-Benito \inst{1}
     \and
     J. Bland-Hawthorn\inst{6}
      \and
     E. P\'erez \inst{1}
      \and
     R. Gonz\'alez Delgado\inst{1}
     \and 
     B. Husemann\inst{7}
     \and
      \'A. R. L\'opez-S\'anchez\inst{8,9}
     \and
     R. Cid Fernandes\inst{10}
      \and
     C. Kehrig \inst{1}
     \and
     C.J. Walcher\inst{7}
     \and
     A. Gil de Paz\inst{4}
      \and
     S. Ellis\inst{8,6}
          }

   \institute{
        Instituto de Astrof\'{\i}sica de Andaluc\'{\i}a (CSIC), Glorieta de la Astronom\'\i a s/n, Aptdo. 3004, E18080-Granada, Spain\\ \email{sanchez@iaa.es}.
        \and
        Centro Astron\'omico Hispano Alem\'an, Calar Alto, (CSIC-MPG),
        C/Jes\'{u}s Durb\'{a}n Rem\'{o}n 2-2, E-04004 Almer\'{\i}a, Spain.
       \and
       Departamento de F\'isica Te\'orica, Universidad Aut\'onoma de Madrid, 28049 Madrid, Spain.
       \and
CEI Campus Moncloa, UCM-UPM, Departamento de Astrof\'{i}sica y CC$.$ de la Atm\'{o}sfera, Facultad de CC$.$ F\'{i}sicas, Universidad Complutense de Madrid, Avda.\,Complutense s/n, 28040 Madrid, Spain.
        \and
        Institute of Astronomy, University of Cambridge, Madingley Road, Cambridge CB3 0HA, UK.
        \and
        Sydney Institute for Astronomy, School of Physics A28, University of Sydney, NSW 2006, Australia.
        \and
        Leibniz-Institut f\"ur Astrophysik Potsdam (AIP), An der Sternwarte 16, D-14482 Potsdam, Germany.
        \and
        Australian Astronomical Observatory, PO BOX 296, Epping, NSW 1710, Australia.
\and
Department of Physics and Astronomy, Macquarie University, NSW 2109, Australia.
        \and
        Departamento de F\'\i sica - CFM - Universidade Federal de Santa Catarina, PO Box 476, 88040-900, Florian\'opolis, SC, Brazil.
     \thanks{Based on observations collected at the Centro Astron\'omico
      Hispano Alem\'an (CAHA) at Calar Alto, operated jointly by the Max-Planck
      Institut f\"ur Astronomie and the Instituto de Astrof\'{\i}sica de Andaluc\'{\i}a (CSIC).}
              }

   \date{Received ----- ; accepted ---- }

 
  \abstract{

In this work we analyze the spectroscopic properties of a large number
of \ion{H}{ii} regions, $\sim$2600, located in 38 galaxies. The sample
of galaxies has been assembled from the face-on spirals in the PINGS
survey and a sample described in M\'armol-Queralt\'o (2011, henceforth
Paper I).
All the galaxies were observed using Integral Field
Spectroscopy with a similar setup, covering their optical extension
up to $\sim$2.4 effective radii within a wavelength range from $\sim$3700 to
$\sim$6900\AA.

We develop a new automatic procedure to detect \ion{H}{ii} regions,
based on the contrast of the H$\alpha$ intensity maps extracted from
the datacubes. Once detected, the procedure provides us with the
integrated spectra of each individual segmented region. In total, we
derive good quality spectroscopic information for $\sim$2600
independent \ion{H}{ii} regions/complexes. This is by far the largest
nearby 2-dimensional spectroscopic survey presented on this kind of
regions up-to-date. Even more, our selection criteria and dataset
guarantee that we cover the regions in an unbiased way, regarding the
spatial sampling.

A well-tested automatic decoupling procedure has been applied to
remove the underlying stellar population, deriving the main
properties (intensity, dispersion and velocity) of the strongest
emission lines in the considered wavelength range (covering from
[\ion{O}{ii}]~$\lambda$3727 to [\ion{S}{ii}]~$\lambda$6731). A final catalogue of the spectroscopic
properties of these regions has been created for each
galaxy. Additional information regarding the morphology, spiral
structure, gas kinematics, and surface brightness of the underlying
stellar population has been included in each catalogue. 

In the current study we focused on the understanding of the average
properties of the \ion{H}{ii} regions and their radial distributions. We found
that, statistically, there is a significant change of the ionization
conditions across the optical extent of the galaxies. The fraction of \ion{H}{ii}
regions located in the intermediate ionization range in a classical BPT
diagram is larger for the central regions ($r< 0.25 r_e$), than in the outer
ones. This is somehow expected, if the origin of this shift is
due to the contamination of non-starforming ionization sources (e.g., AGN,
Shocks, post-AGB stars, etc.), that occur more frequently in the center of the
galaxies.


We find that the gas-phase oxygen abundance and the H$\alpha$
equivalent width present negative and positive gradient,
respectively. The distribution of slopes is statistically compatible 
with a random Gaussian distribution around the mean value, 
if the radial distances are measured in units of the respective effective radius.
No difference in the slope is found for galaxies of different morphologies: 
barred/non-barred, grand-design/flocculent. Therefore, the effective radius 
is a universal scale length for gradients in the evolution of galaxies. 

Other properties have a larger variance across each object, and galaxy
by galaxy (like the electron density), without a clear characteristic value, 
or they are well described by the average value either galaxy by galaxy or
among the different galaxies (like the dust attenuation).

}

   \maketitle


\section{Introduction}

\begin{figure*}
\centering
\includegraphics[width=\textwidth]{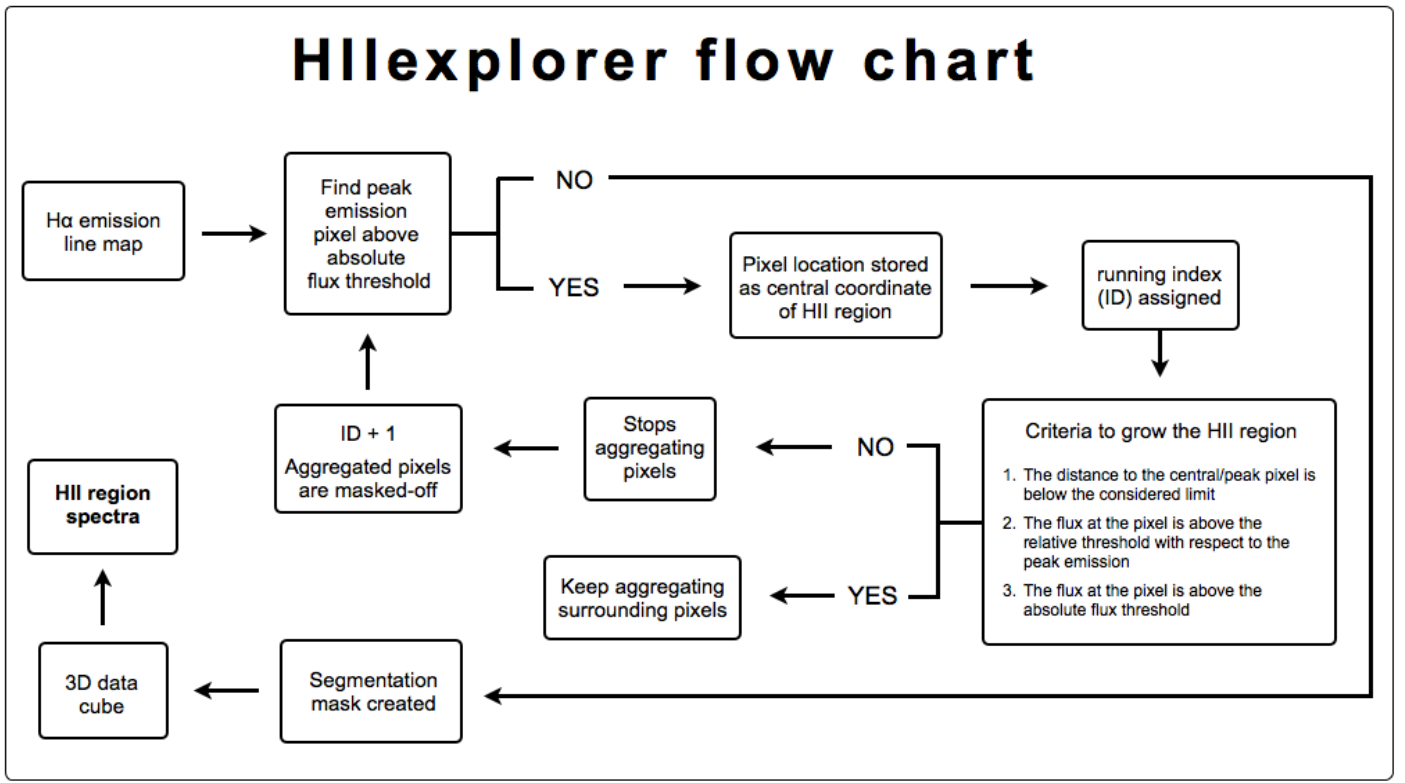}
\caption{
Flow chart of \textsc{HIIexplorer}, the algorithm developed to select, segregate,
and extract the spectra of the individual \ion{H}{ii} regions for the analyzed
galaxies.
  \label{fig:chart}}
\end{figure*}

Nebular emission lines from bright-individual \ion{H}{ii} regions have been,
historically, the main tool at our disposal for the direct measurement
of the gas-phase abundance at discrete spatial positions in galaxies. A 
good observational understanding of the distribution of element abundances 
across the surface of nearby galaxies is necessary to place constrains 
on theories of galactic chemical evolution. The same information is crucial to 
derive accurate star formation histories of and obtain information on the stellar 
nucleosynthesis in normal spiral galaxies.

Several factors dictate the chemical evolution in a galaxy, including
the primordial composition, the content and distribution of molecular
and neutral gas, the star formation history (SFH), feedback, the
transport and mixing of gas, the initial mass function (IMF), etc
\cite[e.g.][and references therein]{angel10,angel10b}. All these
ingredients contribute through a complex process to the evolutionary
histories of the stars and the galaxies in general. Accurate
measurements of the present chemical abundance constrain the different
possible evolutionary scenarios, and thus the importance of
determining the elemental composition in a global approach, among
different galaxy types.

\begin{table*}
 \caption{Main properties of the analyzed galaxies}
 \label{table.gal.prop}
 \begin{center}
 \begin{tabular}{lcccrrrrrrrrrrrrr}        
 \hline\hline                 
 Name & RA & Dec & Class & $z$ & B & B-V & mod & M$_{B}$ & v$_{\rm rot}$ & incl & PA & r$_e$  & Type & Arms \\
 (1) & (2) & (3) & (4) & (5) & (6) & (7) & (8)  & (9 )   & (10 ) & (11) & (12) & (13) & (14) & (15) \\
  \hline       
2MASXJ1319+53   & 13:19:35.9 & +53:30:09.8 & Sb       & 0.0248 & 15.92 & 0.72 & 35.3 & -19.3 &  51  & 28.0 &  55 &  9.7 & L   & NC  \\
CGCG\,071-096   & 13:00:33.2 & +10:07:47.8 & Sb(r)    & 0.0239 & 14.69 & 0.79 & 35.1 & -20.5 & 125  & 45.0 & 220 & 10.7 & I/L & NC  \\
CGCG\,148-006   & 07:44:57.4 & +28:55:39.0 & Sc       & 0.0234 & 15.16 & 0.75 & 35.1 & -19.9 & 180  & 40.5 & 250 & 10.3 & L   & NC  \\
CGCG\,293-023   & 02:30:21.5 & +56:47:29.5 & SBb      & 0.0156 & 15.51 & 0.99 & 34.3 & -18.8 & 325  & 19.6 & 100 &  6.9 & L   & NC  \\
CGCG\,430-046   & 23:00:46.2 & +13:37:07.9 & Sc       & 0.0243 & 14.49 & 0.86 & 35.1 & -20.7 & 168  & 42.1 & 160 & 11.0 & L   & NC  \\
IC\,2204        & 07:41:18.1 & +34:13:55.9 & Sab      & 0.0155 & 15.71 & 0.93 & 34.2 & -18.5 &  49  & 24.1 &  58 & 15.5 & I/E & AGN \\
MRK\,1477       & 13:16:14.7 & +41:29:40.1 & SBa(r)   & 0.0207 & 14.83 & 0.47 & 34.9 & -20.0 & 144  & 59.0 & 100 &  7.9 & L   & NC  \\
NGC\,99         & 00:23:59.5 & +15:46:13.7 & Sc       & 0.0177 & 13.71 & 0.62 & 34.5 & -20.7 &  79  & 33.7 &  40 & 19.9 & L   & NC  \\
NGC\,3820       & 11:42:04.9 & +10:23:03.3 & Sbc      & 0.0203 & 14.97 & 0.95 & 34.8 & -19.8 &  95  & 41.6 & 205 &  8.5 & L   & NC  \\
NGC\,4109       & 12:06:51.1 & +42:59:44.3 & Sa       & 0.0235 & 14.79 & 0.98 & 35.1 & -20.7 & 132  & 44.0 & 220 &  7.5 & E   & NC  \\
NGC\,7570       & 23:16:44.7 & +13:28:58.8 & Sa       & 0.0157 & 13.40 & 0.81 & 34.2 & -20.8 &  81  & 65.7 & 130 & 21.6 & I/E & NC  \\
UGC\,74         & 00:08:44.7 & +04:36:45.1 & Sc       & 0.0131 & 13.51 & 0.85 & 33.8 & -20.3 & 135  & 38.3 & 130 & 24.6 & L   & (s) \\
UGC\,233        & 00:24:42.7 & +14:49:28.8 & SBbcD    & 0.0176 & 14.35 & 0.62 & 34.4 & -20.1 & 161  &  4.1 & 150 &  6.9 & L   & NC  \\
UGC\,463        & 00:43:32.4 & +14:20:33.2 & SABc     & 0.0148 & 12.93 & 0.85 & 34.1 & -21.1 & 134  & 40.0 &  40 & 20.6 & L   & (s) \\
UGC\,1081       & 01:30:46.6 & +21:26:25.5 & Sc       & 0.0104 & 13.54 & 0.83 & 33.3 & -19.8 &  82  & 22.0 & 130 & 31.2 & L   & (s) \\
UGC\,1087       & 01:31:26.6 & +14:16:39.0 & Sc       & 0.0149 & 14.56 & 0.69 & 34.1 & -19.5 &  74  & 21.6 &  60 & 19.1 & L   & (s) \\
UGC\,1529       & 02:02:31.0 & +11:05:35.1 & Sc       & 0.0155 & 13.67 & 1.01 & 34.2 & -20.5 & 166  & 45.4 & 132 & 17.8 & L   & (s) \\
UGC\,1635       & 02:08:27.7 & +06:23:41.7 & Sbc      & 0.0115 & 14.29 & 0.96 & 33.5 & -19.2 &  71  & 32.6 & 220 & 24.0 & L   & NC  \\
UGC\,1862       & 02:24:24.8 & -02:09:44.5 & SABc     & 0.0046 & 13.41 & 0.83 & 31.5 & -18.1 &  68  & 49.4 &  25 & 67.5 & L   & (s) \\
UGC\,3091       & 04:33:56.1 & +01:06:49.5 & SABc     & 0.0184 & 14.67 & 0.88 & 34.5 & -19.8 & 236  & 39.1 & 100 &  5.2 & L   & (s) \\
UGC\,3140       & 04:42:54.9 & +00:37:06.9 & Sc       & 0.0154 & 12.84 & 0.97 & 34.1 & -21.3 &  68  & 31.2 &  40 & 16.7 & L   &AC09 \\
UGC\,3701       & 07:11:42.7 & +72:10:09.5 & Sc       & 0.0097 & 14.76 & 0.87 & 33.3 & -18.5 &  76  &  0.0 &  85 & 22.6 & L   & (s) \\
UGC\,4036       & 07:51:54.7 & +73:00:56.5 & SABb     & 0.0116 & 12.81 & 0.98 & 33.7 & -20.9 &  73  & 24.6 & 125 & 30.6 & L   &AC12 \\
UGC\,4107       & 07:57:01.9 & +49:34:02.5 & Sc       & 0.0117 & 13.64 & 1.06 & 33.7 & -20.9 &  84  & 19.7 &  40 & 19.2 & L   & (s) \\
UGC\,5100       & 09:34:38.6 & +05:50:29.9 & SBb      & 0.0184 & 14.15 & 1.02 & 34.5 & -20.4 & 326  & 69.8 & 100 & 11.1 & I/L & (s) \\
UGC\,6410       & 11:24:05.9 & +45:48:39.9 & SABc     & 0.0187 & 14.32 & 0.78 & 34.7 & -20.3 & 185  & 44.2 & 100 & 16.5 & L   & NC  \\
UGC\,9837       & 15:23:51.7 & +58:03:10.6 & SABc     & 0.0089 & 13.65 & 0.57 & 33.2 & -19.5 &  90  & 23.2 & 145 & 30.8 & L   & (s) \\
UGC\,9965       & 15:40:06.7 & +20:40:50.2 & Sc       & 0.0151 & 14.06 & 0.72 & 34.2 & -20.1 & 137  & 22.4 &  73 & 17.8 & L   & (s) \\
UGC\,11318      & 18:39:12.2 & +55:38:30.5 & Sbc      & 0.0196 & 13.27 & 0.72 & 34.8 & -21.5 &  77  & 21.0 &  72 & 28.7 & I/L & (s) \\
UGC\,12250      & 22:55:35.9 & +12:47:25.1 & SBb      & 0.0242 & 13.78 & 1.04 & 35.1 & -21.4 & 435  & 55.8 &  73 & 26.2& I/L & NC  \\
UGC\,12391      & 23:08:57.2 & +12:02:52.9 & SABc     & 0.0163 & 13.26 & 0.77 & 34.3 & -21.0 & 112  & 27.2 &  40 & 24.9 & L   & AC2 \\
\hline
\multicolumn{13}{c}{PINGS galaxies}\\
\hline
NGC\,628        & 01:36:41.7 & +15:47:01.0 & Sc       & 0.00219 &  9.35 & 0.56 & 29.7 & -20.7 & 101  & 34.9 & 100 & 122.2 & L & AC9 \\
NGC\,1058       & 02:43:30.0 & +37:20:29.0 & Sc       & 0.00173 & 11.24 & 0.62 & 29.9 & -18.5 & 115  & 19.6 & 100 &  45.0 & L & AC3 \\
NGC\,1637       & 04:41:28.2 & -02:51:29.0 & Sc       & 0.00239 & 11.27 & 0.64 & 30.4 & -18.5 & 114  & 31.1 & 131 &  48.7 & L & AC5 \\
NGC\,3184       & 10:18:16.8 & +41:25:27.0 & SABc     & 0.00194 & 10.31 & 0.58 & 30.2 & -20.0 & 108  & 24.2 & 117 & 109.3 & L & AC9 \\
NGC\,3310       & 10:38:45.8 & +53:30:12.0 & SABb     & 0.00331 & 11.15 & 0.35 & 31.2 & -20.1 & 158  & 31.2 & 237 &  37.9 & L & AC1 \\
NGC\,4625       & 12:41:52.7 & +41:16:26.0 & SABm     & 0.00203 & 12.73 & 0.57 & 30.5 & -17.8 & 105  & 46.1 &  73 &  25.5 & L & AC4 \\
NGC\,5474       & 14:05:01.6 & +53:39:44.0 & Sc       & 0.00098 & 11.11 & 0.49 & 29.2 & -18.0 &  40  & 50.2 & 100 &  58.5 & L & AC2 \\
\hline                                   
\end{tabular}
 \end{center}

{ Notes:}
(1): Galaxy name from NED, the NASA$/$IPAC Extragalactic Database
(\url{http://nedwww.ipac.caltech.edu/}). (2)-(3): Right ascension and
declination coordinates in J2000 Equinox, expressed in units of RA
(hh mm ss) and Dec (dd mm ss). (4): Morphologycal types obtained from NED and
Hyperleda \citep[][, \url{http://leda.univ-lyon1.fr}]{patu03} following the
Third Reference Catalogue of Bright Galaxies (RC3) classification \citep[][,
\url{http://vizier.u-strasbg.fr/viz-bin/VizieR?-source=VII/155}]{corw94}. (5):
Redshift values from NED. (6): Dust corrected B-band magnitude calculated from
Hyperleda. (7): (B-V) colors obtained from Hyperleda expressed in
magnitudes. { (8) Distance modulus extracted from Hyperleda. } 
{ (9) M$_{\rm B}$: Dust corrected B magnitude extracted from
Hyperleda. } (10): Rotation velocity in units of km\,s$^{-1}$ from NED. { (11): Absolute value of the inclination angle derived from Hyperleda.} (12): Position angle values from our own analysis as described in
the text, units are degrees. (13): Effective radius from this work, units are arcsecs. (14):
Disks classification (L:\,Late, E:\,early, I:\,intermediate) according to the \cite{lauri10}
criteria. (15): Spiral arm classification following the classes
proposed by \cite{elme87}, when not avalaible we use the RC3 classification.

 \end{table*}


Previous spectroscopic studies have unveiled some aspects of the complex
processes at play between the chemical abundances of galaxies and
their physical properties. Although these studies have been successful
in determining important relationships, scaling laws and systematic
patterns (e.g. luminosity-metallicity, mass-metallicity, and surface
brightness vs. metallicity relations
\citealt{leque79,Skillman:1989p1592,VilaCostas:1992p322,zaritsky94,tremonti04};
effective yield vs. luminosity and circular velocity relations
\citealt{Garnett:2002p339}; abundance gradients and the effective
radius of disks \citealt{diaz89}; systematic differences in
the gas-phase abundance gradients between normal and barred spirals
\citealt{zaritsky94,Martin:1994p1602}; characteristic
vs. integrated abundances \citealt{moustakas06}; etc.), they
have been limited by statistics, either in the number of observed \ion{H}{ii}
regions or in the coverage of these regions within the galaxy
surface.

Hitherto, most studies devoted to the chemical abundance of
extragalactic nebulae have only been able to measure the first two
moments of the abundance distribution: the mean metal abundances of
discs and their radial gradients. Indeed, most of the observations
targeting nebular emission have been made with single-aperture or
long-slit spectrographs, resulting in a small number of galaxies
studied in detail, a small number of \ion{H}{ii} regions studied per galaxy,
and a limited coverage of these regions within the galaxy
surface. The advent of Multi-Object Spectrometers and Integral Field
Spectroscopy (IFS) instruments with large fields of view now offers us
the opportunity to undertake a new generation of emission-line
surveys, based on samples of scores to hundreds of \ion{H}{ii} regions and
full two-dimensional (2D) coverage of the discs of nearby spiral
galaxies.

In the last few years we started a major observational program to
understand the statistical properties of \ion{H}{ii} regions, and to
unveil the nature of the reported physical relations, using IFS. This
program was initiated with the PINGS survey (Rosales-Ortega et
al. 2010), which acquired IFS mosaic data of a number of medium size
nearby galaxies. In S\'anchez et al. (2011) and Rosales-Ortega et
al. (2011), we studied in detail the ionized gas and \ion{H}{ii}
regions of the largest galaxy in the sample (NGC\,628). We then
continued the acquisition of IFS data for a larger sample of visually
classified face-on spiral galaxies \citep[][ hereafter
  Paper\,I]{marmol-queralto11}, as part of the feasibility studies for
the CALIFA survey \citep{sanchez12}. The spatially resolved properties
of a typical galaxy in this sample, UGC9837, was presented by
\cite{viir12}. Face-on galaxies are more suitable to study the spatial
distribution of the properties of \ion{H}{ii} regions: (i) they are
identified and segregated more easily; (ii) their spatial distribution
is less prompt to the errors induced by inclination effects; (iii)
they are less affected by dust extinction along the line of sight
within the galaxy and (iv) it is more easy to identify their
association with a particular spiral arm.

In this article we focus on the study of the properties of the
\ion{H}{ii} regions for the face-on spiral galaxies observed so
far. In Sect. \ref{sample} we summarize the main properties of the
analyzed galaxies. In Sect. \ref{HIIanalyzer} we present the automatic
algorithm developed to detect, segregate and extract the integrated
spectra of the different \ion{H}{ii} regions within a
datacube; a comparison between the results derived with this method
and those provided with other published ones is shown in
Sect. \ref{comp}.  In Sect. \ref{spiral} we describe an analytical
method to define the presence and location of spiral arms within a
galaxy. The method has been tested and used to associate the detected
\ion{H}{ii} regions to different spiral arms and/or to the intra-arm
region. The main spectroscopic properties of the catalogued
\ion{H}{ii} regions and the morphological structure of each galaxy are
described in Sect.  \ref{elines} and \ref{struct}. The main results of
our analysis are included in Sect. \ref{result}, where we describe the
statistical properties of the \ion{H}{ii} regions (Sect. \ref{stats}),
and their radial gradients (Sect. \ref{grad}). The conclusions of this
study are presented in Sect. \ref{discuss}. In the appendix, the publicly 
accessible catalogues of the properties derived for the analyzed
\ion{H}{II} regions are described in Sect. \ref{ape}, and an empirical
correction to decontaminate the [\ion{N}{II}] emission on narrow-band
H$\alpha$ images is proposed in Sect. \ref{ape2}.

\begin{figure*}[tb]
\begin{center}
\includegraphics[width=16cm,trim=0 500 120 50,clip=true]{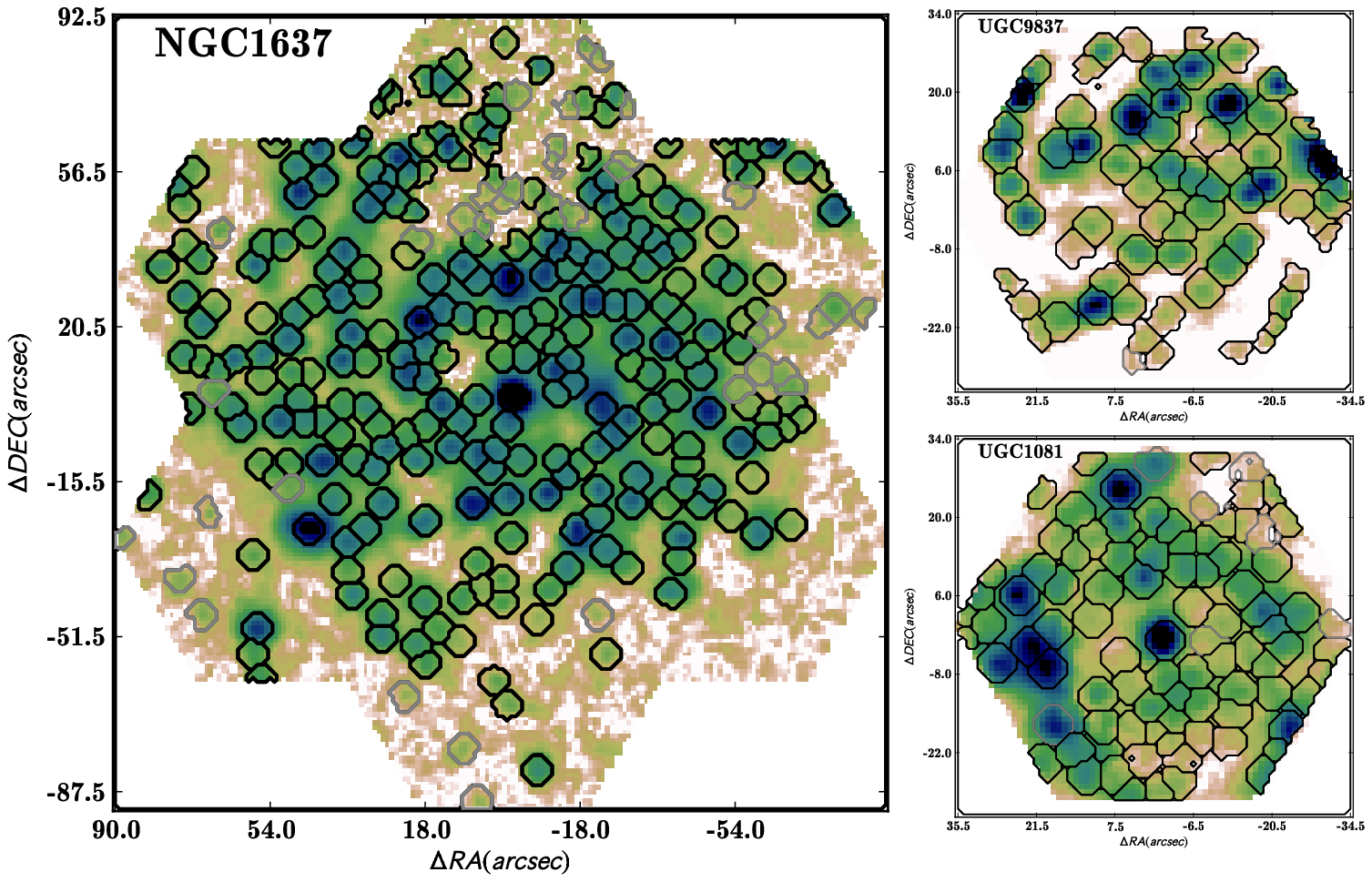}
\caption{IFS-based H$\alpha$ maps derived for three representative galaxies of the
  sample (color images), together with the detected \ion{H}{ii} regions shown as
  black segmented contours. The regions plotted in grey e
  have been removed from the final catalogue after a
  sanity-check due to low S/N, quality problems in the extracted
  spectra, and/or affected by vignetting effects.\label{fig:seg} }
\end{center}
\end{figure*}

\section{Sample of galaxies}\label{sample}

Table \ref{table.gal.prop} lists the sample of galaxies analyzed in the
current study, including for each object, its name, central
coordinates, morphological classification, redshift, and some
additional information that we will describe later. The sample of
galaxies has been selected from two available datasets: (1) the IFS
survey of nearby galaxies described in Paper\,I, which
comprises $\sim$85\% of the galaxies analyzed here, and (2) galaxies
selected from the PINGS survey \citep{rosales-ortega10}, accounting for the
remaining $\sim$15\% of the galaxies. In both cases we selected visually 
classified face-on galaxies, with a relatively unperturbed spiral structure.

The sample is dominated by late-type spirals (31 out of 38), according
to the classification criteria by \cite{lauri10}, shown in Table
\ref{table.gal.prop}. Therefore, we lack the statistics required to
analyze possible differences in the properties of \ion{H}{ii} regions
between late/early-type spirals. About 40\%\ of the galaxies show
evidence of a bar (based on its visual classification listed in Table
\ref{table.gal.prop}), although only in a few of them this feature
clearly dominates the morphology of the galaxy (e.g. UGC\,5100).
Regarding the structure of their spiral arms, the sample includes a
mix of grand-design spirals (e.g. NGC\,628), or clearly flocculent
ones \citep[e.g., UGC\,9837 ][]{viir12}. Although it is by no means a
statistically well defined sample, we consider that it is representative
of the average population of spiral galaxies in the local Universe.

Both sub-samples of galaxies were observed using similar techniques
(IFS), using the same instrument \citep[PMAS in the PPAK
  configuration, ][]{roth05,kelz06}, covering a similar wavelength range
($\sim$3700-6900\AA), with similar resolutions and integration
times. The data reduction was performed using the same procedure
\citep[R3D, ][]{sanchez06a}, as described in Paper\,I and
\cite{rosales-ortega10}. The main difference is that for the first
sample a single pointing strategy using a dithering scheme was
applied, while for the largest galaxies of the PINGS survey, a mosaic
comprising different pointings was required. This is due to the
differences in projected size, considering the different redshift range of
both samples: for the first one corresponds to $\sim$0.01-0.025,
while for the second one corresponds to $\sim$0.001-0.003. Therefore,
in both cases the covered field-of-view (FoV) corresponds to a similar optical
size, $\sim$2 effective radii, in general (the effective radius is classically defined as the radius at which one half of the total light of the system is emitted).

The observational setups allow us to cover the optical wavelength
range, sampling many of the most important emission lines for \ion{H}{ii}
regions, from [\ion{O}{ii}]~$\lambda$3727 to [\ion{S}{ii}]~$\lambda$6731. Details on the
observing strategy, setups, reduction and main
characteristics of the dataset are described in Paper\,I and
Rosales-Ortega et al. (2010). The final dataset comprises 38
individual datacubes, one per galaxy, with a final spatial sampling of
1$\arcsec$/pixel for most of the galaxies.{ The datacubes were created using 
the interpolation scheme described in \cite{sanchez12}, developed for the CALIFA survey.
Despite of the original fiber size (2.7$\arcsec$/fiber), the three pointing dithering scheme
allows to increase slightly the final resolution. The selected offsets, with values corresponding to a
fraction of the fiber-size, allows to cover the gaps between adjacent fibers too. In
average natural seeing conditions of $\sim$1$\arcsec$ \citep{sanchez07a}, this technique allows 
to provide a final spatial resolution of FWHM$\sim$2$\arcsec$ for this instrument \citep{sanchez07b}. The size of the spaxel was selected as the largest convient pixel to the sample this resolution element, 2 pixels per FWHM, i.e., 1$\arcsec$/pixel.}
Due to the large size of
the IFS mosaics of NGC\,628 and NGC\,3184, the two largest galaxies
observed with PINGS, { and the fact that they were not observed
using a dithering scheme for all the pointings },
we set the spatial sampling to
2$\arcsec$/pixels. { In this case, the final resolution is larger than the original size of the fiber, due to the interpolation kernel. A rough estimation indicate that the final spatial resolution is FWHM$\sim$3.5-4$\arcsec$. }

On average the physical spatial sampling ranges
between a few hundreds of parsecs (for the nearest galaxies) to almost
1 kpc (for the more distant ones). { To derive this physical scale it is required
to adopt a certain distance modulus. Consistently with values reported in Table \ref{table.gal.prop},
we adopted the distance modulus provided by Hyperleda, $mod$, defined as:

\begin{equation}
mod = 5 \log (D_L) + 25
\end{equation}

and $D_L$ is the luminosity distance in Mpc\footnotetext{The adopted modulus for each galaxy is included in the final catalogs, described in the Apendix \ref{ape}}.} The derived scale can be compared to the
physical diameter of a well-known \ion{H}{ii} region in our Galaxy, i.e. the
Orion nebula (D$\sim$8 pc), or to the extent of those which are
considered prototypes of extragalactic giant \ion{H}{ii} regions, such as 30
Doradus (D$\sim$200 pc), NGC\,604 (D$\sim$460 pc) or NGC\,5471 (D$\sim$1 kpc) as reported by
\cite{oey03} and \cite{rgb2011}. Thus, given the undersampling in the physical size of the \ion{H}{ii}
regions in our data, we cannot use it to derive direct estimates of the
optical extension of these regions. In other aspects, like the depth,
covered extension of the galaxy, projected resolution and wavelength coverage,
the data provided by both samples are very similar.

\begin{figure}
\centering
\includegraphics[width=8.5cm]{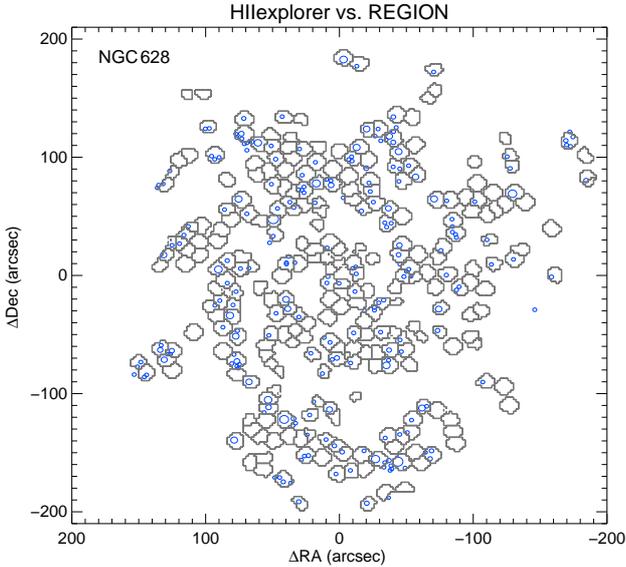}
\caption{
  Comparison between the \ion{H}{ii} region catalogue of NGC\,628 obtained by 
   {\sc HIIexplorer} using IFS data (grey contours, 286 regions) and the (modified)
   \citet{fathi07} catalogue obtained with {\sc Region} using a traditional H$\alpha$
   narrow-band image (blue circles, 180 regions). Only those regions within
   the FoV of the IFS data have been considered.
  \label{fig:region}}
\end{figure}

\begin{table}
 \caption{Detected HII regions within our sample.}
 \label{table.HII}
 \begin{center}
 \begin{tabular}{lrrrrrrr}        
 \hline\hline                 
Galaxy & N$_{\rm HII}$ & N$_{\rm HII}^*$ &  N$_{\rm H\beta}$ & F$_{\rm H\alpha}$ & $\sigma_{\rm H\alpha}$ & r$_{\rm HII}$ & $\sigma_{\rm r}$\\
 (1)  &    (2)       & (3)              & (4)     & (5) & (6) & (7)  & (8)       \\
\hline                    
2MASXJ1319+53            & 19 & 18& 14 &1.25& 0.93& 16.48& 1.42\\
CGCG\,071-096            & 27 & 27& 22 &1.65& 1.70& 15.38& 1.45\\
CGCG\,148-006            & 40 & 29& 26 &1.52& 1.33& 14.15& 1.37\\
CGCG\,293-023            & 19 & 18&  8 &1.36& 1.25& 10.29& 1.02\\
CGCG\,430-046            & 33 & 24& 23 &1.83& 1.58& 15.33& 0.96\\
IC\,2204                 & 59 & 59& 30 &1.57& 1.74& 10.17& 0.95\\
MRK\,1477                & 8  & 8&   7 &2.46& 2.77& 14.52& 0.68\\
NGC\,99                  & 63 & 63& 62 &1.77& 1.89& 11.43& 1.01\\
NGC\,3820                & 18 & 16& 15 &1.62& 1.47& 12.56& 1.04\\
NGC\,4109                & 15 & 15& 15 &1.75& 1.75& 15.68& 0.85\\
NGC\,7570                & 50 & 50& 27 &1.65& 2.04& 10.25& 0.78\\
UGC\,74                  & 91 & 78& 65 &1.28& 1.03& 8.01& 0.79\\
UGC\,233                 & 49 & 49& 26 &1.89& 2.02& 10.99& 1.20\\
UGC\,463                 & 85 & 83& 80 &1.79& 1.77& 9.14& 0.87\\
UGC\,1081                & 90 & 87& 81 &1.29& 1.32& 6.25& 0.92\\
UGC\,1087                & 82 & 81& 76 &1.21& 0.92& 8.91& 0.92\\
UGC\,1529                &116 & 77& 52 &1.65& 1.34& 9.19& 0.94\\
UGC\,1635                & 85 & 84& 84 &0.91& 0.54& 6.64& 0.72\\
UGC\,1862                & 56 & 55& 53 &1.19& 1.06& 2.69& 0.20\\
UGC\,3091                & 68 & 66& 61 &1.09& 0.73& 11.08& 1.18\\
UGC\,3140                & 87 & 86& 86 &1.62& 1.51& 9.28& 0.96\\
UGC\,3701                & 80 & 69& 53 &1.46& 1.24& 6.55& 0.51\\
UGC\,4036                &104 &104& 79 &1.67& 1.73& 7.48& 0.84\\
UGC\,4107                & 68 & 68& 61 &1.45& 1.41& 7.65& 0.68\\
UGC\,5100                & 28 & 28& 20 &1.61& 1.71& 11.89& 1.08\\
UGC\,6410                & 62 & 61& 60 &1.33& 1.20& 11.98& 1.02\\
UGC\,9837                & 65 & 64& 64 &1.37& 1.41& 5.95& 0.85\\
UGC\,9965                & 68 & 67& 65 &1.55& 1.45& 9.81& 0.91\\
UGC\,11318               & 76 & 75& 62 &1.56& 1.41& 12.53& 1.65\\
UGC\,12250               & 81 & 41& 21 &1.46& 1.12& 14.71& 1.06\\
UGC\,12391               & 91 & 84& 71 &1.54& 1.45& 10.21& 0.89\\
\hline
\multicolumn{7}{c}{PINGS}\\
\hline
NGC\,628                 & 373 & 366& 282 &2.70& 2.89& 2.87& 0.33\\
NGC\,1058                & 331 & 258& 179 &2.69& 2.83& 1.47& 0.11\\
NGC\,1637                & 297 & 297& 251 &2.46& 2.89& 1.35& 0.14\\
NGC\,3184                & 169 & 169& 124 &2.86& 2.94& 3.31& 0.36\\
NGC\,3310                & 203 & 130& 121 &3.93& 4.25& 2.53& 0.33\\
NGC\,4625                &  66 &  49&  46 &3.10& 3.11& 1.83& 0.25\\
NGC\,5474                & 122 & 121&  95 &2.74& 2.91& 1.13& 0.14\\
\hline                    
 \end{tabular}
 \end{center}

{ Notes}:
(1) Name of the galaxy used along this article; (2) number of detected HII
regions; (3) number of HII regions with good quality spectra, as described in
the text; { (4) number of HII regions with H$\beta$ emission line detected at 3$\sigma$ significance}; (5) median value of the H$\alpha$ intensity derived from the
narrow-band images in \Funits; (6) standard deviation of the H$\alpha$
intensity in the same; (7) median value of the estimated radii of the HII
regions in units of 100 pc (we need to note that the size derived by {\sc
  HIIexplorer }
is an ill-defined parameter for our dataset) ; (8) standard deviation of the
previous estimated radii, in the same units.

\end{table}


\section{Extraction of the \ion{H}{ii} regions}\label{HIIanalyzer}

The segregation of \ion{H}{ii} regions and the extraction of the corresponding
spectra is performed using a semi-automatic procedure, named 
{\sc HIIexplorer}\footnote{\url{http://www.caha.es/sanchez/HII_explorer/}}. 
The procedure is based on some basic assumptions: (a) \ion{H}{ii} regions are
peaky/isolated structures with a strong ionized gas emission, clearly
above the continuum emission and the average ionized gas emission
across the galaxy; (b) \ion{H}{ii} regions have a typical physical size of
about a hundred or a few hundreds of parsecs \citep[e.g.][]{rosa97,lopez2011,oey03}, which
corresponds to a typical projected size at the distance of the
galaxies of a few arcsec.

The algorithm requires a set of input parameters: (i) a
line emission map, with the same world-coordinate system (WCS) and
resolution as the input datacube (preferentially an H$\alpha$ emission
line map); (ii) a flux intensity threshold for the peak emission of
each \ion{H}{ii} region; (iii) a maximum distance to the peak location for a
pixel associated with each \ion{H}{ii} region; (iv) a relative threshold with
respect to this peak emission of the minimum intensity of each \ion{H}{ii}
region; (v) an absolute threshold of the minimum intensity
corresponding to each \ion{H}{ii} region. All these parameters can be derived
from either a visual inspection and/or a statistical analysis of the
H$\alpha$ emission line map. The algorithm produces as an output a
segmentation FITS file describing the pixels associated to each \ion{H}{ii}
region, designated with a running index starting with 1 (e.g. the primary \ion{H}{ii}
region ID in this article), with the zero reserved to areas not
associated with any \ion{H}{ii} region (i.e. regions free of emission or below
the absolute threshold described above).

The segregation algorithm is based on a simple iterative procedure,
summarized in the flow chart shown in Fig. \ref{fig:chart}. As a
first step the algorithm looks for the brightest pixel within the 
emission line map. 
Its location is stored as the peak/central coordinate of a new \ion{H}{ii} region,
associated with a certain running index (ID number). After this, the adjacent
pixels are aggregated to this \ion{H}{ii} region if all of the following criteria
are fulfilled: (i) the distance to the central/peak pixel is below the
selected limit; (ii) the flux at the pixel is above the
relative threshold with respect to the peak emission; (iii) the flux
at the  pixel is above the absolute flux threshold described
before. Whenever any of these criteria are not fulfilled
the aggregation procedure stops, the ID number is increased by one,
all the aggregated pixels are masked-off, and the peak-identification
procedure is repeated. The overall procedure stops whenever no new
peak is detected above the selected peak-intensity threshold.

The outcome of the procedure is illustrated in Fig. \ref{fig:seg} where we
show: (i) the input emission line map, in this case the H$\alpha$ map
corresponding to a set of galaxies and (ii) the corresponding derived
segmentation map. Despite the simplicity of the described algorithm
it is clearly seen that (1) it is able to detect all the \ion{H}{ii} regions
that can be identified by eye, and (2) it produces a reliable
segmentation map. The black segmented areas indicate those regions with
good quality spectra, while the grey ones indicate those with poor extracted
spectra. The actual procedure to detect and reject those ones is described
later. In addition, the procedure provides with a mask where all the \ion{H}{ii} 
regions are flagged out. This mask is important to define the areas where it is 
possible to study the diffuse gas emission.

\begin{figure*}[tb]
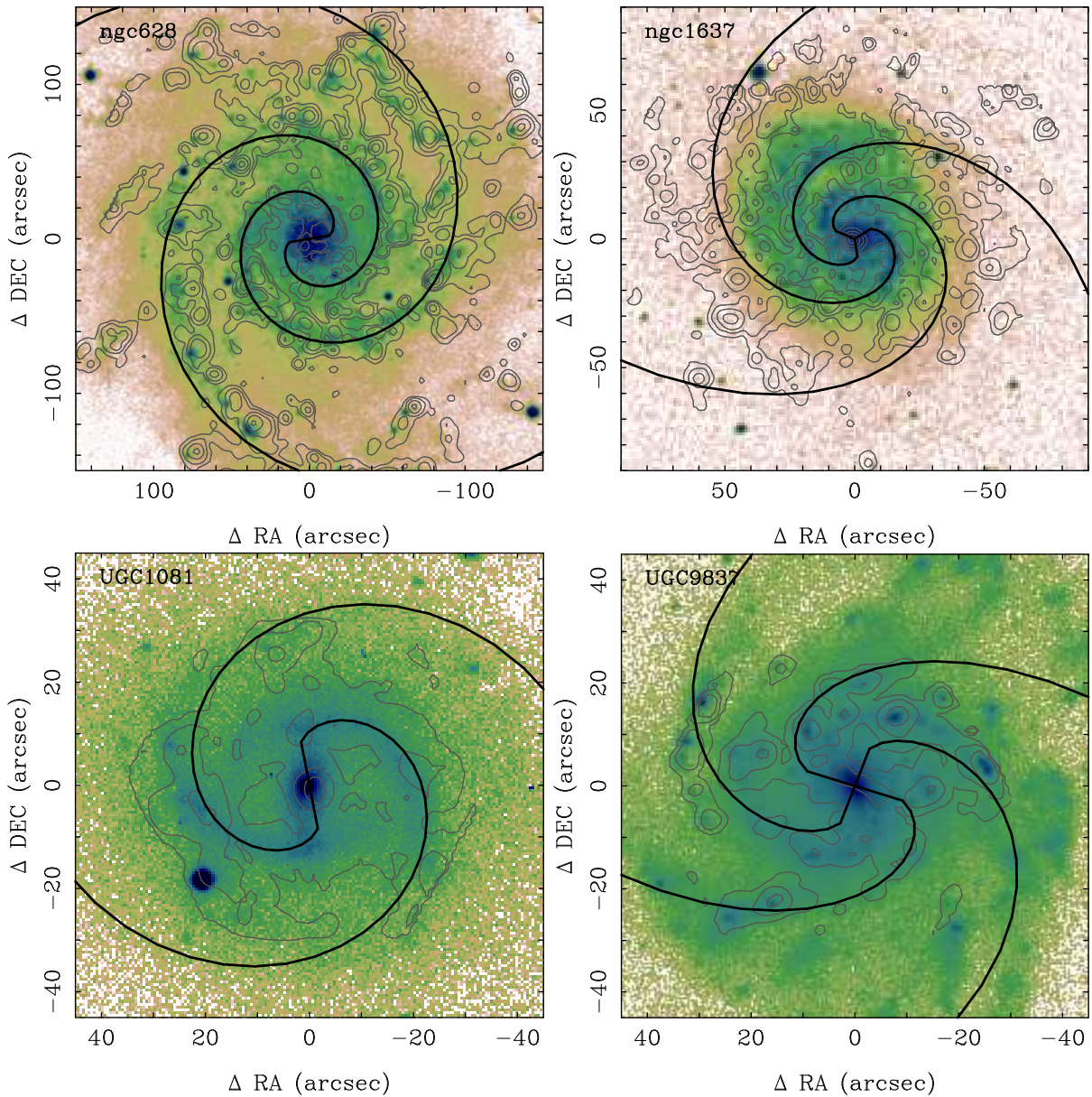

\begin{center}
\includegraphics[width=8cm,angle=270,clip=true]{figs/spiral.img.ngc628.ps}
\includegraphics[width=8cm,angle=270,clip=true]{figs/spiral.img.ngc1637.ps}
\includegraphics[width=8cm,angle=270,clip=true]{figs/spiral.img.UGC1081.ps}
\includegraphics[width=8cm,angle=270,clip=true]{figs/spiral.img.UGC9837.ps}
\caption{Continuum map of four representative galaxies of the full sample
  (color images), together with a contour plot of the H$\alpha$ intensity map
  (both in arbitrary units). The solid lines show the location of the
  identified spiral arms as a result of the fitting procedure. Left side
  panels show galaxies with clearly defined spiral arms, while right side
  panels show galaxies with poorly identified ones. \label{fig:spiral} }
\end{center}
\end{figure*}

There are other publicly accessible packages for the automatic
selection/segregation of \ion{H}{ii} regions in the 
literature (e.g. {\sc HIIphot}, \citealt{thilk00}; {\sc Region}, \citealt{fathi07}), 
that in principle could be adapted
for the main purpose of the current study. However, these packages are
strongly focused on the analysis of narrow-band images, of much higher
spatial resolution, where the \ion{H}{ii} regions are {\it clearly} resolved.
In some cases the procedure requires a detailed
knowledge of the observational procedure (number of frames co-added,
ADU of the CCD, etc.). For example, {\sc HIIphot} uses object recognition
techniques to make a first guess at the shapes of all sources and then allows
for departure from such idealized seeds through an iterative growing
procedure. In essence, this algorithm is similar to the one used by
{\sc SExtractor} \citep{bert96}, for the detection and segregation of
galaxies in crowed fields. We experimented with these packages before
developing our own code, but we did not get any optimal solution. The
main reasons were that (1) our data have a much coarser resolution than
the one provided by narrow-band imaging (even from ground-based
telescopes); (2) reconstructed IFU map have a strong
cross-correlated noise among nearby interpolated pixels and (3) none
of the preceding codes provide a final segmentation map usable to
extract the integrated spectra of the \ion{H}{ii} regions from the datacubes
in a convenient way.

We experimented with the use of {\sc HIIexplorer} on a H$\alpha$
narrow-band image provided by the SINGS legacy survey for NGC\,628
\citep{kenn03}. A visual inspection of the selected regions with those
shown by \cite{thilk00}, indicates that although we detect similar
regions, {\sc HIIexplorer} tends to define regions of mostly {\it equal} 
size. This is expected, since for the spatial
resolution the maximum size allowed for each region is reached before
that imposed by the ratio of local to peak intensity. Our code was never 
meant to provide a particularly reliable measure of the projected size, 
as in our data this parameter is ill-defined. Specifically, \ion{H}{ii} regions 
can be significantly smaller than the resolution element size. 
In Sect. \ref{comp} we present a quantitative 
comparison with methods available in the literature.

Once tested the procedure, we applied it to our IFS data. First, we
create a H$\alpha$ intensity map for each object by co-adding the flux
intensity within a square-shaped simulated filter centered at the wavelength of
H$\alpha$ (6563\AA), with a width of 60~\AA. The adjacent continuum for each pixel
was derived by averaging the flux intensity within two similar bands red-
and blue-shifted 100~\AA\ from the center of the initial one. This continuum
intensity is then subtracted from the H$\alpha$ intensity to derive a continuum-subtracted 
emission line map. The central wavelength of all these bands has been
shifted to the observed frame taking into account the redshift of the
object. The separations between the filters and the filter
widths are large enough to avoid any possible error in the derivation of
the H$\alpha$ intensity map due to kinematic shifts.

\begin{table}
 \caption{Derived parameters for the modelled spiral arms.}
 \label{table.results}
 \begin{center}
 \begin{tabular}{lccrrr}        
 \hline\hline                 
Galaxy & QF & N$_{\rm A}$ & A & B & C\\
 (1)    &    (2)       & (3)                & \multicolumn{3}{c}{(4)}\\
\hline                    
2MASXJ1319+53          & 1& 2&  10&  3.0& 11.0\\
CGCG\,071-096             & 1& 2&  26&  1.0&  6.7\\
CGCG\,148-006            & 1& 2&   9&  1.0&  6.1\\
CGCG\,293-023            & 0& 2&  27&  1.0&  7.7\\
CGCG\,430-046            & 0& 2&   6&  1.0&  4.7\\
IC\,2204                 & 1& 2&  18&  1.5&  7.7\\
MRK\,1477                & 0& 2&  26&  0.2&  1.6\\
NGC\,99                  & 1& 2&  20&  1.0&  4.5\\
NGC\,3820                & 0& 4&  17&  1.0&  2.6\\
NGC\,4109                & 0& 2&   8&  1.0&  5.0\\
NGC\,7570                & 1& 2&  50&  0.9&  9.0\\
UGC\,74                  & 1& 2&  22&  1.0&  4.2\\
UGC\,233               & 1& 2&  22&  1.0&  6.7\\
UGC\,463                 & 0& 3&  26&  1.0&  3.6\\
UGC\,1081                & 1& 2&  36&  1.0&  5.3\\
UGC\,1087                & 0& 2&  20&  1.0&  6.0\\
UGC\,1529                & 0& 2&  13&  1.0&  4.2\\
UGC\,1635                & 0& 2&  16&  1.0&  5.2\\
UGC\,1862                & 1& 2&  56&  1.0&  7.7\\
UGC\,3091                & 0& 5&  29&  1.0&  1.4\\
UGC\,3140                & 1& 2&  26&  1.0&  4.7\\
UGC\,3701                & 0& 2&  20&  1.0&  2.9\\
UGC\,4036                & 1& 2&  27&  1.0&  3.1\\
UGC\,4107                & 1& 3&  20&  1.0&  4.2\\
UGC\,5100               & 1& 2&  55&  2.0& 16.6\\
UGC\,6410               & 1& 2&  27&  1.0&  6.7\\
UGC\,9837                & 0& 4&  27&  1.0&  2.6\\
UGC\,9965                & 0& 2&  29&  1.0&  5.3\\
UGC\,11318               & 1& 2&  25&  1.0&  5.8\\
UGC\,12250               & 1& 2&  17&  1.0&  6.2\\
UGC\,12391               & 0& 2&  30&  1.0&  4.5\\
\hline
NGC\,628                 & 1& 2&  58&  5.0& 44.0\\
NGC\,1058                & 0& 2&  50&  1.0&  4.1\\
NGC\,1637                & 0& 3&  26& 15.0& 54.0\\
NGC\,3184                & 1& 2& 170& 15.0& 70.0\\
NGC\,3310                & 1& 2&  14&  1.1&  2.5\\
NGC\,4625                & 0& 1&  23&  1.0&  9.7\\
NGC\,5474                & 0& 2&  60&  1.0&  7.6\\
\hline                    
 \end{tabular}
 \end{center}

{ Notes}:
(1) Galaxy name used along this article; (2) quality flag of the
 analysis of the spiral arms: 1\,=\,well defined arms, 0\,=\,arms not
 well-defined; (3) number of arms detected with our modelling; 
 (4) {\it A, B, C} parameter of the spiral model, described in
 Sect. \ref{spiral}, equation \ref{eq_spiral} by \cite{ring09}. $A$ is in units of arcsec, $B$ is a dimensionless
 parameter, and $C$ is in units of radians$^{-1}$.
\end{table}


However, this H$\alpha$ intensity map is contaminated with the adjacent [\ion{N}{ii}]
emission lines, and it is not corrected for the emission of the 
underlying stellar population (see Appendix\ref{ape2} for a discussion on the topic).
A cleaner H$\alpha$ emission map can be derived
using emission-line/stellar population decoupling procedures 
\citep[e.g.,][]{rosales-ortega10,sanchez11,marmol-queralto11,sanchez12}. This
allow us to recover much fainter emission line regions on top of
underlying strong absorption features. However, in most cases these
emission line regions do not correspond with classical \ion{H}{ii} regions
and are associated with other ionization
processes \citep[e.g.][, and references there in]{kehrig12}. On the
other hand, the adopted procedure resembles as much as possible the
classical procedure used to detect \ion{H}{ii} regions, which will allow us to
make a better comparison with previous results.

As the main goal of the current study is to extract the spectroscopic
properties of the \ion{H}{ii} regions, we
applied {\sc HIIexplorer} adopting the following input parameters for
all the galaxies: (i) a minimum flux density for the peak
intensity of an \ion{H}{ii} region of 2~\fuden; (ii) a minimum
relative flux to the peak intensity for associated pixels
corresponding to the same \ion{H}{ii} region of 10\% and (iii) a
maximum distance to the location of the peak of 3.5$\arcsec$
(7$\arcsec$ for NGC\,628 and NGC\,3184). The maximum distance was selected using and iterative process, maximizing the number of detections when compared with a visual inspection, and do not allow to segment clearly single \ion{H}{ii} regions. Then, we extracted a single
spectrum for each region by co-adding all the spectra in the original
cubes with the same identification index (ID) in the the derived
segmentation map. The final spectra are stored in the so-called
row-stacked spectra (RSS) format \cite{sanchez04}, which comprises a
2D spectral image, with the spectrum corresponding to each \ion{H}{ii}
region order by rows (each one corresponding to the considered ID),
and an associated position table which records the barycenter of each
\ion{H}{ii} region (based on the H$\alpha$ intensity). This format
allows us to visualize individual spectra of \ion{H}{ii} regions and 
their spatial distribution using standard techniques \citep[e.g., E3D,][]{sanchez04}. 
The ID is a unique
identification index that will be used to identify the \ion{H}{ii}
regions hereafter (including the tables,
figures and on-line material).

Table \ref{table.HII} summarizes the results from this analysis. It
shows, for each galaxy, the number of \ion{H}{ii} regions detected,
the number of regions with good quality extracted spectra, the median
H$\alpha$ flux intensity as directly measured from the IFS-based
narrow-band image, and its standard deviation. Due to the different
redshift range, there is a wide variance in the median flux intensity
of the H$\alpha$ emission, much larger than the absolute luminosity as
we will describe later. Since we are selecting \ion{H}{ii} regions
which physical size is slightly smaller than our typical resolution
element, this implies that we are actually not aggregating a large
number of {\it real} \ion{H}{ii} regions in each detected complex. Our
comparisons with higher resolution narrow-band images confirms this
suspicion. Finally, we have included in the table the median radius
of the regions, defined as R=$\sqrt{A/\pi}$, where A is the area
within a region (Rosales-Ortega et al. 2011). We should state clearly
here that the physical scale is an ill-defined parameter in our
survey, due to two reasons: 1) the coarse spatial sampling compared to
the expected size of \ion{H}{ii} regions, these can be significantly
smaller than the resolution element size and 2) the adopted procedure
to detect and segregate the regions, namely the introduction of an
angular upper size limit to the continuous emission region.  Only for
the galaxies at lower redshift, the sizes of the \ion{H}{ii} regions
are of the order of the expected one, i.e. $\sim$100 pc.

In practice, our segregated \ion{H}{ii} regions may comprise several classical
ones, in particular for the more distant galaxies. Detailed simulations on
the effect of resolution loss have shown us that on average each selected
region at $z\sim$0.02 may comprise 1-3 \ion{H}{ii} regions from the ones selected from
low redshift galaxies, $z\sim$0.002 (Mast et al., in prep.).
{ Following \cite{lopez2011}, the considered \ion{H}{ii} 
regions would have a size of a few to several hundreds of parsecs, based on their 
H$\alpha$ luminosity, detailed in Section \ref{stats}, Table \ref{table.med.val}. Thererefore,
the results from the simulations are expected, due to the typical size of an extragalactic \ion{H}{ii} region and the lose of physical resolution at the higher redshifts. }
 i.e. we are
selecting \ion{H}{ii}-regions and/or \ion{H}{ii} aggregations (note that throughout this paper
we will refer indistinctively to these segmented regions as {\em \ion{H}{ii} regions}). 
Therefore, our results are not useful to analyze additive/integrated
properties on individual \ion{H}{ii} regions, like the H$\alpha$ luminosity function,
but are perfectly suited for the study of line ratios, chemical abundances and
ionization conditions.

In total, we have detected 3107 \ion{H}{ii} regions, 2573 of them with
good spectroscopic information. To our knowledge this is by far the
largest 2-dimensional, nearby spectroscopic \ion{H}{ii} region survey ever
accomplished.

\section{Comparisons with previous selection methods to detect \ion{H}{ii} regions}\label{comp}

In order to assess quantitatively the degree of segmentation provided
by {\sc HIIexplorer} with respect to other traditional \ion{H}{ii}
region catalogues generators, we performed a comparison between the
\ion{H}{ii} region catalogues of NGC\,628, 
{ NGC\,3184 and NGC\,5474} obtained with {\sc
  HIIexplorer} { and those reported in the literature}.

{ For NGC\,628 we used the} H$\alpha$ image extracted from the IFS data
with a resolution of 2 arcsec/pixel and a FoV of $\sim6$ arcmin, and
the one produced by the {\sc REGION} software in \citet{fathi07}
(hereafter Fa07), obtained from a narrow-band H$\alpha$ image with a
resolution of 0.33 arcsec/pixel and a FoV of $\sim11$ arcmin. Similar
comparisons could be performed with any other package created to
segregate \ion{H}{ii} regions \cite[e.g. {\sc
    HIIPhot},][]{thilk00}. We selected this one because Fa07 provided
publicly accessible catalogues.

The full Fa07 catalogue has 376 regions of which 299 are within the
FoV of our IFS data. However, the public Fa07 catalogue reports only
the position and the full-width at half maximum (FWHM) of each
\ion{H}{ii} region, not the actual shape obtained by the software, and
this leads to significant overlaps when the {\sc REGION} catalogue is
plotted over the galaxy image. Taking this into account and
considering the difference in resolution between the two H$\alpha$
images, we created a modified version of the Fa07 catalogue in order
to make a fair comparison.  The modified Fa07 catalogue was obtained
in an iterative way. First we took the first region of the catalogue
and calculated its distance from the rest of the regions. Those
regions for which the distance was less or equal to the sum of their
radii were considered as a single region. In this case, the involved
regions are removed from the original catalogue and a new entry is
added with coordinates and size corresponding to the luminosity
weighted mean of the merged regions, the process is repeated for the
rest of the catalogue entries in an iterative manner. We obtain 180
regions in the modified Fa07 catalogue of NGC\,628.

Fig. \ref{fig:region} shows the comparison between the modified Fa07
catalogue and the {\sc HIIexplorer} segmentation map. The blue circles
correspond to the modified Fa07 catalogue, while the grey contours to
the 286 segmented regions obtained by {\sc HIIexplorer} for NGC\,628
based on the IFS data.  We note that, a) {\sc HIIexplorer} detects and
segments more regions than Fa07, except for those cases in which the
difference in spatial resolution (0.33 vs. 2 arcsec/pixel) prevents
further segmentation; b) There is a nearly 1:1 correspondence of
regions detected in Fa07 with respect to {\sc HIIexplorer}, the
incompleteness of Fa07 with respect to {\sc HIIexplorer} is 5\%; c)
19\% of the regions in {\sc HIIexplorer} have 2 or more regions of the
modified Fa07 catalogue, which is simply due to the difference in
resolution. We have checked visually the extracted spectra of the
additional \ion{H}{ii} regions detected by our algorithm, and
inspected the original narrow band image and they seem to be real
\ion{H}{ii} regions, clearly distingued from the low surface
brightness diffuse gas. The performance of {\sc HIIexplorer} compared
with {\sc REGION} is remarkable, considering both that the narrow-band
H$\alpha$ image used to generate the Fa07 catalogue is deeper than the
image extracted from the IFS data, and that {\sc HIIexplorer} runs in
a completely automated way.

{ The \ion{H}{ii} regions of NGC\,3184 and NGC\,5474 were studied by 
\cite{brad2006} using REGION (hereafter B06). For NGC\,3184, the catalogue obtained 
by B06 contains 576 \ion{H}{ii} regions of which 209 are within the
FoV of our IFS data. Like the Fathi et al. (2007) case for NGC628, the
B06 catalogues report only the offset from the galaxy centre and the
total area of the region, not the actual shape obtained by the
software, which leads to significant overlaps when the REGION B06
catalogue is plotted over a RA vs. Dec plane using an effective radius
derived from the B06 catalogue. Therefore, we applied the same
methodology for a fair comparison obtaining a modified B06 catalogue
for this galaxy, imposing a H$\alpha$ luminosity
threshold of $\log$(L$_{\mathrm{H}\alpha}$) $>$ 37.96 erg~s$^{-1}$ (the minimum luminosity detected
by {\sc HIIexplorer} at this redshift). The level of completeness is 73\%, i.e. regions detected by
{\sc HIIexplorer} with respect to B06 (note that in the majority of cases
there is a 1:1 correspondence); in 15 cases 2 or more B06 regions are
found within 1 segmented area by {\sc HIIexplorer}. However, in 5 cases 2
\ion{H}{ii} regions by {\sc HIIexplorer} correspond to 1 region found by B06, while
13 regions detected by {\sc HIIexplorer} are not present in the B06
catalogue.

In the case of NGC\,5474, the original B06 catalogue contains 165  \ion{H}{ii}
regions, of which 98 are within the FoV of the IFS data. For this
galaxy, we worked directly with the published catalogue without
further modifications for a better visual comparison. There was no
need to apply a luminosity threshold since all the regions were above
the minimum luminosity observed by the regions segmented by
{\sc HIIexplorer}, $\log$(L$_{\mathrm{H}\alpha}$) $\sim$ 36.6 erg~s$^{-1}$. We note that {\sc HIIexplorer} 
detects and segments more regions than B06,
except for those cases in which the difference in spatial resolution
prevents further segmentation. The level of completeness (regions
detected by {\sc HIIexplorer} compared to the B06 catalogue) is of 90\%
(including 1:1 correspondence and multiple B06 HII regions within one
{\sc HIIexplorer} segmentation), but interestingly 31 regions detected by
{\sc HIIexplorer} are not found in the B06 catalogue, which is surprising
given the that the H$\alpha$ image used to generate the B06 catalogue is
deeper than the H$\alpha$ map extracted from the IFS data.

}

This exercise shows that {\sc HIIexplorer} is capable of performing an
excellent \ion{H}{ii} region extraction for the resolution of our IFS data, 
and that the generated catalogues are comparable (and even
more efficient) than those generated in a traditional way based on
narrow-band H$\alpha$ imaging.

\begin{table*}
  \caption{General properties derived for the 10 brighest HII regions within UGC9837.}
 \label{table.coords}
 \begin{center}
 \begin{tabular}{lccrrrrrrccrrrrrrr}
 \hline\hline                 
 ID &
 RA &
 Dec&
 X$_{\mathrm{obs}}$&
 Y$_{\mathrm{obs}}$&
 X$_{\mathrm{res}}$&
 Y$_{\mathrm{res}}$&
 R&
 $\theta$&
 N$_a$&
 F&
 D$_{\mathrm{arm}}$&
 D$_{\mathrm{sp}}$&
 $\theta_{sp}$&
 $v_{\mathrm{rot}}$&
 L$_{\rm H\alpha}$\\
(1) & 
(2) & 
(3) & 
(4) & 
(5) & 
(6) & 
(7) & 
(8) & 
(9) & 
(10) &
(11) &
(12) & 
(13) & 
(14) &
(15) &
(16) \\
\hline
 UGC\,9837-001       &  230.9689&  58.0583&     24.0&    19.3&      0.4&     6.8&      6.8&    86.5&   2&  1&     2.5&   244.2&     4.6&   2635.0&  39.75\\
 UGC\,9837-002       &  230.9611&  58.0549&    -29.7&     6.9&      4.6&    -4.5&      6.5&   315.6&   4&  1&     1.9&   123.2&    -3.0&   2715.1&  39.56\\
 UGC\,9837-003       &  230.9636&  58.0579&    -12.5&    17.9&      4.5&     0.0&      4.5&     0.0&   4&  1&     6.0&    29.5&   -13.4&   2723.2&  39.42\\
 UGC\,9837-004       &  230.9661&  58.0574&      4.6&    16.0&      2.1&     2.9&      3.6&    53.3&   3&  1&     0.9&    66.2&     2.9&   2680.2&  39.35\\
 UGC\,9837-005       &  230.9650&  58.0580&     -2.5&    18.1&      3.3&     1.8&      3.8&    29.1&   3&  1&     4.7&   121.9&     4.6&   2694.1&  39.24\\
 UGC\,9837-006       &  230.9670&  58.0479&     10.9&   -18.3&     -4.3&    -0.3&      4.3&   184.6&   1&  1&     3.0&   220.5&     5.9&   2602.2&  39.17\\
 UGC\,9837-007       &  230.9673&  58.0559&     13.1&    10.8&      0.3&     3.7&      3.8&    85.8&   3&  0&     7.1&    35.0&   -20.5&   2639.4&  39.27\\
 UGC\,9837-008       &  230.9625&  58.0540&    -20.1&     4.0&      3.0&    -3.1&      4.4&   313.8&   4&  0&     6.4&    78.4&    11.7&   2702.2&  38.96\\
 UGC\,9837-009       &  230.9623&  58.0588&    -21.0&    21.1&      6.0&    -1.1&      6.1&   349.3&   3&  1&     6.3&   220.5&     7.9&   2737.6&  38.99\\
 UGC\,9837-010       &  230.9644&  58.0596&     -6.6&    24.0&      4.8&     1.9&      5.1&    21.3&   3&  1&     1.6&   143.2&     3.8&   2712.7&  39.13\\
\hline                                   
 \end{tabular}
 \end{center}

{ Notes}:
(1) Unique ID of the HII region; (2) Right ascension of the HII region, in
degrees, for the J1200 equinox; (3) Declination of the HII region, in degrees,
for the J1200 equinox; (4) Relative right ascension from the center of the
galaxy, in arcsecs; (5) Relative declination from the center of the galaxy, in
arcsecs; (6) Deprojected X coordinate from the center, in kpc; (7) Deprojected
Y coordinate from the center, in kpc; (8) Deprojected distance to the center,
in kpc; (9) Deprojected position angle, in degrees; (10) ID of the nearest
spiral arm, used to associate a HII region with a particular arm; (11) Flag
indicating if the HII region is clearly associated to the corresponding spiral
arm (1) or not (0); (12) Minimum distance to the nearest spiral arm, in
arcsec; (13) Spiralcentric distance, i.e. distance along the nearest spiral
arm from the center, in arcsec; (14) Angular distance to the nearest spiral
arm, in degrees; { (15) Velocity of the ionized gas derived the fitting to the
H$\alpha$ emission line of the HII region, in km/s; (16)  
Decimal logarithm of the dust corrected absolute luminosity of H$\alpha$, in units of erg~s$^{-1}$.
{ The full catalogue of \ion{H}{ii} regions for this object and the remaining ones discussed along this article are listed in Appendix \ref{ape}, including the errors of the velocity and the H$\alpha$ luminosity.}
}

 \end{table*}


\section{Analytical characterization of the spiral arms}\label{spiral}

A fundamental question regarding the star-forming regions in galaxies
is whether their distribution and properties depend on their association
(or not) with a particular spiral arm. Two main questions are directly
connected with this one: (i) whether there are azimuthal variations 
within the spectroscopic properties of the \ion{H}{ii} regions, which would 
possibly reflect non-radial differences in the galaxies evolution, maybe induced by non secular processes, and (ii) whether the properties of the
inter-arm \ion{H}{ii} regions are different than those of the \ion{H}{ii} intra-arms
ones, which will reflect a possible differential evolution associated
with ram pressure in the spiral arms. The lack of a sample with a 
statistically large number of \ion{H}{ii} regions, with homogeneously 
derived spectroscopic properties, and with a good characterization 
of the structure of the spiral arms has not allowed to give a conclusive 
answer to these questions so far.

In the following we attempt to give a good description of the structure (number, shape, radial 
path) of the spiral arms and to define a procedure to associate \ion{H}{ii} 
regions with each spiral arm and/or classify them as inter-arm ones. We adopt the prescription proposed by
\cite{ring09} to describe the general shape of the spiral arms. This formalism
describes the radial path of any spiral arm using the formula:

\begin{equation}
r(\theta)= \frac{A}{{\rm log} ({B}\\ {\rm tan} (\theta/2{C}))}
\end{equation}\label{eq_spiral}

This function intrinsically generates a bar in a continuous, fixed
relationship relative to an arm of arbitrary winding sweep. $A$ is
simply a scale parameter while $B$, together with $C$, determines the
spiral pitch. Roughly, larger $C$ results in tighter winding. Greater
$B$ results in larger arm sweep and smaller bar/bulge, while smaller
$B$ fits larger bar/bulge with a sharper bar/arm junction. Thus $B$
controls the ``bar/bulge-to-arm'' size, while $C$ controls the tightness
much like the Hubble scheme. Special shapes such as ring galaxies with
inward and outward arms are also described by the analytic
continuation of the same formula, which is particularly useful to
analyze the diversity of spiral structures within our sample.

The previous formula describes the radial path in the physical plane
of the disk for each spiral arm. To describe the full observed spiral
structure, for a galaxy with $N_A$ arms, it is required to project it at
the observed plane (taking into account the inclination and position
angle), and to add $N_A$ copies of the considered arm, rotated by
an angle of 360/$N_A$ degrees with respect to the precedent one.

The optimal parameters that describe the current spiral structure
($N_A$, $A$, $B$ and $C$) have been derived using an interactive
fitting algorithm that is based on two simple assumptions: (1) the
spiral arms trace the location of the stronger H$\alpha$ emission and
continuum emission. i.e., the integrated intensity along the arm
should be maximized; and (2) the \ion{H}{ii} regions are more
frequently clustered around the spiral arms. i.e., the distance from
each region to the nearest spiral arm should be minimized. The
analytical parameters of the spiral arms are then derived based on
these assumptions and using as inputs (i) a broad band image of the
galaxy. The SDSS $g$-band image in most of the cases \citep[extracted from the SDSS imaging survey, ][, and Paper\,I]{york00}, and when not
feasible the $V$-band one, from our own observations (Paper I), or
literature data (Rosales-Ortega et al. 2010); (ii) the spatial distribution of
\ion{H}{ii} regions derived by {\sc HIIexplorer}; and (iii) a few
simple assumptions of the number of arms and the
scale-length of the possible bar and/or the initial ring, based on the
visual inspection of the images. In general, we 
selected the spiral structure with the smallest possible number of spiral
arms that fulfill the criteria.

We are aware that the formalism adopted here to describe the analytical
structure of the spiral arms is clearly not the most mathematically exact
one. However it is useful for the ultimate goal of our study, i.e. to
determine how many spiral arms are in a considered galaxy and if a
certain \ion{H}{ii} region belongs to an arm or not. A more analytical
description of the spiral arms is clearly out of the scope of the
current study.

\begin{figure*}[tb]
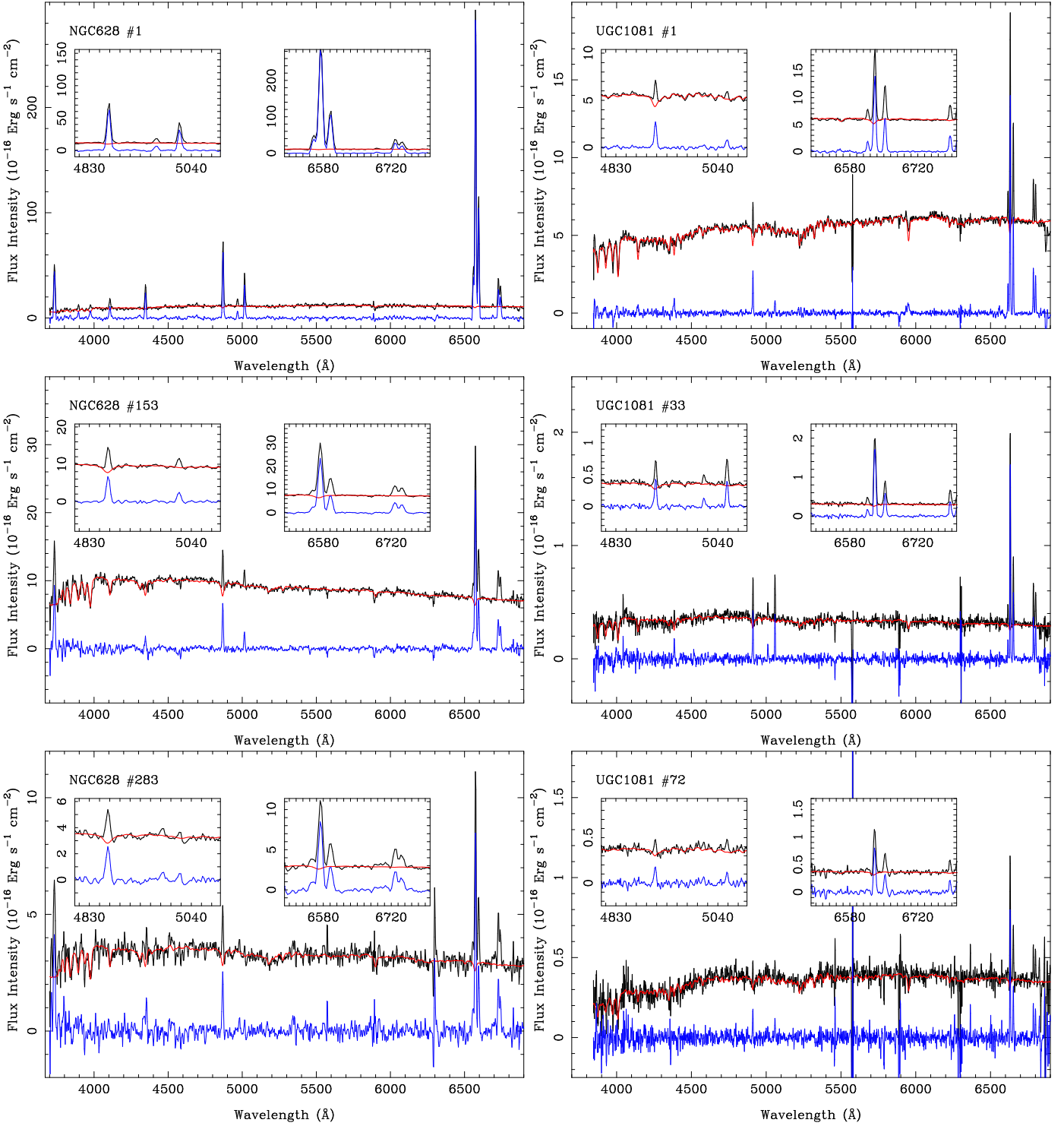

\begin{center}
\includegraphics[width=6.5cm,angle=270,clip=true]{figs/HII.fit.ngc628.1.ps}
\includegraphics[width=6.5cm,angle=270,clip=true]{figs/HII.fit.UGC1081.1.ps}
\includegraphics[width=6.5cm,angle=270,clip=true]{figs/HII.fit.ngc628.2.ps}
\includegraphics[width=6.5cm,angle=270,clip=true]{figs/HII.fit.UGC1081.2.ps}
\includegraphics[width=6.5cm,angle=270,clip=true]{figs/HII.fit.ngc628.3.ps}
\includegraphics[width=6.5cm,angle=270,clip=true]{figs/HII.fit.UGC1081.3.ps}
\caption{Spectra of three typical \ion{H}{ii} regions for two different galaxies,
  NGC\,628 (left panels), and UGC\,1081 (right panels). The panels correspond,
  from top to bottom, to the brightest, average and faintest \ion{H}{ii} regions with
  good quality data for each particular object. The spectral range shows all
  the emission lines analyzed in this article \label{fig:HII_spec}. From top to bottom: each panel
  shows the integrated spectra of
  the considered \ion{H}{ii} region (solid-black), together with the best model for
  the underlying stellar population (solid-red), and the pure-gas spectra
  (solid-blue). The two boxes show a zoom of the same plots at the H$\beta$
  and H$\alpha$ wavelength range.}
\end{center}
\end{figure*}

Table \ref{table.results} lists the results of this analysis,
including, for each galaxy, its name, the number of derived spiral
arms, a flag indicating the reliability of the results, and the parameters of the radial path for each arm, according to
the described  formula: $A$, $B$ and $C$. The quality flag is 1 for those
galaxies with a clearly distinguised spiral structure and a well
defined set of parameters to describe them, and for those
galaxies without a well defined spiral structure (flocculent). Fig.
\ref{fig:spiral} shows four examples of the derived spiral structure
for the galaxies in our sample, including two cases with well defined
spiral structure (NGC\,628 and UGC\,1081), one without a clear defined spiral
structure (NGC\,1637) and a clear flocculent case (UGC\,9837).

We associated \ion{H}{ii} regions to the nearest spiral arm by
computing the minimum distance between the centroid of the \ion{H}{ii}
region and the radial distribution of the considered arm. The mean of
these distances is then used as a scale-length to separate between
\ion{H}{ii} regions clearly associated with an arm, and possible
inter-arm ones. A final flag is included in the corresponding
catalogue table describing the coordinates of the detected \ion{H}{ii}
regions, indicating the nearest arm and the relative distance with
respect to the median one. Table \ref{table.coords} illustrates the
result of this analysis. It shows, for one single galaxy (UGC\,9837),
the absolute, relative, polar and deprojected coordinates of the 10
brightest \ion{H}{ii} regions, together with identification of the
nearest spiral arm, the Cartesian and angular distance to this arm and
the {\it spiralcentric} distance (i.e. the distance to the center
along the spiral arm). { In this context brightness means the peak
  intensity within a certain \ion{H}{ii}, as considered by
  \textsc{HIIexplorer} }. In addition, we included in this table the
systemic velocity and absolute luminosity of H$\alpha$ for each
\ion{H}{ii}, derived on basis of the emission line fitting described
in Section \ref{elines}.  Similar parameters are derived for all the
\ion{H}{ii} regions in the different galaxies, as indicated in
Appendix \ref{ape}. To our knowledge, this is the first attempt to
perform an analytical association of \ion{H}{ii} regions to a
particular arm and/or to an inter-arm area in a survey mode.

\begin{table*}
  \caption{Emission line ratios with respect to H$\beta$, and H$\beta$ line intensity  for the 10 brighest HII regions within UGC9837.}
 \label{table.flux}
 \begin{center}
 \begin{tabular}{lrrrrrrrrr}
 \hline\hline                 
 ID &
F(H$\beta$) &
$\frac{\mathrm{[OII]3727}}{\mathrm{H\beta}}$ &
$\frac{\mathrm{[OIII]5007}}{\mathrm{H\beta}}$ &
$\frac{\mathrm{[OI]6300}}{\mathrm{H\beta}}$ &
$\frac{\mathrm{H\alpha}}{\mathrm{H\beta}}$ &
$\frac{\mathrm{[NII]6583}}{\mathrm{H\beta}}$ &
$\frac{\mathrm{HeI6678}}{\mathrm{H\beta}}$ &
$\frac{\mathrm{[SII]6717}}{\mathrm{H\beta}}$&
$\frac{\mathrm{[SII]6731}}{\mathrm{H\beta}}$\\
\hline
 UGC9837-001       &    20.60$\pm$    0.75&    1.20$\pm$   0.05&    4.44$\pm$   0.20&    0.08$\pm$   0.02&    4.44$\pm$   0.27&    0.26$\pm$   0.11&    0.06$\pm$   0.01&    0.34$\pm$   0.03&    0.25$\pm$   0.03\\
 UGC9837-002       &    36.03$\pm$    0.66&    2.46$\pm$   0.07&    2.07$\pm$   0.06&    0.08$\pm$   0.01&    3.31$\pm$   0.17&    0.37$\pm$   0.12&    0.05$\pm$   0.01&    0.48$\pm$   0.03&    0.33$\pm$   0.03\\
 UGC9837-003       &    29.28$\pm$    0.63&    2.97$\pm$   0.13&    1.03$\pm$   0.04&    0.08$\pm$   0.02&    3.19$\pm$   0.21&    0.55$\pm$   0.16&    0.05$\pm$   0.01&    0.65$\pm$   0.04&    0.45$\pm$   0.04\\
 UGC9837-004       &    25.17$\pm$    0.57&    3.14$\pm$   0.13&    1.27$\pm$   0.05&    0.11$\pm$   0.02&    3.18$\pm$   0.21&    0.55$\pm$   0.15&    0.02$\pm$   0.01&    0.62$\pm$   0.04&    0.44$\pm$   0.04\\
 UGC9837-005       &    13.09$\pm$    0.51&    3.04$\pm$   0.26&    0.85$\pm$   0.07&    0.11$\pm$   0.04&    3.57$\pm$   0.31&    0.69$\pm$   0.19&    0.06$\pm$   0.01&    0.67$\pm$   0.08&    0.52$\pm$   0.08\\
 UGC9837-006       &    19.40$\pm$    0.54&    3.04$\pm$   0.13&    1.77$\pm$   0.08&    0.08$\pm$   0.04&    3.04$\pm$   0.21&    0.38$\pm$   0.13&    0.02$\pm$   0.01&    0.50$\pm$   0.05&    0.37$\pm$   0.04\\
 UGC9837-007       &    13.90$\pm$    0.54&    2.93$\pm$   0.19&    1.48$\pm$   0.10&    0.10$\pm$   0.03&    3.58$\pm$   0.29&    0.50$\pm$   0.17&    0.07$\pm$   0.01&    0.61$\pm$   0.06&    0.45$\pm$   0.06\\
 UGC9837-008       &    12.07$\pm$    0.45&    3.53$\pm$   0.27&    0.97$\pm$   0.07&    0.13$\pm$   0.04&    3.04$\pm$   0.28&    0.59$\pm$   0.19&    0.03$\pm$   0.01&    0.71$\pm$   0.07&    0.51$\pm$   0.06\\
 UGC9837-009       &     9.31$\pm$    0.48&    2.10$\pm$   0.14&    3.52$\pm$   0.23&               \nodata&  3.35$\pm$   0.27&    0.27$\pm$   0.11&    0.04$\pm$   0.01&    0.38$\pm$   0.06&    0.29$\pm$   0.06\\
 UGC9837-010       &    13.46$\pm$    0.51&    3.52$\pm$   0.23&    1.35$\pm$   0.09&    0.18$\pm$   0.05&    3.30$\pm$   0.29&    0.52$\pm$   0.19&    0.04$\pm$   0.01&    0.73$\pm$   0.07&    0.52$\pm$   0.06\\
\hline                                   
 \end{tabular}
 \end{center}

{ Notes}: H$\beta$ fluxes are in units of
10$^{-16}$ erg s$^{-1}$ cm$^{-2}$; All line intensities have been derived after
subtracting the underlying stellar population, but without any further
correction. { The full catalogue of emission line ratios for the \ion{H}{ii} regions analyzed in this object and the remaining ones discussed along this article are listed in Appendix \ref{ape}.}

 \end{table*}


\begin{table*}
  \caption{Emission line equivalent with of the strongest emission lines analyzed, for the 10 brighest HII regions within UGC9837.}
 \label{table.EW}
 \begin{center}
 \begin{tabular}{lrrrrrrrr}
 \hline\hline                 
 ID & \multicolumn{8}{c}{Equivalent Width}\\
\hline
  & 
[OII] &
H$\beta$ &
[OIII]&
[OI] &
H$\alpha$ &
[NII]&
HeI &
[SII]\\
  & 
$\lambda 3727$ &
  &
$\lambda 5007$ &
$\lambda 6300$ &
 &
$\lambda 6583$ &
$\lambda 6678$ &
$\lambda 6717+6731$ \\
\hline
 UGC9837-001       &    -80.6$\pm$     22.1&    -83.8$\pm$     17.6&   -367.1$\pm$    208.1&     -8.0$\pm$      1.2&   -503.6$\pm$    497.3&    -29.6$\pm$     33.4&     -6.3$\pm$      0.3&    -69.5$\pm$      7.1 \\
 UGC9837-002       &   -102.7$\pm$     40.3&    -53.9$\pm$      6.4&   -113.7$\pm$     19.5&     -6.0$\pm$      0.5&   -269.6$\pm$     94.5&    -30.2$\pm$     13.4&     -3.7$\pm$      0.2&    -69.2$\pm$      5.5 \\
 UGC9837-003       &    -66.8$\pm$     19.1&    -22.7$\pm$      2.1&    -24.1$\pm$      1.2&     -2.2$\pm$      0.2&    -96.5$\pm$     11.9&    -16.5$\pm$      3.2&     -1.4$\pm$      0.1&    -34.0$\pm$      1.3 \\
 UGC9837-004       &    -66.8$\pm$     23.8&    -23.2$\pm$      2.2&    -31.0$\pm$      1.9&     -3.2$\pm$      0.3&   -101.6$\pm$     13.4&    -17.9$\pm$      3.6&     -0.8$\pm$      0.1&    -34.6$\pm$      1.4 \\
 UGC9837-005       &    -42.9$\pm$      6.7&    -14.0$\pm$      1.4&    -12.2$\pm$      0.6&     -1.9$\pm$      0.3&    -64.1$\pm$      7.8&    -12.3$\pm$      2.3&     -1.1$\pm$      0.2&    -22.1$\pm$      0.8 \\
 UGC9837-006       &    -76.2$\pm$     29.8&    -30.0$\pm$      2.5&    -54.0$\pm$      4.3&     -3.1$\pm$      0.7&   -129.2$\pm$     31.3&    -16.3$\pm$      5.5&     -0.7$\pm$      0.3&    -38.1$\pm$      1.8 \\
 UGC9837-007       &    -51.2$\pm$     10.7&    -21.1$\pm$      2.4&    -31.7$\pm$      2.3&     -2.6$\pm$      0.4&    -97.7$\pm$     18.6&    -13.5$\pm$      3.7&     -1.8$\pm$      0.1&    -30.2$\pm$      1.3 \\
 UGC9837-008       &    -55.0$\pm$     18.5&    -15.2$\pm$      1.7&    -15.1$\pm$      0.7&     -2.4$\pm$      0.3&    -59.9$\pm$      5.9&    -11.8$\pm$      2.0&     -0.5$\pm$      0.1&    -25.1$\pm$      0.8 \\
 UGC9837-009       &    -40.2$\pm$      7.3&    -22.9$\pm$      2.9&    -80.9$\pm$     11.1&     -2.6$\pm$      1.4&   -109.0$\pm$     12.6&     -8.5$\pm$      1.9&     -1.3$\pm$      0.3&    -21.5$\pm$      1.2 \\
 UGC9837-010       &    -53.4$\pm$     16.2&    -16.5$\pm$      1.7&    -22.7$\pm$      1.5&     -4.0$\pm$      0.5&    -72.0$\pm$      7.9&    -11.4$\pm$      2.3&     -0.9$\pm$      0.2&    -28.1$\pm$      1.4 \\
\hline                                   
 \end{tabular}
 \end{center}

{ Notes}: { All the listed equivalent widths are in units of \AA. The full catalogue of equivalent widths for the \ion{H}{ii} regions analyzed in this object and the remaining ones discussed along this article are listed in Appendix \ref{ape}.}

 \end{table*}


\section{Deriving the main spectroscopic properties of the \ion{H}{ii} regions}\label{specHII}

\subsection{Decoupling the emission lines from the underlying stellar population.}\label{decouple}

To extract the nebular physical information of each individual \ion{H}{ii} region, the
underlying continuum must be decoupled from the emission lines for
each of the analyzed spectra. Several different tools have been
developed to model the underlying stellar population, effectively
decoupling it from the emission lines \citep[e.g.,][]{cappellari04,cid-fernandes05,ocvrik2006,sarzi2006,sanchez07a,koleva2009,macarthur2009,walcher11,sanchez11}.
Most of these tools are based on the same principles, i.e., they assume
that the stellar emission is the result of the combination of
different (or a single) simple stellar populations (SSP), and/or the
result of a particular star-formation history, whose emission is
redshifted due to a certain systemic velocity, broadened and smoothed
due a certain velocity dispersion and attenuated due to a certain dust
content. 

For the particular case of the \ion{H}{ii} regions, the main purpose of this analysis
is to provide a reliable subtraction of the underlying
stellar population. For doing so, we performed a simple but robust
modeling of the continuum emission. We use the routines described in
\cite{sanchez11} and \cite{rosales-ortega10}, which provided us with certain
parameters describing the physical components of the stellar
populations (e.g., luminosity-weighted ages, metallicities and dust
attenuation, together with the systemic velocity and velocity
dispersion) and a set of parameters describing each of the analyzed
emission lines (intensity, velocity and velocity dispersion). A simple SSP
template grid was adopted, consisting of three ages (0.09, 1.00 and 17.78
Gyr) and two metallicities (0.0004 and 0.03).  The models were extracted from
the SSP template library provided by the MILES project
\citep{vazdekis10}. { The two considered metallicities are the most metal poor and most
  metal rich with the largest coverage range in ages, within the
  considered library. The oldest stellar population was selected to
  reproduce the reddest possible underlying stellar population, mostly
  due to larger metallicities than the one considered in our
  simplified model, although it is clearly older than the accepted
  cosmological time of the Universe. Our youngest stellar population
  is the 2nd youngest in the MILES library with both extreme
  metallicities. No appreciable difference was found between using
  this one or the youngest one ($\sim$80 Myr). Finally, we selected an
  average stellar population, of $\sim$1Gyr, required to reproduce the
  intermediate-to-blue stellar populations, and to produce more
  reliable corrections of the underlying stellar absorptions
  (Paper\,I). } This library is { clearly} insufficient to describe
in detail the nature of all the stellar populations and star formation
histories. However, it covers the parameter space of possible stellar
populations well enough to describe them at to 1st order, providing a
clean residual information of the ionized gas.  { Evenmore, with a
  combination of the considered templates it is possible to
  reconstruct any of the SSP of the full MILES library within an
  accuracy similar to our photometric uncertainty ($\sim$10\%,
  Paper\,I). Therefore, to include any other template is redundant for
  the main purpose of this analysis.}

Fig. \ref{fig:HII_spec} illustrates the results of the fitting
procedure for three \ion{H}{ii} regions (the brightest, the average and the
faintest one, in terms of H$\alpha$ luminosity) extracted from two
typical galaxies (NGC\,628 and UGC\,1081). The figure shows for each \ion{H}{ii}
region the extracted spectrum (black line), together with the best
multi-SSP model for the stellar population (red line), and the
pure nebular emission spectrum (blue line).

\subsubsection{Deriving the main properties of the emission lines}\label{elines}

To derive the properties of the stronger emission lines detected in the
{\it stellar-population subtracted} spectra, each line was fitted with
a single Gaussian, coupled with the systemic velocity and velocity
dispersion of different emission lines when needed (e.g., for
doublets and triplets). This procedure provide us with the intensity,
systemic velocity and velocity dispersion for each emission line. Note that by
subtracting a stellar continuum model derived with a set of SSP templates, we
are already taking into account (and correcting for, to a first order) the
contribution of underlying absorption, which is particularly important in the
H$\alpha$ and H$\beta$ lines.

\begin{figure*}
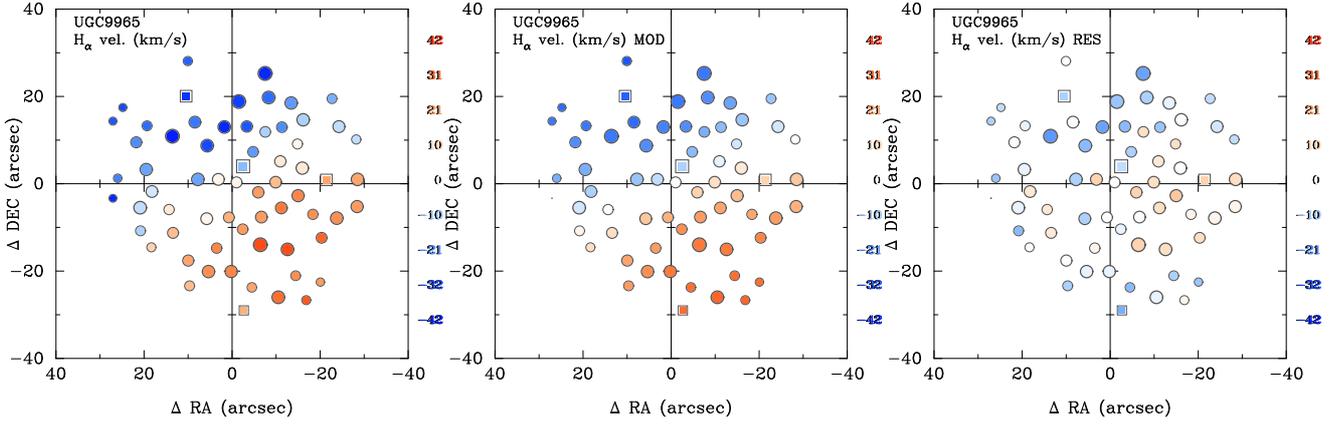

\centering
\includegraphics[width=5.5cm,angle=270]{figs/UGC9965.vel_map.ps}\includegraphics[width=5.5cm,angle=270]{figs/UGC9965.vel_mod.ps}\includegraphics[width=5.5cm,angle=270]{figs/UGC9965.vel_res.ps}
\caption{{\it Left-panel:} H$\alpha$ velocity field derived from the
  analysis of the \ion{H}{ii} regions for UGC\,9965. {\it Central-panel:} Best model
  fitted to the velocity map using a simple arctan velocity curve, and fitting
  the inclination, position angle and maximum rotational velocity. {\it
    Right-panel:} Residual of the subtraction of the model to the velocity
  map. Each plotted symbol in each panel represent an individual \ion{H}{ii}
  region. The circles represent those \ion{H}{ii} regions below the \cite{kauffmann03}
  demarcation line in the {\it classical} BPT diagram \citep{baldwin81}, and the squares
  corresponds to those ones located in the intermediate region between that
  line and the \cite{kewley01} one, as described later in the text. The size
  of the symbols are proportional to the H$\alpha$ intensity.\label{fig:vel}}
\end{figure*}

As discussed in Paper\,I, there are different
issues that may affect the final quality of the individual spectra in
the datacubes (uncleaned cosmic rays, trace problems, low transmission
fibers, spectra near to the edge of the FoV and vignetting). These
quality issues, that affect a reduced number of spaxels, propagate
along the segregation, extraction and ``stellar continuum'' cleaning
processes, and therefore they may affect the final quality of the
``pure emission'' spectra of the \ion{H}{ii} regions. In order to minimize the
impact of these issues on the final sample of \ion{H}{ii} spectra, we have
performed an automatic quality check. Only those ``pure emission''
spectra fulfilling the following criteria are flagged as good quality
data:

\begin{enumerate}

\item The derived intensity for H$\alpha$ is above zero or below three
 times that of the brightest \ion{H}{ii} region based on the narrow-band
  image intensity. The contrary may happen in case of problems
  with the fitting procedure, or problems with one or a few spectra of those 
  that were co-added to derive the integrated spectrum (like a cosmic rays).

\item The fraction of spectral pixels with negative values in the original spectra, 
  i.e.~prior to the subtraction of the underlying stellar continuum,  is
  lower than 10\%. The contrary may happen in the outer regions of the
  galaxies, if there is any problem with the sky subtraction.

\item The fraction of spectral pixels in the ``pure emission'' spectra of the
  \ion{H}{ii} region with a value below the median flux 1$\sigma$ within the
  wavelength range between 3900 and 6500\AA, is at maximum three times lower
  than the median of this fraction for all the \ion{H}{ii} regions in the same
  galaxy. This criterion is used to remove spectra strongly affected by the
  vignetting effect, which affects only $\sim$30\%~ of the data (Paper\,I).

\item The derived intensity for H$\alpha$ is more than five times above the
  background noise ($\sigma_{back}$), estimated as $\sigma_{back} =
   \sigma_{6300-6500}~ FWHM_{line}$, where (i) $\sigma_{6300-6500}$ is the
   standard deviation of the continuum intensity once subtracted the
   underlying stellar component for the wavelength range between 6300 and
   6500\AA (i.e., a continuum adjacent to H$\alpha$); and (ii) $FWHM_{line}$
   is the full width at half maximum derived for the emission line, as described before.

\end{enumerate}

\begin{figure*}
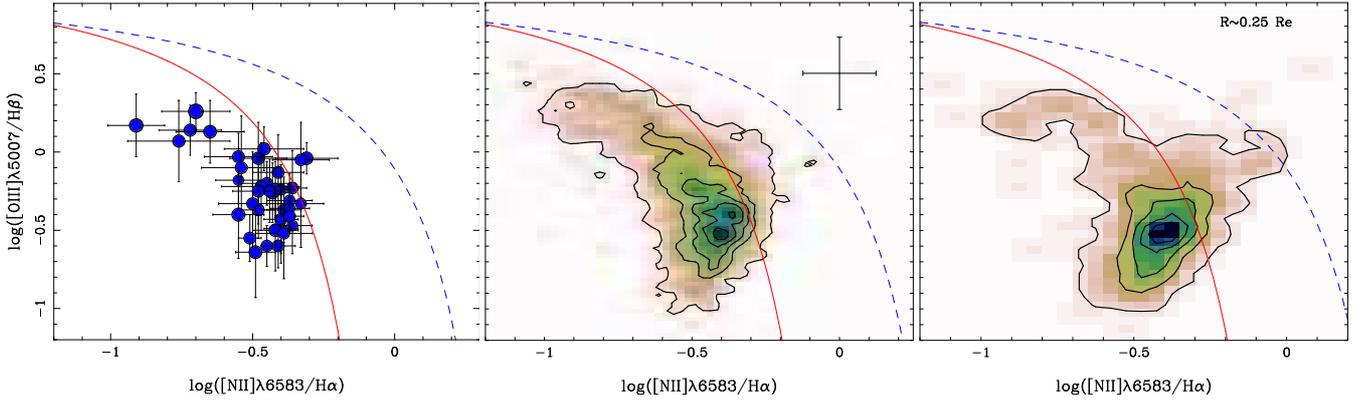

\centering
\includegraphics[width=5.2cm,clip,trim=0 0 0 2.2,angle=270]{figs/diag_med_val.ps}
\includegraphics[width=5.2cm,clip,trim=0 59.2 0 0,angle=270]{figs/diag_O3_N2.all.ps}
\includegraphics[width=5.2cm,clip,trim=0 59.2 0 0,angle=270]{figs/diag_O3_N2.center.ps}
\caption{{\it Left panel:} [\ion{O}{iii}]~$\lambda$5007/H$\beta$
  vs. [\ion{N}{ii}]~$\lambda$6583/H$\alpha$ diagnostic diagram for the average
  properties of the \ion{H}{ii} regions galaxy by galaxy listed in Table
  \ref{table.med.val}. The error bars indicate the standard deviation from the
  mean value. {\it Middle panel:} Similar diagnostic diagram for all the
  emission line regions detected by the described analysis with good quality
  measurements of both parameters (2230 regions). The color image and
  contours show the density distribution of these regions. The first contour
  indicates the mean density, with each consecutive one increasing by four
  times this mean value. The \cite{kauffmann03} (red solid-line) and
  \cite{kewley01} (blue dashed-line) demarcation curves are usually invoked to
  distinguish between star-forming regions (below the red solid-line), and
  other source of ionization, like AGN/shocks/post-AGB (above the blue
  line). Regions between both lines are considered intermediate ones,
  indicating a mixed origin for the ionization. The error-bars at the top-left
  indicate the typical (mean) errors for the considered line ratios. {\it
    Right panel:}  Similar diagram, including only the 124 regions at the core
  of each galaxy ($r<0.5 r_e$, i.e., at $\sim 0.25 r_e$ in average). The fraction of regions in the { intermediate}
  location is clearly higher.\label{fig:diag}}
\end{figure*}

The remaining regions are flagged out, masked, and the corresponding
spectra are set to zero. The criteria were based on iterative
experiments on the data, and visual inspections of hundreds of
spectra, before and after subtracting the underlying
continuum. Although the fraction of flagged-out/rejected spectra
change from object to object, on average this final cleaning affects
$\sim$15\% of the \ion{H}{ii} regions, as can be seen in Table
\ref{table.HII}.  { Finally, only those \ion{H}{ii} regions with
  measured H$\beta$ emission line detected at $>$3$\sigma$
  significance were considered for further analysis \cite[e.g.][]{mari11}, although they
  were not masked out and their spectra were not set to zero. This
  criteria was included to consider only those regions with good line
  diagnostic ratios and well defined Balmer ratio, both required in
  further analysis. It further reduces the number of selected
  \ion{H}{ii} regions by $\sim$5\% on average, although is some cases
  the fraction is much larger (see Table \ref{table.HII})}. Due to the
size of our original sample and the pseudo-random selection of
\ion{H}{ii}-regions that are affected by these issues, we consider
that this last cleaning process will have little effect in the overall
statistical significance of our survey.

{ Once derived the emission line intensities, we estimate their
  corresponding equivalent width for each \ion{H}{ii} region and line.
  For doing so, instead of using the classical procedure (i.e.,
  measuring the flux within a narrow-band wavelength range centred in
  the line and in two adjacent ones corresponding to the continuum),
  we make use of the results from our fitting analysis. We derive the equivalent width
  by dividing the emission line integrated intensities by the underlysing continuum
  flux density. We estimated the continuum as the median intensity in a band-width of
  100$\AA$, centred in the line, using the gas-subtracted spectra
  provided by our fitting procedure. With this method we can estimate
  the equivalent width of nearby lines, which contaminate the
  measurements of this parameter using the classical method.  }

Table \ref{table.flux} illustrates the result of this analysis. It
shows, for a sub-set of \ion{H}{ii} regions in a particular galaxy
(UGC\,9837) the H$\beta$ line intensity and relative flux of some of the
most prominent emission lines.{ Table \ref{table.EW} reports the equivalent widths
for the corresponding emission lines and regions}. The same parameters are derived
for all the \ion{H}{ii} regions in the different galaxies, as indicated in
Appendix \ref{ape}.

\subsubsection{Structural parameters of the galaxies}\label{struct}

To understand the fundamental properties of the \ion{H}{ii} regions and their
relation with the overall evolution of galaxies, it is required to characterize the 
main structural parameters of these galaxies. We have collected the available 
information in public collections like NED (\url{http://ned.ipac.caltech.edu/}) and 
Hyperleda (\url{http://leda.univ-lyon1.fr/}). Table \ref{table.gal.prop} already contains
the most relevant parameters for the current study, including the
morphological type, the redshift, the integrated $B$-band magnitude and the B-V
color. All the galaxies in the sample are spirals by selection, but
different kinds of spiral galaxies are covered, including galaxies with and
without bars, galaxies showing rings, etc. The observed $B$-band magnitude is
$\sim$14 mag, in average. However, the galaxies selected from the
PINGS sample are in general brighter ($B\sim$11 mag). As expected,
galaxies have blue colors ($B-V\sim$0.8 mag), although with a
considerable dispersion. The covered absolute magnitudes range 
from $M_V\sim -$23 mag to $\sim -$18 mag. In summary, the
sample covers typical members of the so-called blue cloud, from typical
standard spiral galaxies to almost dwarfs.

The listed information was complemented with additional
parameters, like the maximum rotation velocity, the inclination, the
position angle and the effective radius (defined as the radius at which one half of the total light of the system is emitted), derived from the analysis of
the data presented here. The first three parameters were derived from
the modeling of the gas velocity pattern extracted from the the
H$\alpha$ emission line fitting for the \ion{H}{ii} regions, described in
previous sections. The wide spatial coverage and high S/N of the
H$\alpha$ emission line in the integrated spectra for each region
guarantee a good determination of the velocity pattern. The rotation 
curve was fitted using a simple arctan model \citep{stav90},

\begin{equation}
v(r) = v_{sys} + \frac{2~v_{\rm rot}}{\pi} {\rm arctan} (s\cdot r-c)
\end{equation}

\noindent where $v_{sys}$ is the systemic velocity of the gas and
$v_{\rm rot}$ is the asymptotic rotation speed of the disc, $s$ characterizes
the slope of $v(r)$ in the inner part of the galaxy, $r$ is the
distance to the rotational center, and $c$ is the parameter that
characterizes any offset in the rotation axis of the galaxy. A model
of the velocity map was created by re-projecting the best fitting
arctan function, taking into account the position angle and
inclination of the galaxy. We fitted to the data following a
similar procedure as the one described in \cite{sanchez12}, using a
$\chi^2$-minimization algorithm included in FIT3D \citep{sanchez06b}.

As an initial guess for the fitting, the position angle and
inclination were derived from the isophotal analysis described later
in this section, and the maximum rotational velocity was set to half
of the maximum difference in velocity from receding to approaching
velocities. For all the galaxies the rotational center is fixed to the
location of the peak intensity in the $V$-band image created from the
datacubes. The parameter $c$ is fixed to zero (i.e., it is assumed
that there is no offset between the rotation and photometric centers).
The $v_{sys}$ is fixed to the median value of the gas velocities for
those \ion{H}{ii} regions located in the inner regions ($r<$0.5$r_{max}$,
where $r_{max}$ is the maximum distance to the center for all the \ion{H}{ii}
regions).

Finally $v_{\rm rot}$ and $s$ are fitted, together with the position angle
and inclination of the galaxies, $v_{rot}$ is fitted within a range
between $\sim$0.3 and 6 times the maximum velocity difference among
the \ion{H}{ii} regions, and $s$ is fitted between 0.1 and 10 arcsec$^{-1}$.
Fig. \ref{fig:vel} illustrates a typical result of this analysis,
showing, for a particular galaxy (UGC\,9837) the H$\alpha$ velocity map,
the best fit model, and the residual. Despite the low
inclination of the galaxies, in most of the cases it is possible to
obtain a good model. In most of the cases the residual
velocities ranges between $\pm$15 km s$^{-1}$, $\sim$15\% of the
maximum rotational velocity. This is expected due to random motions 
in the galaxies, compare e.g. \citet{andersen08, neumayer11}.

The effective radius was derived based on an analysis of the azimuthal
surface brightness (SB) profile, derived based on
elliptical isophotal fitting of the ancillary $g$-band images
collected for the galaxies \citep[extracted from the SDSS imaging survey,][and Paper\,I]{york00}. When these ancillary images 
were { not} available 
we used the $B$-band (Paper\,I). In order to homogenize the
dataset, both sets of SBs were transformed to the $B$-band using the
average $g-B$ color for each galaxy. When both band images where
available a comparison between the directly derived and the estimated
surface brightness profile was performed, finding no significant
differences in the average gradient. We note here that the observed
$B$ and $g$-bands sample a range of wavelengths between
$\sim$4150-4750\AA\ and $\sim$4450-5210\AA, respectively, due to the
redshift range of the sample. Therefore, there is an inherent
imprecision in the intrinsic wavelength range in this analysis.

The surface brightness profile was then fitted using a pure exponential
profile, following the classical formula,

\begin{equation}
I = I_0 \exp[ - (r/r_d)]
\end{equation}

\noindent
where $I_0$ is the central intensity, and $r_d$ is the disk
scale-length \citep{free70}, using a simple polynomial regression
fitting. Prior to this analysis, a visual inspection is performed to remove
the inner-most values of the SB profile, strongly affected by seeing effect,
and/or not following a linear relation due to the presence of other components
like the bulge and/or bars.

The scale-length is used to derive the effective radius, defined as
the radius at which the integrated flux is half of the total one,
by integrating the previous formula, and deriving the relation:

\begin{equation}
r_e = 1.67835 r_d 
\end{equation}

\noindent
The results of these analyses are included in Table
\ref{table.gal.prop}. In average, the derived inclination agrees with
the visual selection of the galaxies as face-on spirals. The average
inclination is $\sim$33$\degree$, and only two galaxies have an
inclination larger than 60$\degree$ (NGC\,7570 and UGC\,5100).
This confirms our visual classification as face-on galaxies. The
average maximum rotational velocity is $\sim$100 km s$^{-1}$, with a
wide range of values, between $\sim$50 km s$^{-1}$ and $\sim$300 km
s$^{-1}$ \citep[values which are typical for spiral galaxies, e.g.,][]{persic96}. 
Note that the effective radius ranges between $\sim$1.5 and $\sim$5.5 kpc.

\section{Analysis and results}\label{result}

In this section we analyze both the mean statistical properties of the
\ion{H}{ii} regions and explore the possible regular patterns in their radial
variations.

\subsection{Statistical properties of the \ion{H}{ii} regions}\label{stats}

Despite the many different spectroscopic studies in extragalactic
\ion{H}{ii} regions, we still do not have the understanding of which are the
statistical spectroscopic properties of these common star-forming
regions. This is a fundamental problem that it is mostly due to the lack of
big statistical samples, and the reduced number of coherent compilations.
The lack of a well defined set of {\it normal} values for the most
frequent parameters, like the diagnostic line ratios (e.g.,
[\ion{O}{iii}]/H$\beta$ and/or [\ion{N}{ii}]/H$\alpha$), ionization strength, dust
attenuation and/or electron density is a clear limitation to
understand if a particular set of \ion{H}{ii} regions is different from the
average, and if different at which significance level. To address this
question a statistically significant, large sample of \ion{H}{ii} regions is required, 
with well derived spectroscopic parameters, over a large sample of
star-forming galaxies of different types. An additional requirement is 
good spatial coverage, not biased towards the outer (bright) \ion{H}{ii} regions, which is
a common bias in this kind of studies. Despite the large number of
\ion{H}{ii} regions catalogued in this work, the current sample is
still incomplete to address this fundamental question. We will require
a sample as the one that will be provided by a survey like CALIFA
(S\'anchez et al. 2012), without selection effects by galaxy types.
However, the current catalogue of \ion{H}{ii} regions is good enough to derive
the statistical properties of these regions for a sub-set of galaxies:
quiescent/non highly disturbed, field, average luminosity spiral
galaxies.

\begin{table*}
 \caption{{ Median} physical parameters of the ionized gas derived for the considered galaxies}
 \label{table.med.val}      
 \begin{center}
 \begin{tabular}{lrrrrrrrrr}        
 \hline\hline                 
Galaxy    & L$_{\rm H\alpha}$ &  EW$_{\mathrm{H\alpha}}$ 
          & A$_V$ 
          & $\frac{\mathrm{[OIII]}}{\mathrm{H\beta}}$ 
          & $\frac{\mathrm{[NII]}}{\mathrm{H\alpha}}$ 
          & U 
          & $O/H$ 
          & r$_{\rm HII}$ & n$_{e}$ \\
  & (1)
  & {(2)}
& {(3)}
& {(4)}
& {(5)}
& {(6)}
& {(7)}
& {(8)}
& {(9)}
\\
  \hline       
2MASXJ1319+53      &40.3 $\pm$ 0.6     &  -28.8 $\pm$    6.7 &    1.4 $\pm$    0.3 &    0.14 $\pm$    0.16 &   -0.72 $\pm$    0.11 &   -3.84 $\pm$    0.34 &    8.48 $\pm$    0.08 &  2.8 $\pm$  6.1 &     \nodata  \\
CGCG\,071-096      &40.0 $\pm$ 0.6      &  -35.0 $\pm$    9.2 &    1.4 $\pm$    0.5 &   -0.04 $\pm$    0.23 &   -0.48 $\pm$    0.10 &   -3.71 $\pm$    0.21 &    8.59 $\pm$    0.10 &  1.9 $\pm$  1.8 &     \nodata  \\
CGCG\,148-006      &40.0 $\pm$ 0.5      &  -24.0 $\pm$    5.4 &    1.5 $\pm$    0.6 &   -0.13 $\pm$    0.24 &   -0.41 $\pm$    0.09 &     \nodata &    8.66 $\pm$    0.10 &  \nodata &     \nodata  \\
CGCG\,293-023      &39.7 $\pm$ 0.5      &  -20.7 $\pm$    4.5 &    1.5 $\pm$    0.2 &   -0.20 $\pm$    0.14 &   -0.45 $\pm$    0.10 &   -4.06 $\pm$    0.15 &    8.67 $\pm$    0.05 &  1.6 $\pm$  1.5 &     0.1 $\pm$   2.6  \\
CGCG\,430-046      &40.5 $\pm$ 0.5      &  -31.4 $\pm$    7.6 &    1.5 $\pm$    0.4 &   -0.26 $\pm$    0.19 &   -0.43 $\pm$    0.07 &     \nodata &    8.70 $\pm$    0.08 &     \nodata &     \nodata  \\
IC\,2204           &39.7 $\pm$ 0.6      &  -20.6 $\pm$    4.9 &    1.6 $\pm$    0.5 &   -0.23 $\pm$    0.24 &   -0.36 $\pm$    0.05 &     \nodata &    8.75 $\pm$    0.08 &     \nodata &  2.0 $\pm$   4.5  \\
MRK\,1477          &41.0 $\pm$ 1.4      &  -39.1 $\pm$   17.7 &    1.4 $\pm$    0.4 &   -0.04 $\pm$    0.10 &   -0.31 $\pm$    0.11 &   -3.45 $\pm$    0.31 &    8.64 $\pm$    0.02 &  8.1 $\pm$  3.5 &     \nodata  \\
NGC\,99            &39.6 $\pm$ 0.6      &  -41.0 $\pm$   12.3 &    1.2 $\pm$    0.5 &    0.13 $\pm$    0.20 &   -0.65 $\pm$    0.12 &     \nodata &    8.47 $\pm$    0.10 &     \nodata &     \nodata  \\
NGC\,3820          &40.2 $\pm$ 0.3      &  -19.0 $\pm$    3.2 &    0.9 $\pm$    1.2 &   -0.60 $\pm$    0.13 &   -0.45 $\pm$    0.02 &   -3.78 $\pm$    0.13 &    8.79 $\pm$    0.04 &  3.5 $\pm$  1.0 &     \nodata  \\
NGC\,4109          &40.4 $\pm$ 0.7      &  -17.7 $\pm$    3.0 &    1.9 $\pm$    0.4 &   -0.37 $\pm$    0.21 &   -0.39 $\pm$    0.06 &   -3.93 $\pm$    0.13 &    8.78 $\pm$    0.09 &  3.8 $\pm$  2.0 &     \nodata  \\
NGC\,7570          &39.5 $\pm$ 0.7      &  -16.4 $\pm$    3.6 &    1.4 $\pm$    0.7 &   -0.24 $\pm$    0.16 &   -0.40 $\pm$    0.04 &     \nodata &    8.71 $\pm$    0.04 &     \nodata &  1.6 $\pm$   9.4  \\
UGC\,74            &39.2 $\pm$ 0.5      &  -12.5 $\pm$    3.5 &    1.3 $\pm$    0.5 &   -0.47 $\pm$    0.18 &   -0.36 $\pm$    0.07 &     \nodata &    8.77 $\pm$    0.05 &     \nodata &  2.7 $\pm$   3.8  \\
UGC\,233           &40.0 $\pm$ 0.8      &  -34.0 $\pm$   10.4 &    1.3 $\pm$    0.5 &    0.02 $\pm$    0.14 &   -0.46 $\pm$    0.14 &     \nodata &    8.58 $\pm$    0.08 &     \nodata &     \nodata  \\
UGC\,463           &39.7 $\pm$ 0.5      &  -23.6 $\pm$    4.9 &    1.6 $\pm$    0.4 &   -0.60 $\pm$    0.14 &   -0.41 $\pm$    0.06 &     \nodata &    8.79 $\pm$    0.04 &     \nodata &  0.4 $\pm$   3.6  \\
UGC\,1081          &38.7 $\pm$ 0.4      &  -12.9 $\pm$    3.6 &    1.1 $\pm$    0.4 &   -0.31 $\pm$    0.17 &   -0.37 $\pm$    0.08 &     \nodata &    8.73 $\pm$    0.07 &     \nodata &  0.4 $\pm$   1.5 \\
UGC\,1087          &39.1 $\pm$ 0.3      &  -21.9 $\pm$    5.8 &    1.0 $\pm$    0.5 &   -0.24 $\pm$    0.18 &   -0.42 $\pm$    0.08 &     \nodata &    8.67 $\pm$    0.08 &     \nodata &  3.0 $\pm$   4.7  \\
UGC\,1529          &40.0 $\pm$ 0.5      &  -12.6 $\pm$    2.6 &    1.7 $\pm$    0.5 &   -0.49 $\pm$    0.22 &   -0.40 $\pm$    0.06 &     \nodata &    8.77 $\pm$    0.08 &     \nodata & 11.2 $\pm$  20.4  \\
UGC\,1635          &38.8 $\pm$ 0.4      &  -11.2 $\pm$    2.2 &    1.4 $\pm$    0.6 &   -0.36 $\pm$    0.16 &   -0.37 $\pm$    0.07 &     \nodata &    8.73 $\pm$    0.06 &     \nodata &  2.3 $\pm$   3.5  \\
UGC\,1862          &38.1 $\pm$ 0.4      &  -10.6 $\pm$    3.0 &    1.5 $\pm$    0.5 &   -0.18 $\pm$    0.19 &   -0.55 $\pm$    0.08 &     \nodata &    8.61 $\pm$    0.06 &     \nodata &  1.4 $\pm$   1.8  \\
UGC\,3091          &39.3 $\pm$ 0.5      &  -25.4 $\pm$    7.6 &    1.3 $\pm$    0.4 &   -0.22 $\pm$    0.28 &   -0.47 $\pm$    0.14 &     \nodata &    8.67 $\pm$    0.14 &     \nodata &  1.1 $\pm$   2.4  \\
UGC\,3140          &39.8 $\pm$ 0.5      &  -23.2 $\pm$    5.2 &    1.6 $\pm$    0.4 &   -0.43 $\pm$    0.22 &   -0.40 $\pm$    0.06 &     \nodata &    8.76 $\pm$    0.05 &     \nodata &  1.8 $\pm$   4.3  \\
UGC\,3701          &39.3 $\pm$ 0.3      &  -26.3 $\pm$    8.6 &    1.6 $\pm$    0.5 &   -0.03 $\pm$    0.15 &   -0.55 $\pm$    0.12 &     \nodata &    8.57 $\pm$    0.07 &     \nodata &  1.7 $\pm$   2.0  \\
UGC\,4036          &39.8 $\pm$ 0.5      &  -12.3 $\pm$    3.7 &    1.5 $\pm$    0.5 &   -0.33 $\pm$    0.28 &   -0.33 $\pm$    0.08 &     \nodata &    8.78 $\pm$    0.06 &     \nodata &  2.7 $\pm$   3.3  \\
UGC\,4107          &39.7 $\pm$ 0.5      &  -19.5 $\pm$    3.5 &    1.5 $\pm$    0.5 &   -0.25 $\pm$    0.20 &   -0.44 $\pm$    0.04 &     \nodata &    8.69 $\pm$    0.06 &     \nodata &  2.8 $\pm$   2.1  \\
UGC\,5100          &39.9 $\pm$ 0.7      &  -19.5 $\pm$    4.6 &    1.2 $\pm$    0.4 &   -0.05 $\pm$    0.24 &   -0.33 $\pm$    0.10 &   -3.58 $\pm$    0.39 &    8.72 $\pm$    0.10 &  2.9 $\pm$  2.5 &     \nodata  \\
UGC\,6410          &39.7 $\pm$ 0.5      &  -25.4 $\pm$    5.7 &    1.3 $\pm$    0.5 &   -0.25 $\pm$    0.21 &   -0.48 $\pm$    0.09 &   -3.68 $\pm$    0.27 &    8.65 $\pm$    0.09 &  1.2 $\pm$  0.7 &     \nodata  \\
UGC\,9837          &38.7 $\pm$ 0.5      &  -29.5 $\pm$   12.0 &    0.7 $\pm$    0.4 &    0.07 $\pm$    0.26 &   -0.76 $\pm$    0.18 &   -3.53 $\pm$    0.45 &    8.48 $\pm$    0.14 &  0.4 $\pm$  0.2 &  0.5 $\pm$   2.7   \\
UGC\,9965          &39.7 $\pm$ 0.3      &  -27.3 $\pm$    8.9 &    1.7 $\pm$    0.5 &   -0.10 $\pm$    0.33 &   -0.54 $\pm$    0.14 &   -3.76 $\pm$    0.29 &    8.62 $\pm$    0.14 &  1.4 $\pm$  0.7 &  4.5 $\pm$  12.1  \\
UGC\,11318         &40.1 $\pm$ 0.5      &  -34.3 $\pm$    7.9 &    1.8 $\pm$    0.4 &   -0.50 $\pm$    0.26 &   -0.42 $\pm$    0.08 &   -3.68 $\pm$    0.24 &    8.76 $\pm$    0.07 &  2.0 $\pm$  1.1 &     \nodata  \\
UGC\,12250         &39.6 $\pm$ 0.5      &  -21.0 $\pm$    3.7 &    1.2 $\pm$    0.5 &   -0.41 $\pm$    0.10 &   -0.37 $\pm$    0.05 &     \nodata &    8.75 $\pm$    0.03 &     \nodata &     \nodata  \\
UGC\,12391         &39.7 $\pm$ 0.3      &  -21.5 $\pm$    4.7 &    1.4 $\pm$    0.4 &   -0.37 $\pm$    0.18 &   -0.48 $\pm$    0.06 &     \nodata &    8.70 $\pm$    0.07 &     \nodata &  3.9 $\pm$  21.8  \\\hline
NGC\,628           &38.7 $\pm$ 0.5      &  -50.4 $\pm$   26.7 &    1.1 $\pm$    0.5 &   -0.40 $\pm$    0.28 &   -0.55 $\pm$    0.09 &   -3.75 $\pm$    0.29 &    8.68 $\pm$    0.10 &  0.5 $\pm$  0.5 &   0.5 $\pm$   3.1  \\
NGC\,1058          &38.7 $\pm$ 0.5      &  -24.8 $\pm$   11.7 &    1.0 $\pm$    0.4 &   -0.33 $\pm$    0.33 &   -0.50 $\pm$    0.12 &   -3.82 $\pm$    0.27 &    8.70 $\pm$    0.12 &  0.5 $\pm$  0.3 &   0.4 $\pm$   2.3  \\
NGC\,1637          &38.8 $\pm$ 0.5      &  -19.0 $\pm$    6.7 &    1.3 $\pm$    0.4 &   -0.52 $\pm$    0.29 &   -0.39 $\pm$    0.10 &   -3.84 $\pm$    0.25 &    8.79 $\pm$    0.08 &  0.5 $\pm$  0.8 &  1.5 $\pm$   2.6  \\
NGC\,3184          &39.2 $\pm$ 0.4      &  -43.0 $\pm$   15.7 &    1.3 $\pm$    0.5 &   -0.64 $\pm$    0.29 &   -0.49 $\pm$    0.07 &   -3.79 $\pm$    0.27 &    8.80 $\pm$    0.08 &  1.0 $\pm$  0.6 &  0.7 $\pm$   2.7  \\
NGC\,3310          &39.8 $\pm$ 0.8      & -133.0 $\pm$   51.8 &    0.2 $\pm$    0.2 &    0.26 $\pm$    0.12 &   -0.70 $\pm$    0.12 &   -3.31 $\pm$    0.44 &    8.42 $\pm$    0.07 &  1.0 $\pm$  1.4 &  0.5 $\pm$   1.8  \\
NGC\,4625          &39.3 $\pm$ 0.4      &  -24.6 $\pm$    6.0 &    0.6 $\pm$    0.4 &   -0.55 $\pm$    0.15 &   -0.51 $\pm$    0.04 &   -3.81 $\pm$    0.16 &    8.75 $\pm$    0.05 &  1.0 $\pm$  0.3 &  0.3 $\pm$   1.8  \\
NGC\,5474          &38.4 $\pm$ 0.4      &  -64.6 $\pm$   28.0 &    0.7 $\pm$    0.4 &    0.17 $\pm$    0.20 &   -0.91 $\pm$    0.10 &   -3.49 $\pm$    0.44 &    8.39 $\pm$    0.08 &  0.3 $\pm$  0.2 &  0.2 $\pm$   3.0  \\
\hline
Mean               &39.5 $\pm$ 0.6      &  -28.4 $\pm$  20.8  &   1.3  $\pm$   0.4 &   -0.25  $\pm$   0.23  &  -0.47  $\pm$   0.13  &  -3.71 $\pm$   0.18  &   8.67 $\pm$   0.10  & 1.9  $\pm$  1.8  &  1.9  $\pm$   2.3  \\
\hline                    
 \end{tabular}
 \end{center}

{ Notes}: The table shows the median values and standard deviations of the
following physical parameters: (1) Decimal logarithm of the dust corrected luminosity of H$\alpha$ in erg~s$^{-1}$; (2) Equivalent width of the pure H$\alpha$
emission line expressed in \AA (values decontaminated from the underlying
stellar absorption); (3) Dust attenuation derived from the H$\alpha$/H$\beta$
line ratio; (4) [OIII]$\lambda$5007/H$\beta$ line ratio, in decimal logarithm
scale; (5) [NII]$\lambda$6583/H$\alpha$ line ratio, in decimal logarithm
scale; (6) Ionization parameter; (7) Oxygen abundance, 12\,+\,log(O/H), derived
using the O3N2 indicator; (8) St\"omgren radius calculated for the HII regions
in units of 100 parsecs; (9) Electron density in units of 100 electrons per
cm$^{3}$.

\end{table*}


Among the several different spectroscopic parameters describing \ion{H}{ii}
regions, we have analyzed a set of them based on the strongest
detected emission lines: (i) EW$_{H\alpha}$, the equivalent width of
the H$\alpha$ emission line. This parameter is directly related to the
fraction of very young stars ($\sim$ 10 Myr), and can be used to
estimate the aging process of the ionizing population; (ii) A$_{\rm V}$, 
the dust attenuation derived from the H$\alpha$/H$\beta$ Balmer
decrement. To derive it 
the extinction law by \cite{cardelli89} was assumed, with $R_{\rm V}=$3.1, 
and the theoretical value for the unobscured line ratio for
case B recombination of  H$\alpha$/H$\beta=2.86$, for $T_e$=10,000\,K 
and $n_e$=100\,cm$^{-3}$ \citep{osterbrock89}; (iii) two typical
diagnostic line ratios, [\ion{O}{iii}]~$\lambda$5007/H$\beta$ and
[\ion{N}{ii}]~$\lambda$6583/H$\alpha$, that define the nature of the
ionization source; (iv) the ionization parameter, estimated as ${\rm log}_{10}
U = -3.02 - 0.80 {\rm log}_{10} ([\ion{O}{ii}]/[\ion{O}{iii}])$, a measurement of the
strength of the ionization radiation \citep{diaz00}; (v) the oxygen
abundance, 12\,+\,log(O/H), derived using the O3N2 indicator defined by
\cite{pettini04}, that estimates the gas enrichment; (vi) r$_{\rm HII}$, 
the Str\"omgren radius of the \ion{H}{ii} regions \citep{osterbrock89}, i.e.,
a hint of the size of the \ion{H}{ii} regions based on pressure equilibrium
considerations, and (vii) $n_e$, the electron density derived from the [\ion{S}{ii}]
doublet line ratio, i.e., a proxy to the density of the ionized gas
\citep[e.g.,][]{osterbrock89}.

\begin{figure*}
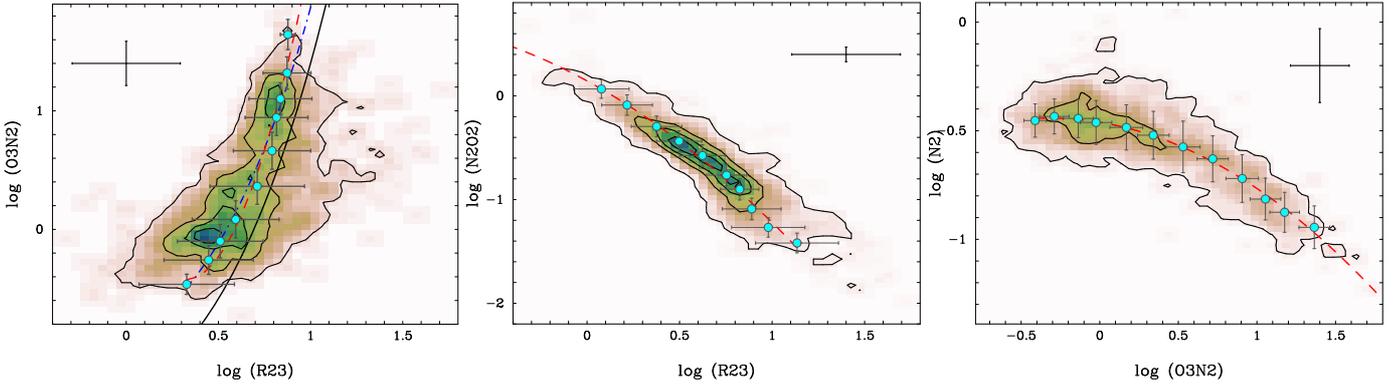

\centering
\includegraphics[width=5.0cm,angle=270]{figs/R23_O3N2.ps}
\includegraphics[width=5.0cm,angle=270]{figs/R23_N2O2.ps}
\includegraphics[width=5.0cm,angle=270]{figs/plot_O3N2_N2.ps}
\caption{{\it Left-panel:} Distribution of the O3N2 vs. R$_{23}$ line ratios
  for all the 1142 \ion{H}{ii} regions with detected [\ion{O}{ii}]~$\lambda$3727 emission
  line. The image and contours show the density distribution of both
  parameters. The first contours is at the mean density, with a regular
  spacing of four times this value for each consecutive contour. The
  red-dashed line shows the best fit found using a 2nd-order polynomial
  function between the two considered parameters. The light-blue solid-circles
  indicate the median values of both parameters, with their corresponding
  standard deviations represented as error bars, for consecutive bins of 0.10
  in $\Delta$O3N2. The black solid line shows the relation expected between
  both parameters when they are used to derive the same oxygen
  abundance, assuming the \cite{tremonti04} fit for R$_{23}$ and the
  \cite{pettini04} calibration for O3N2. The dot-dashed blue line shows the
  same relation, if the oxygen abundance derived based on R$_{23}$ was 70\% of
  the one derived using the \cite{tremonti04} fit. {\it Central
    panel:} Similar plot for the distributions of the N2O2 line ratio vs. 
  R$_{23}$, for the same \ion{H}{ii} regions. In this case the light-blue
  solid-circles indicate the median values and corresponding standard
  deviations of the parameters for consecutive bins of 0.10 in
  $\Delta$N2O2. {\it Right panel:} Similar plot for the distribution of the N2
  line ratio vs. O3N2. The blue solid circles indicate the median
  values of both parameters, with their corresponding standard deviations, for
  consecutive bins of 0.10 in $\Delta$O3N2. The red dashed-line shows the best
  fit found using a 2nd-order polynomial function between both parameters. In the
  three panels the typical/median errors of the represented parameters are
  represented as a black error-bar.
  \label{fig:R23}}
\end{figure*}

Table \ref{table.med.val} shows the mean values and standard
deviations of the considered parameters for the \ion{H}{ii} regions of each
galaxy, together with the mean value for all the galaxies. Based on
this analysis, it is possible to describe two kind of behaviors: on
one hand, some of the parameters have a large variation object by
object (e.g., EW[H$_\alpha$], r$_{\rm HII}$), reflecting the different
physical conditions of \ion{H}{ii} regions in individual galaxies. On the other
hand, there are parameters with a well defined mean value and little
variation object-by-object, and even region by region (e.g., the dust
attenuation, the ionization parameter and the oxygen abundance). i.e.,
despite of the many differences between the considered galaxies
(luminosity, morphology, color), and the physical conditions in each
\ion{H}{ii} region, it is possible to define a statistically meaningful 
{\it standard} mean value for certain spectroscopic parameters. 

Due to the particular sample of galaxies studied here, and the large
number of \ion{H}{ii} regions explored, we consider that these values define
the average physical conditions of \ion{H}{ii} regions for spiral galaxies in the
Local Universe. They can either be used to determine whether a
particular (spiral) galaxy deviates from the average population (i.e., it is
metal rich or poor and/or it is more or less dusty), or as the anchor point of
chemical evolution of ionized gas along cosmological times.
{ The \ion{H}{ii} regions discussed here have a
range of H$\alpha$ luminosities between 10$^{37}$ and 10$^{41}$ erg~s$^{-1}$, with a well
defined bell-like shape centred at 10$^{39.5\pm 0.6}$ erg~s$^{-1}$. Therefore, most of our 
regions correspond to intermediate/luminous ones, as indicated in Section \ref{sample}.
These \ion{H}{ii} regions are expected to have a typical size of few to several hundreds of parsecs,
at the edge of our spatial resolution or below.} 

For some parameters the mean value is well defined (i.e., it shows a
dispersion around the mean value of the order of the estimated
error). Thus it is a good characterization of the considered property
of the ionized gas. However, it is important to remember that the
variations within the considered distributions reflect changes in the
physical conditions of the ionized nebulae, either within each galaxy,
or galaxy by galaxy. To illustrate this effect we present in
Fig. \ref{fig:diag} three panels showing the classical
[\ion{O}{iii}]/H$\beta$ vs. [\ion{N}{ii}]/H$\alpha$ BPT
\citep{baldwin81} diagnostic diagram for (i) the average values shown
in Table \ref{table.med.val}; (ii) all the emission line regions
detected using the procedure described in previous sections; and (iii)
those emission line regions located at the center of the galaxies
($r<0.5 r_e$). In each panel, we include the Kauffmann et al. (2003)
and Kewley et al. (2001) demarcation curves. These curves are usually
invoked to distinguish between star-forming regions, and other sources
of ionization, like AGN/Shocks/post-AGBs. The location within both
curves is normally assigned to a mix origin for the ionization and/or
contamination by different sources of ionization. Since our sample is
dominated by \ion{H}{ii} regions, most of the regions lie in the
demarcation region corresponding to star-forming areas, although a few
of them are located in the so-called intermediate region. There are
clear differences galaxy by galaxy, reflecting changes in the average
oxygen abundance and ionization strength. On the other hand, it is
clear that for the regions located in the center of the galaxies the
fraction of emission line regions lying in the intermediate region is
larger. This is expected since these are the regions more likely to be
contaminated by other ionization sources than star-formation: e.g.,
central low-intensity AGNs, Shocks, post-AGB stars. { It is beyond
  the scope of this article to determine the real nature of this
  ionization in the central regions.  This will require a more detail
  analysis and the use of other more suitable diagnostic diagrams like
  [NII]/H$\alpha$ vs. [SII]/H$\alpha$ and/or [NII]/H$\alpha$
  vs. [SII]6717,6731/$H\alpha$, to distinguish between HII regions,
  SNR and PNe. We just want to emphasize that if there is any possible
contamination, this will more likely affect only the central regions. }
This { clearly} indicates a change in
the ionization conditions galaxy by galaxy and across the optical extension of each
galaxy. One may wonder whether the described differences are induced by the 
errors in the represented parameters. However, most probably this is not the case. 
The typical error is represented as an error-bar in the central panel of the
figure, showing that it is smaller than the dispersion of values found (in
particular for the range of [\ion{N}{ii}]/H$\alpha$ values sampled).

{ Another possible explanation is that the large range of redshifts/distance 
has an effect in variation of some properties due to our coarse resolution. This could be the case for the
differences found for the St\"omgren radius or the luminosity of H$\alpha$ for some galaxies, like NGC\,628, at a distance of $\sim$9 Mpc
and Mrk\,1477, at a distance of $\sim$90 Mpc. However, this cannot be the only explanation, since there are other galaxies with large
difference in both parameters at lower relative cosmological distance, like UGC\,6410 (at $\sim$80 Mpc), and NGC\,4109 (at $\sim$100 Mpc).}

In summary, although the mean values listed in
Table \ref{table.med.val} are somehow representative of the average
ionization conditions in each galaxy, they do not show the complete
picture, since there are clear/expected variations of the ionizing
condition across the optical extension of the galaxies.

\subsubsection{Strong-line calibrators of the oxygen abundance}\label{empiOH}

One of the fundamental parameters that can be derived from the
spectroscopic analysis of \ion{H}{ii} regions is the oxygen abundance. Oxygen
is one of the easiest elements to measure in \ion{H}{ii} regions, due to the
strength of its emission lines in the optical wavelength range. This
is fortunate, since O is an $\alpha$-process element made directly in
short-lived massive stars (dominant in \ion{H}{ii} regions). It is a good
proxy of all heavy elements, comprising $\sim$50\% of all the metals
by mass in all the Universe. Therefore, it is a fundamental parameter
to understand the evolution of the stellar populations
galaxy-by-galaxy and at different locations within the same galaxy.

Accurate abundance measurements for the ionized gas in galaxies
require the determination of the electron temperature (T$_e$) in this
gas which is usually obtained from the ratio of auroral to nebular
line intensities, such as [\ion{O}{iii}]~$\lambda$4363/[\ion{O}{iii}]~$\lambda\lambda$4959,5007
\citep{osterbrock89}. It is well known that this procedure is
difficult to carry out for metal-rich galaxies since, as the
metallicity increases, the electron temperature decreases (as the
cooling is via metal lines) and the auroral lines eventually become
too faint to measure. Therefore, calibrators
based on strong emission lines are used. \cite{angel12} has recently presented
a revision on the different methods, showing their main problems, and
illustrating in which range/conditions then can be applied. In
particular they describe the most widely used methods in large galaxy
surveys, that rely on the measurement of different strong emission
lines (and line ratios) in the \ion{H}{ii} region spectrum and empirical
calibrations with regions of well-known oxygen abundance.

One of the first strong emission line methods was proposed by
\cite{pagel79}. It relies upon the ratio of [\ion{O}{ii}]~$\lambda$3727 
and [\ion{O}{iii}]~$\lambda\lambda$4959,5007 with
respect to H$\beta$, the so-called R$_{23}$ ratio:

\begin{equation}
{\rm R_{23}} = \frac{I([\ion{O}{ii}]~\lambda3727)+I([\ion{O}{iii}]~\lambda\lambda4959,5007)}{I({\rm H}\beta)}
\end{equation}

\noindent
However, this index is double valued, with two different
calibrations for low metallicity (12\,+\,log(O/H)$<$8.1) and high
metallicity (12\,+\,log(O/H)$>$8.4) \ion{H}{ii} regions. There is an ill-defined
regime where regions with the same R$_{23}$ ($>0.7$) value have oxygen
abundances that differ by almost and order of magnitude \citep[e.g., see Figure~A1 in ][]{angel10b}. Ratios such
as O3N2 \citep{allo79,pettini04}, N2O2
\citep{Zee1998,dopita00}, N2 \citep{Zee1998,pettini04,deni02}, and many
others, were introduced in an attempt to solve this ambiguity in the
derived abundances:

\begin{eqnarray}
 {\rm O3N2} &=& \frac{I([\ion{O}{iii}]~\lambda5007)/I({\rm H}\beta)}{I([\ion{N}{ii}]~\lambda6584)/I({\rm H}\alpha)} \\
 {\rm N2O2} &=&  \frac{I([\ion{N}{ii}]~\lambda6584)}{I([\ion{O}{ii}]~\lambda3727)} \\
   {\rm N2} &=&  \frac{I([\ion{N}{ii}]~\lambda6584)}{I(H\alpha)}
\end{eqnarray}

\noindent
Some of these indicators are strongly affected by the dust
attenuation, like {\rm N2O2} and ${\rm R_{23}}$, and should be
determined after correcting it. Others are less sensitive to dust
attenuation. Furthermore, empirical calibrations based on direct
estimates of the electron temperature of the ionized gas
systematically provide oxygen abundances which are systematically
0.2-0.4~dex lower than those derived using calibrations based on
photoionization models \citep[e.g.][]{angel12}.

To derive the main statistical properties of the oxygen abundance, we
adopted the O3N2 ratio. This ratio is less dependent on the adopted
correction for the dust attenuation, and uses emission lines covered
by our wavelength range for all the galaxies in the current sample.
Like the N2O2 indicator, it relies on the use of the intensity of
[\ion{N}{ii}]. In the high metallicity regime --12+log(O/H)$\gtrsim$8.3-- nitrogen is produced by both
massive and intermediate-mass stars \citep[e.g.][]{pilyugin03} and hence
nitrogen is essentially a secondary element in this metallicity regime.
The [\ion{N}{ii}]/H$\alpha$ ratio will therefore become
stronger with on-going star formation and evolution of galaxies, until
very high metallicities are reached, i.e. 12\,+\,log(O/H)$>$9.0, where
[\ion{N}{ii}] starts to become weaker due to the very low electron temperature
caused from strong cooling by metal ions. However, {\rm N2} is an indirect
estimator that may depend on the ionization strength.

We should remember here that for $\sim$50\%\ of our targets
[\ion{O}{ii}]~$\lambda$3727 was not covered by the observed wavelength
range, and therefore, { the fraction of \ion{H}{ii} regions with
  R$_{23}$ and N2O2 line ratios is just half of the total number of
  regions, since both rely on this emission line}. The
direct estimate based on the electron temperature is of less use. The
fraction of \ion{H}{ii} regions with reliable measurements of the the
auroral [\ion{O}{iii}]~$\lambda$4363 is just $\sim$10\%. This line is
either too faint or is blended and/or strongly affected by the
Hg\,I\,4348 atmospheric light pollution emission line, and even
H$\gamma$, in most of the cases. Due to that we consider O3N2 the best
estimator for the purpose of the current study. In general it is
considered that this estimator is valid for the range between
$-$1$<\log {\rm O3N2}<$1.9 \citep[e.g.][]{yin07}. However it becomes
less reliable for metallicities lower than 12\,+\,log(O/H)$<$8.3
\citep[e.g.][]{yin07,angel12}.


It is { beyond  the scope} of the current study to make a detailed
comparison of the oxygen abundances derived using the different
proposed methods, in a similar way as it was presented by
\cite{kewley08,angel12}. However, it is a good sanity check to compare between
some of the most frequently used ones. This comparison will help us
to understand (1) if the \ion{H}{ii} regions described here are in the valid
range for the adopted metallicity indicator; (2) which are the real
differences in using one indicator or another; and (3) how accurate is
the adopted correction for the dust attenuation, since we can compare
indicators with different degree of dependency on the adopted
correction.

\begin{figure}
\centering
\includegraphics[width=6.5cm,angle=270]{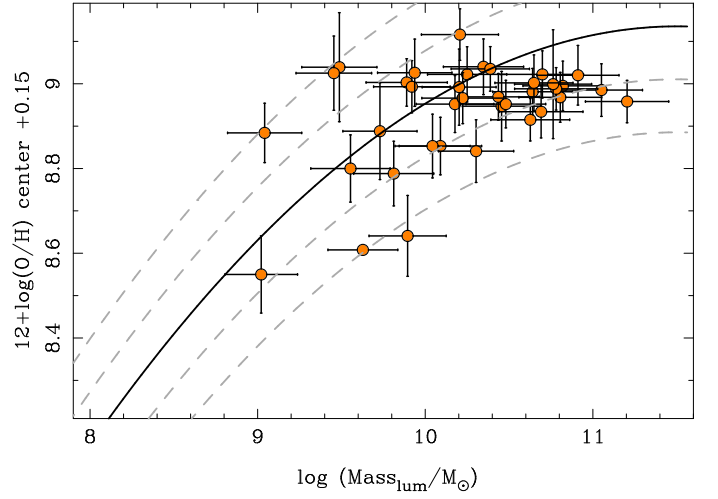}
\caption{Relation between the stellar mass, in units of the solar masses, and
  the oxygen abundance at the central regions ($r<0.2~r_e$), using the O3N2 calibration, for the galaxies considered in
  the current study (solid circles). The black-solid line shows the relation
  found by \cite{tremonti04} for the star-forming galaxies in the SDSS
  sample, corrected for the mismatch between their suggested calibrator and the currently used here (i.e., $-$0.15 dex). The grey-dashed lines show the $\sim$1$\sigma$ and $\sim$2$\sigma$
  range around this relation. \label{fig:MassOH}}
\end{figure}

Fig. \ref{fig:R23}, shows the comparison between O3N2, N2O2 and N2
indicators, all of them using the [\ion{N}{ii}]~$\lambda$6584 line, with respect to the solely
oxygen-based R$_{23}$ ratio. Only the \ion{H}{ii} regions with detected
[\ion{O}{ii}]~$\lambda$3727 emission line have been included in this plot (1142
individual regions). In each panel it is shown the image and contours of the
density distribution of both parameters. The red-dashed lines show the best
fits found using a 2nd-order polynomial function between the two considered
parameters. The light-blue solid-circles indicate the median values of both
parameters, with their corresponding standard deviations represented as error
bars, for consecutive equally spaced bins of 0.10 dex of the parameters shown. In all cases the distributions are similar to the ones
previously described in the literature between the considered estimators
and/or their derived abundance
\citep[e.g.][]{Zee1998,yin07,rosales11}. However, in none of the
precedent studies the distributions are shown with such a large statistical
sample.

In the three cases there is a clear correlation between the
median/high dense distribution of both parameters. The correlation is
stronger for N2O2 vs. R$_{23}$ and for N2 vs. O3N2 than for the
remaining one. { The first of this relations is well described by a 2rd order polynomial function:

\begin{eqnarray}
\log {\rm N2O2} &=& 0.15 -0.97 \log {\rm R_{23}} -0.39 [\log {\rm R_{23}}]^2 
\end{eqnarray}

The dispersions of each parameter around the best fit curve are
$\sigma_{\rm N2O2}\sim$0.11 and $\sigma_{\rm R23}\sim$0.17, respectively.
The dispersion in the y-axis is very similar to the typical (mean) errors
of the corresponding line ratio, $e_{\rm N2O2}\sim0.07$. This indicate that
most of the dispersion is due to the errors in the derived line ratios, and
not in a physical dependence with a third parameter.

The second relation is also well described by a 2rd order polynomical function:

\begin{eqnarray}
\log {\rm N2} &=& 0.46 -0.12 \log {\rm O3N2} -0.19 [\log {\rm O3N2}]^2 
\end{eqnarray}

\noindent

For this distribution, the dispersions of each parameter around the best fit curve are
$\sigma_{N2}\sim$0.10 and $\sigma_{\rm O3N2}\sim$0.32, respectively. Once more,
the dispersion in the y-axis is very similar to the typical (mean) errors
of the corresponding line ratio, $e_{\rm N2}\sim0.17$. Again, this indicates
that most of the dispersion is due to the errors in the measured line ratios, rather
than a unknown relation with a third parameter. 

Finally, the correlation is less tight for the the O3N2 and R$_{23}$
parameters, with a broader dispersion and a tail for R$_{23}>0.7$. In this case
a 2nd order polynomial function describe will enough }
\begin{equation}
\log {\rm O3N2} = 0.18 -3.76 \log {\rm R_{23}} +5.88 [\log {\rm R_{23}}]^2 
\end{equation}

\noindent
However, the dispersion around the best curve is larger for the O3N2
parameter, $\sigma_{\rm O3N2}\sim$0.62, but similar for the R$_{23}$
one, $\sigma_{\rm R23}\sim$0.17. This dispersion cannot be explained
only by the larger typical (mean) error of the represented line ratios
($e_{\rm O3N2}\sim0.2$ and $e_{\rm R23}\sim0.3$, respectively), and
most probably indicates a real difference { and/or a dependence
  with a third parameter (like the ionization strength).}

\cite{tremonti04} proposed a non linear fit of the oxygen
abundance for the R$_{23}$ indicator that is nowadays widely
used. Assuming that this calibration derives a similar oxygen
abundance than the one proposed by \cite{pettini04} is is possible to derive the 
following relation between both parameters:

\begin{equation}
y = -1.406 +0.987 x +0.825 x^2 + 1.396 x^3
\end{equation}

\noindent
where: $ y = \log {\rm O3N2}$ and $ x = \log {\rm R_{23}}$. 
This relation has a similar shape than the one described before for the
range of values corresponding to 0.5$<$ R$_{23}<$1,
however it is { offset} by $\sim$0.15 dex in R$_{23}$ and $\sim -$0.5 dex in
O3N2. The oxygen abundance derived using both indicators agrees if
it is assumed that the \cite{tremonti04} relation overestimate the
abundance by $\sim$30\%. The difference is due to the photoionization models
adopted by \cite{tremonti04}, as discussed in \cite{angel12}.
 Correcting for that difference, the derived relation between
O3N2 and R$_{23}$ agrees with the one found with our dataset (blue
dashed-dotted line in Fig. \ref{fig:R23}). Therefore, the oxygen
abundance derived using the \cite{pettini04} and \cite{tremonti04}
calibrators do not provide consistent results, as already pointed out
in previous studies \citep[e.g.][]{angel12}. However, the discrepancy
is in the zero-point of the adopted calibration, rather than in the
actual values for the indicators. The dispersion is tighter for
R$_{23}<0.7$, i.e., the parameter range for which the oxygen abundance is
not double valued. For values of R$_{23}$ larger than 0.7 we found that
the discrepancy with the derived relation shows a trend with the
ionization parameter, being larger at lower values of $U$.

\cite{kewley08} performed a comparison between the oxygen abundance
derived using different strong line indicators, based on analysis of
large sample of SDSS spectra of star-forming galaxies. They derived a
set of transformation for the different analyzed indicators, including
R$_{23}$ and O3N2. These transformation are consistent with the ones
presented here, for the range of abundance for which their are valid
\citep[see Table 3 in ][]{kewley08}.

\begin{table*}
 \caption{Results of the analysis of the radial gradients of a set of properties.}
 \label{table.grad}      
 \begin{center}
 \begin{tabular}{lrrrrrrrrr}        
 \hline\hline                 
 Parameter & N$_{\rm gal}$ & r$_{\rm cor}$  & $\sigma_{r}$
 & med$_{\rm a}$ & $\sigma_{\rm a}$ & err$_{\rm a}$ 
 & med$_{\rm b}$ & $\sigma_{\rm b}$ & err$_{\rm b}$ \\
  (1)     & (2) & (3) & (4)
& (5)     & (6) & (7)
& (8)     & (9) & (10) \\
\hline                    
$\mu_{\rm B}$                        & 25  &   0.98  &   0.02 &  21.64  &   0.84 &   0.26 &   1.68 &   0.01 &   0.23\\
$B-V$                               & 25  &   0.50  &   0.22 &   0.87  &   0.20 &   0.30 &  -0.20 &   0.15 &   0.27\\
\hline
r$_{\rm HII}$ (100 pc)                & 11  &   0.61  &   0.24 &   2.57  &   0.25 &   0.30 &  -0.48 &   0.20 &   0.26 \\
12\,+\,log$_{\rm 10}$(OH)                & 25  &   0.60  &   0.25 &   8.84  &   0.13 &   0.30 &  -0.12 &   0.11 &   0.27\\
log$_{\rm 10}$($|$EW$_{H\alpha}$$|$)   & 25  &   0.57 &   0.21 &   0.95  &   0.32 &   0.25 &   0.51 &   0.28 &   0.22\\
log$_{\rm 10}$($[$NII$]$/H$\alpha$)  & 25  &   0.54  &   0.24 &  -0.41  &   0.13 &   0.30 &  -0.10 &   0.13 &   0.27\\
log$_{\rm 10}$($[$OIII$]$/H$\beta$   & 25  &   0.53  &   0.28 &  -0.59  &   0.32 &   0.25 &   0.29 &   0.26 &   0.22\\
log$(U)$                            & 11  &   0.35  &   0.24 &  -3.93  &   0.29 &   0.04 &   0.09 &   0.12 &   0.08 \\
n$_e$ (100 cm$^{-3}$)                & 22  &   0.21  &   0.13 &   2.48  &   7.79 &   0.26 &   0.31 &   0.67 &   0.24 \\
A$_V$ (mag)                         & 25  &   0.17  &   0.17 &   1.60  &   0.88 &   0.25 &  -0.29 &   0.52 &   0.22 \\
\hline                    
 \end{tabular}
 \end{center}

{ Notes}:
(1) Parameter which gradient is analyzed, eg., $PARAM = a + b * R/Re$; (2)
Number of galaxies fulfilling the criteria (n$_{\rm HII}>$30 regions); (3)
median value of the derived correlation coefficients for the analyzed
gradient; (4) standard deviation of the derived correlation coefficients; (5)
median value for derived zero-points for each galaxy; (6) standard deviation of
the derived zero-point; (7) median error of the derived zero-points; (8) median
value of the derived slopes for each galaxy; (9) standard deviation of the
derived slopes; (10) median error of the derived slopes.

\end{table*}


In summary, it seems that (i) N2O2 and R$_{23}$ indicators provide
statistically similar empirical information on the oxygen abundance,
i.e. measuring one or the other in the considered range of parameters
will yield a similar estimation of the oxygen abundance, when a
consistent calibration is used. On the other hand, O3N2 provides
different information than the other two indicators, when dealing with
individual regions, even when consistent calibrators are used; 
(ii) the range of values covered by the O3N2 indicator
for the \ion{H}{ii} regions of our sample is within the values where
this indicator is reliable (as indicated before); and (iii) the lack
of outliers in the N2O2 vs. R$_{23}$ distribution indicates that we
have adopted a good correction for the dust attenuation.

\begin{figure*}
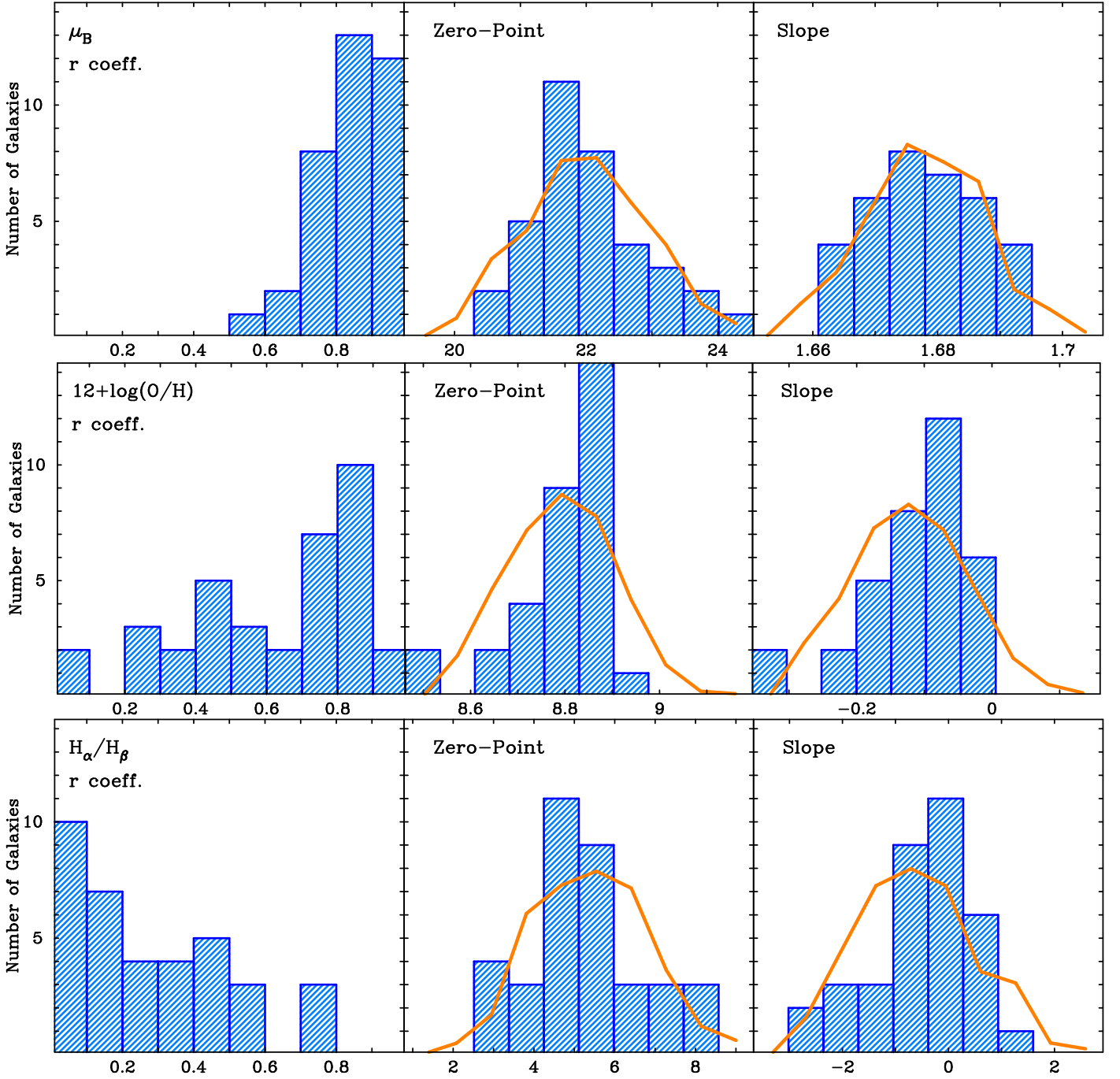

\centering
\includegraphics[width=6cm,angle=270]{figs/93ana_get_plot_name.ps}
\includegraphics[width=6cm,angle=270]{figs/84ana_get_plot_name.ps}
\includegraphics[width=6cm,angle=270]{figs/19ana_get_plot_name.ps}
\caption{Summary of the analysis of the gradients of a set of physical
  properties of the \ion{H}{ii} regions in the considered galaxies. Each panel shows,
  from left to right, (i) the distribution of correlations coefficients of a
  each of the considered parameter (from top to bottom: the surface
  brightness, the oxygen abundance and the dust attenuation) with respect to
  the radial distance; (ii) the distribution of the zero-points of the linear
  regression and (iii) the distribution of the slope of the same
  regression. The orange solid line represents, for each of the last two
  histograms, the expected histogram in case of a Gaussian distribution of the
  data, assuming the mean and standard-deviation of the distribution of each analyzed parameter,
  and sampled with same bins.
    \label{fig:stats}}
\end{figure*}

\subsubsection{Mass-Metallicity Relation}

Stellar mass and metallicity are two of the most fundamental physical
properties of galaxies. Both are metrics of the galaxy evolution
process, the former reflecting the amount of gas locked up into stars,
and the later reflecting the gas reprocessed by stars and any
exchange of gas between the galaxy and its environment. Understanding
how these quantities evolve with time and in relation to one another
is central to understanding the physical processes that govern the
efficiency and timing of star formation in galaxies.

\cite{tremonti04} found a tight correlation between stellar mass and
gas-phase metallicity (MZR) spanning over 3 orders of magnitude in stellar
mass and a factor of 10 in metallicity. This correlation has been
interpreted as a direct evidence of the anticorrelation between metal
loss with baryonic mass, and the ubiquity of galactic winds and their
efficiency in removing metals from low-mass galaxies. Additional
interpretations consider that the observed mass-metallicity relation
is due to a more general relation between stellar, gas-phase
metallicity and star formation rate \citep{lara10,mann10}.

Despite the strength of the considered correlation and how it
matches with our current understanding on galaxy evolution, there are
a few caveats on how the MZR is derived: (i) The oxygen abundance used
by \cite{tremonti04} was derived assuming a Bayesian approach
based on simultaneous fits to all the most prominent lines assuming
the results provided by photoionization models. However, as explained
before these methods systematically overestimate by 0.2--0.4~dex the
oxygen abundances derived using the Te-based calibrations\citep{angel12}; (ii) the
calibrator was derived for \ion{H}{ii} regions, but applied to integrated
apertures in galaxies. Although it is widely used, it is is not
firmly tested if this procedure is valid; (iii) the apertures of the
considered spectra cover a different optical extent for each galaxy,
due to the fixed aperture of the SDSS fiber spectra (3$\arcsec$).

So far, there has been no major effort to test the reported relation
using spatially resolved spectroscopic information, like the one
provided by the current analysis. The number/statistics of our
analysis, galaxy-by-galaxy, is reduced, and we can only test if our
results are consistent with the derived relation. However, it will be
a first step towards a stronger statistical result, that could be
provided with the large sample currently under observation by the
CALIFA survey \citep{sanchez11}.

We derive the luminosity mass of the considered galaxies using the
integrated $B$-band magnitudes, $B-V$ colors and the average
mass-luminosity ratio ($M/L$) described in \cite{bell01}. This
derivation is less accurate than the one described in
\cite{tremonti04}. However, it should be enough for the purpose of
this analysis, considering that the derived $M/L$ has an expected
scatter of $\sim$0.15 dex \citep{bell01}. As an additional
cross-check we also derived the dynamical mass. We use the
effective radius ($r_e$), the maximum rotational velocity 
($v_{\rm rot}$) listed in Table \ref{table.gal.prop}, and the classical
formula \footnote{\url{http://www.astro.virginia.edu/class/whittle/astr553/Topic05/Lecture_5.html}}:

\begin{equation}
M_{\rm dyn} = \frac{{r_e}{v_{\rm rot}}^2}{G}.
\end{equation}

\noindent
We note here that this mass is indeed the dynamical mass at a
particular radius (roughly twice the effective one), and therefore,
the derived values should be taken with caution for any further
interpretation. As expected the dynamical mass is slightly larger than
the stellar mass, since the first considers the total mass of the
galaxy. { However, both masses agree within $\pm$0.2 dex for
  $\sim$90\% of the objects}. Similar results have been recently
reported for the same sort of galaxies, using more sophisticated
derivations of the stellar masses \citep{gers12}. This consistent
result strengthens the validity of our derived masses.

Figure \ref{fig:MassOH} shows the distribution of the oxygen abundance
at the center of the galaxies vs. the stellar masses described
before. The oxygen abundance has been derived extrapolating the radial
gradients found for each galaxy (see Sec. \ref{grad}), to the
center. For comparison, we estimate the central abundance using the
average of the metallicity values within 0.2$r_e$, using the O3N2
calibrator, for those galaxies with \ion{H}{ii} regions at this
distance (20 out of 38). Both values agree within the errors. We
prefer to use the extrapolation of the derived gradient, since they
have lower dispersion, and they can be derived for more galaxies
(i.e., all but MRK\,1477). We include in Fig \ref{fig:MassOH} the
relation between both quantities derived by \cite{tremonti04},
corrected for the mismatch between their calibrator and the currently
used here (i.e., 0.15 dex). The derived central abundance and stellar
masses for the galaxies analyzed here are consistent with this
relation, showing a similar distribution: about 75\% of the galaxies
are located at less than $\sim$1$\sigma$ of the scaled relation.

\begin{figure*}
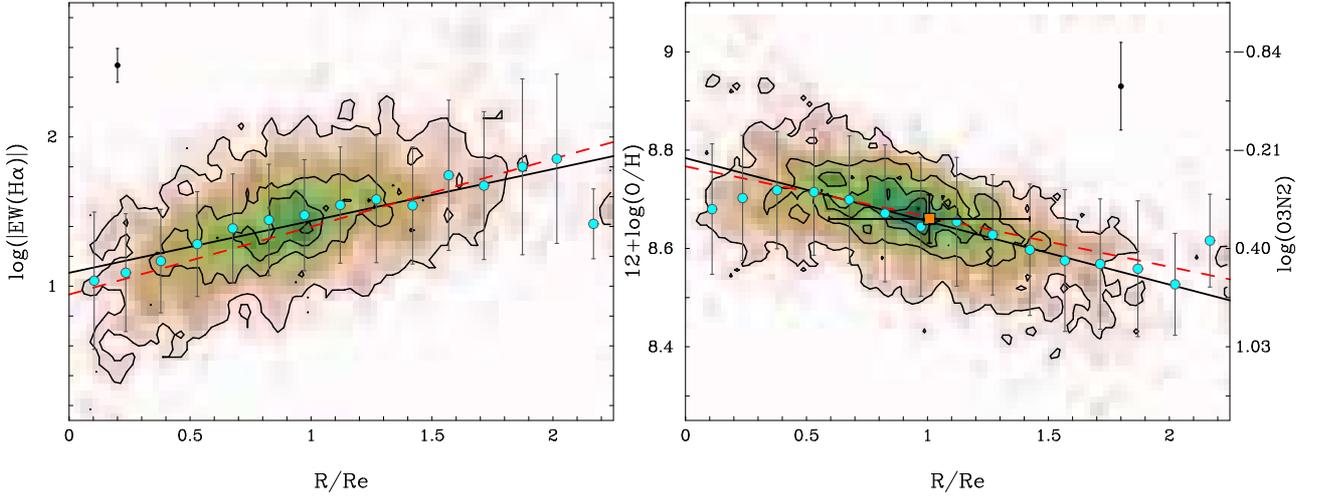

\centering
\includegraphics[width=6.5cm,angle=270]{figs/Re_EW_Ha_all.ps}
\includegraphics[width=6.5cm,angle=270]{figs/Re_OH_all.ps}
\caption{{\it Left panel:} Radial distribution of the equivalent width of
  H$\alpha$ (in logarithm scale of its absolute value), once scaled to the
  average value at the effective radius for each galaxy. The image and
  contours show the density distribution of \ion{H}{ii} regions in this parameter
  space. The first contour indicates the mean density, with a regular spacing
  of four times this value for each consecutive contour. The light-blue
  solid-circles indicate the mean value (with its corresponding standard
  deviation represented as error bars), for each consecutive radial bin of
  $\sim$0.15 R$_e$. The solid line shows the average linear regression found
  for each individual galaxy, as listed in Table \ref{table.grad}. The red
  dashed line shows the actual regression found for all the \ion{H}{ii} regions
  detected for all the galaxies. The average error of the equivalent width of
  H$\alpha$ for the \ion{H}{ii} regions represented in this plot is shown by a single
  error bar located at the top-left side of the panel. {\it Right panel:}
  Similar radial distribution for the oxygen abundance derived using the O3N2
  indicator, once scaled to the average value at the effective radius for each
  galaxy, following the same format. The average error of the derived oxygen
  abundance (without considering systematic errors) is shown by a single error
  bar located a the top-right side of the panel. The right-side scale shows
  the corresponding value for the O3N2 indicator. The solid-orange square
  indicate the average abundance of the solar neighborhood, at the distance
  of the Sun to the Milky-Way galactic center (in effective radius). \label{fig:gradOH}}
\end{figure*}

\subsection{Radial gradients}\label{grad}

It is well known that different spectroscopic properties of
\ion{H}{ii} regions show strong variations across the area of disk
galaxies. In particular, some of these parameters (e.g., oxygen
abundance, EW[H$\alpha$], etc.), show a strong radial gradient, that
in average indicates that more evolved, metal rich, stellar populations
are located in the center of galaxies, and less evolved, metal
poor ones are in the outer ones \citep[e.g.][and references
  therein]{zaritsky94,rosales11}. This observational result is
consistent with the our current understanding of the formation and
evolution of spiral galaxies \citep[e.g.][, and references therein]{tsuj10}. Gas accretion brings gas into the inner
region, where it first reaches the required density to ignite
star formation. Thus the inner regions are populated by older stars,
and they have suffered a faster gas reprocessing, and galaxies
experience an inside-out mass growth \citep[e.g.][]{matt89,bois99}.
Both the extinction-corrected color gradients in nearby galaxies,
\cite{mun07}, and weak dependence of the mass-size relation with
redshift \citep{truj04,bard05,truj06} support an inside-out scenario
for the evolution of disks.

Pure inside-out growth is affected by internal and external
events, like bars and/or galaxy interaction and mergers, which induce
radial migration. In principle, radial migration affects the stellar populations 
more strongly, since they are dynamically more
susceptible to the effects that produce migration (interactions, bars, etc.). The
overall result is a change in the slope of the surface brightness
profile and colors of these galaxies \citep[e.g.][]{bakos08}.
However, gas clumps, like \ion{H}{ii} regions, may also be affected
\citep[e.g.][]{min12}. The net effect of these processes is a possible
flattening of the radial gradient and/or a truncation at outer
regions, where some of these properties (e.g. oxygen abundance) show
values that are expected at more inner regions. This change in the
slope of the radial gradient of oxygen abundance has been reported by
several authors \cite[e.g.][]{bresolin09,mari11,bresolin12}. If true,
migration is an efficient mechanism to intermix the gas and stellar
population properties of galaxies, and even to enrich the intergalactic 
medium \citep[][ for a different explanation to the surface-brightness truncation]{patri09}.

Despite of the several different studies describing these
observational events, there is a large degree of discrepancy between
the actual derived parameters describing the gradients: (i) slope of
the gradient; (ii) average value and dispersion of the zero-point and
(iii) scale-length of the truncation. In general, this is mostly due
to different observational biases and historical methods to perform
the analysis: (i) most of the studies lack a proper statistical
number of analyzed \ion{H}{ii} regions per galaxy; (ii) in many cases the
samples cover mostly the outer regions (R$>$R$_e$), with a lack of \ion{H}{ii}
regions in the inner ones; (iii) there is no uniform method to analyze
the gradients. In some cases the physical scales of the
galaxies (i.e., the radii in kpc) are used, \citep[e.g.][]{mari11}. In others the
scale-length are normalized to the R$_{25}$ radii, i.e., the radii at
which the surface brightness in the $B$-band reach the value of 25
mag/arcsec$^2$ \citep[e.g.][]{rosales11}. Finally, a much more
reduced number of studies tries to normalize the scale-length based on
the effective radii (described in Sec. \ref{struct}). \cite{diaz89}, in one of
the earliest compilations of \ion{H}{ii} regions already showed that the effective
radius is the best one to normalize the abundance gradients. This is
somehow expected since it is directly related with the mass
concentration and reflects how fast the gas is recycled in different
galaxies. Despite its benefits, it is the least used one, which has
produced an ill-defined set of definitions for these gradients. It is
important to remember here that both the physical scale of the radial
distance and/or the normalized one to an absolute parameter like the
$R_{25}$ radius, although they are widely used, could not produce
useful gradients to compare galaxy to galaxy, since in both cases the
derived gradient is correlated with either the scale-length of the
galaxy of its absolute intensity.

\subsection{Analysis of gradients galaxy by galaxy}\label{grad_ours}

We used our catalogue of \ion{H}{ii} regions to characterize the radial
gradients of the physical properties of the \ion{H}{ii} regions. For each
galaxy we derive the correlation coefficient, and the slope and
zero-point of a linear regression; for the radial distribution we derived each
of the parameters discussed in the previous section. The radial distance is
normalized to the effective radius of each galaxy, listed in Table
\ref{table.coords}. Only those galaxies with a good sampling of the
radial distribution of \ion{H}{ii} regions have been considered. To do so, we
adopted two criteria: (i) the galaxy should comprises more than 30
individual \ion{H}{ii} regions, and (ii) there should be equally distributed
along the sampled region ($\sim$2.5 $R_e$).

Table \ref{table.grad} summarizes the result of this analysis. It
includes, for each of the considered parameters, the actual number of
galaxies considered($N_{gal}$), the median of the correlation coefficients
($r_{cor}$), with its standard deviation ($\sigma_r$), the median zero-point
($med_a$), with its standard deviation ($\sigma_a$), and the mean error of
each individual zero-point ($err_a$), together with the median slope
($med_b$), its standard deviation ($\sigma_b$), and the mean error of each
individual derived slope ($err_b$). In addition to the physical parameters of the
\ion{H}{ii} regions discussed in the previous section we have included, for
comparison purposes two properties of the underlying stellar
population, that are well known to correlate with the radial distance:
the $B$-band surface brightness ($\mu_{\rm B}$), and the $B-V$ color. The table is
ordered by the strength of the correlation for the physical parameters
of the \ion{H}{ii} regions.

The overall analysis is illustrated is Fig. \ref{fig:stats}, where
we show the distribution of the correlation coefficients, zero-points,
and slopes for each individual galaxy, for three particular
parameters: (i) the surface brightness, as an example of a strongly
correlated parameter; (ii) the oxygen abundance, i.e., the strongest
correlated parameter for the ionized gas; and (iii) the dust
extinction, a parameter with no gradient in most of the cases. For
highly correlated parameters the mean value, shown in Table
\ref{table.med.val} is not enough to characterize their behavior.

There are parameters that show a statistically significant correlation with the
radial distance ($r_{cor}>$0.5, and $\sigma_{cor}<$0.3). These parameters are
the Str\"omgren radii, the oxygen abundance, the equivalent width of H$\alpha$
and the classical ionization diagnostic line rations ([\ion{O}{iii}]/H$\beta$ and
[\ion{N}{ii}]/H$\alpha$). The first correlation may indicate that the \ion{H}{ii}
regions are larger towards the center. However, we have to be cautious
about this statement, since we do not have a direct estimate of the
actual size of the \ion{H}{ii} regions, and the Str\"omgren radii are based on
particular radiation equilibrium conditions \citep{osterbrock89}. We
need to recall here that since many of our {\it so-called} \ion{H}{ii} regions, are
indeed, \ion{H}{ii} aggregates, this radius should be considered as an {\it effective
  radius}. The next two correlations show that the more metal rich and evolved
stellar populations are located in the inner regions, in agreement with 
expectations. The remaining two gradients indicates that in general
the strength of the ionization is larger in the outer than in the
inner regions. Finally, the ionization parameter shows a weaker trend,
in the way that less powerful \ion{H}{ii} regions are located in the
central areas, while the stronger ones are located in the outer
regions. { However, this result has to be taken with care, since the correlations are weak
and the ionization parameter depends both on the electron density and the number of
ionizing photons, which may present also radial gradients. }

Finally, there are parameters with little dependence with the radial
distance, like the dust attenuation and the electron density. If the
radial gradient of the parameters discussed before (e.g., oxygen abundance and/or equivalent width of H$\alpha$) is a consequence of the
time evolution of the \ion{H}{ii} regions (aging, enrichment, lose of
strength and expansion), those later ones are not strongly affected by
this overall evolution.

\subsection{Universal gradients}\label{grad_global}

For those properties showing a strong correlation, it is equally
important to understand if the gradient is universal, within our range
of explored parameters. Inspecting the histograms shown in Fig.
\ref{fig:stats} and the values for the slopes listed in Table
\ref{table.grad}, we can conclude that all the parameters showing
either a clear correlation or a trend show well defined slopes, taking
into account the considered errors for the measured parameters. 
For the equivalent width of H$\alpha$ and the oxygen abundance a
Lilliefors-test indicates that the histogram of the slopes of the
gradients are consistent with a Gaussian distribution \citep{tLIL67a}. This 
implies that we can define a characteristic value for the slope and 
that we do not find a population of galaxies with slopes inconsistent with 
the normal distribution. 
To illustrate this result we show in Fig. \ref{fig:gradOH} the
radial distribution of two of the parameters with stronger correlation: (i)
the equivalent width of H$\alpha$ and (ii) the oxygen abundance. We
include all the \ion{H}{ii} regions in our catalogue ($\sim$2600). 
For doing so, a global offset has been applied galaxy by galaxy
to normalize the equivalent width and the abundance at effective
radius. This offset was derived by subtracting the value derived by
the individual regression to the median value listed in Table
\ref{table.grad}. A new regression is estimated for all the dataset of
\ion{H}{ii} regions. Both the average regression listed in Table
\ref{table.grad} (red dashed-line), and the regression for the full
sample (black solid-line), has been included. It is clear that both
correlations are very similar, which support our claim that there is a
universal slope for the gradient of either the equivalent width and the oxygen
abundance when the radial scale is normalized by the effective radius. 

Fig. \ref{fig:gradOH} shows that the scatter around the described
correlation, $\sim$0.6 dex, is much larger than the average error of
each individual measurement, $\sim$0.15 dex, for the case of the
equivalent width of H$\alpha$. However, in the case of the oxygen abundances
both values are very similar, $\sim$0.2 dex. If we interpret this scatter
as a typical coherent scale-length, we can conclude that stellar populations
with similar star-formation histories are distributed at a wider radial ranges
than the corresponding ionized gas with similar oxygen abundance. Thus, the
possible intermix due to radial migrations (or similar effects) affects strongly
the stars than the ionized gas, as expected. A larger relative scatter in the properties
of the stellar population compared with the oxygen abundance ones was reported even
in our Galaxy \citep[e.g.][]{friel95}.

For the particular case of the oxygen abundance, the same caveats
expressed in Sect. \ref{empiOH} have to be taken into account
here. We derive the abundance using the O3N2 empirical estimator
\citep{pettini04}, calibration which has a certain range of
validity. Due to that we included the corresponding scale of O3N2
values to illustrate that we are in the range where this calibrator
provides reliable results. We repeated the analysis using the R$_{23}$
calibrator, deriving similar results for the radial gradient. However,
there is a larger scatter in this distribution, mostly due to the
larger error bar of our derivation of oxygen abundance based on this
later method, and the lower number of \ion{H}{ii} regions with measured
[\ion{O}{ii}]~$\lambda$3727 emission line ($\sim$50\%\ of the total sample).

Despite these caveats, the main result are valid: i.e., there seems
to be a universal radial gradient for oxygen abundance when
normalized with the effective radii of the galaxies. This result apparently
contradicts previous ones, in which the slope of the radial
gradients show a trend on certain morphological characteristics of
the galaxies. \cite{VilaCostas:1992p322} shows that barred spirals
have a shallower gradient than non-barred ones. However, we must
recall here that this statements are based, in general, in gradients
on physical scales (i.e. dex kpc$^{-1}$), not in normalized ones. A
slight decrease in the scatter between the different slopes is
appreciated by \cite{VilaCostas:1992p322} and \cite{diaz89}, when the
scale-length is normalized by the disk-scale, which agrees with our
results. We conjecture that both results would come into agreement 
if the size-luminosity/morphological type relation was considered. 

In order to make an independent cross-check we analyzed the
distribution of gradients among our galaxies for barred (10 galaxies)
and un-barred ones (28 galaxies). No significant difference is found
between the distributions, or between any of them and the one
comprising the full sample of galaxies; on the basis of a $t$-test
analysis, the probability of being different was 63.1\%\ in the worst
case. In addition to this test we repeated the comparison for those
galaxies with and without well defined spiral arms (see Table
\ref{table.results}). The nature of flocculent spirals is still
unclear, although the star formation in these galaxies (and therefore,
the \ion{H}{ii} regions) may not be connected with density waves, but
rather with differentially rotating galactic disks
\citep[e.g.][]{efre89}. They could be produced also due to
gravitational instabilities in the stars due to interactions
\citep[e.g.][]{toom64}. We found no differences among the
distributions of slopes for both kind of galaxies. 

Finally, for the Grand Design galaxies (i.e., those galaxies with two
clearly defined spiral arms listed in Table \ref{table.results}), we
repeated the analysis of the gradients for the \ion{H}{ii} regions
located at each spiral arm. Despite of the lower number statistics, we
found not significant difference between the gradients for each spiral
arm.

Fig. \ref{fig:gradOH} shows that for the spatial scale which is well
sampled by our catalogue of \ion{H}{ii} regions (r$<2$r$_e$), the mean
value of the analyzed properties follow the described
distribution. For radii larger than $\sim$2 effective radii, the mean
abundance and H$\alpha$ equivalent width seem to be slightly offset,
with values compatible with more inner regions ($\sim1$
r$_e$). Although this seem to agree with previous results reporting a
flattening on the oxygen abundance, and even with the migration as a
possible explanation, we have to take this result with a lot of
caution. There is just a hand-full of \ion{H}{ii} regions sampled by
our catalogue in this spatial regime (less than a 1\%), and therefore
the result is { not significant}. On the other hand, it seems that
there is a decrease of mean oxygen abundance in the inner regions of
the galaxies ($<$0.3 r$_e$), compared with the expected value
extrapolating the reported correlation. This decrease is not due to
the abundance indicator used, since we detect it using either O3N2 or
R$_{23}$ one. Indeed, this decrease was already detected in individual
galaxies, like in NGC 628 \citep[][]{sanchez11,rosales11}.

We explored the possibility of a statistical origin of this effect,
by analyzing the distribution of oxygen abundance galaxy by
galaxy. However, we found that this decrease is present in 20 of the
28 galaxies with \ion{H}{ii} regions detected in the central region. A lower slope
in the abundance gradient may induce this effect. However, we found
no significant difference in their gradients, with respect to the average
values.

Finally, we have explored which would be the actual location of the
solar neighborhood in the proposed abundance gradient shown in
Fig. \ref{fig:gradOH}. For the distance of the Sun to the center of
the Milky way we adopted the value deduced by \cite{eisen03}, of
7.94$\pm$0.42 kpc. The value of the disk scale-length, and therefore
the effective radius, is still controversial, with large discrepancies
between different results. The more likely value is $\sim$4.5 kpc
(i.e., $r_e\sim$7.6 kpc), although a dispersion larger than 1 kpc is
expected \citep{kruit11}. Therefore the Sun would be at $\sim$1.1
effective radius of the Galactic center. We note that the abundance in
the solar neighborhood is also under debate \citep{solar07}. The
current convention is that it is around $\sim$8.7, with a dispersion
of $\sim$0.2 dex \citep{solar06}. Adopting this scale-length and
abundance the actual location of the solar neighborhood would be just
at the average location from our derive radial gradient (orange square
in Fig. \ref{fig:gradOH}). Considering that the oxygen abundance has
been scaled to the average value at the effective radius, it is
interesting to note that our galaxy behaves as the average for typical
spiral galaxies.

\section{Summary and Conclusions}\label{discuss}

In this article we have analyzed the \ion{H}{ii} regions of a sample of 38
nearby galaxies. The sample has been constructed selecting the almost face-on
spiral galaxies presented in Paper\,I, and similar ones extracted from the
PINGS survey \cite{rosales-ortega10}. In both cases, the galaxies have been
observed using IFS, using the same instrument and sampling similar wavelength
ranges with similar resolutions. The IFS data cover most of the optical
extension of these galaxies, up to $\sim$2-2.5 effective radii, with enough S/N
for the purpose of this study.

We presented {\sc HIIexlorer}, an automatic method to detect and segregate \ion{H}{ii}
regions/aggregations based on the contrast of the H$\alpha$ intensity maps
extracted from IFS datacubes. Once detected, the procedure provides us with a
segmentation map that is used to extract the integrated spectra of each
individual \ion{H}{ii} region. The method has been compared successfully with other
codes available in the literature for NGC\,628, the largest galaxy (in
projected size) included in the sample.

In total, we have detected 3107 \ion{H}{ii} regions, 2573 all of them with good
spectroscopic information (i.e. $\sim$60 \ion{H}{ii} regions per galaxy). This is by
far the largest nearby, 2-dimensional spectroscopic survey presented for this
kind of regions up-to-date. Even more, our selection criteria and dataset
guarantee that we cover the galaxies in an unbiased way, regarding the spatial
sampling. Therefore, the final sample is well suited to understand the spatial
variation of the spectroscopic properties of \ion{H}{ii} regions in this kind of
galaxies.

A well-tested automatic decoupling procedure has been applied to
remove the underlying stellar population, and to derive the main
properties (intensity, dispersion and velocity) of the strongest
emission lines in the considering wavelength range (covering from
[\ion{O}{ii}]~$\lambda$3727 to [\ion{S}{ii}]~$\lambda$6731). A final catalogue of the spectroscopic
properties of these regions has been created for each
galaxy. Additional information regarding the morphology, spiral
structure, gas kinematics, and surface brightness of the underlying
stellar population has been added to each catalogue. In particular, a
detailed analysis of the number and structure of the spiral arms has
been presented, associating each \ion{H}{ii} region to the nearest spiral arm,
and classifying them among inter-arms and intra-arms ones. This
analysis is highly experimental, and its main aim is to understand
asymmetries in the spiral distribution of \ion{H}{ii} regions. To our knowledge
this is the first time that an \ion{H}{ii} region catalogue is provided with
two dimensional information and linked to the spiral structure and
intra-arm association. The full capabilities of the derived
catalogues will be presented in subsequent articles.

In the current study we focused on the understanding of the average
properties of the \ion{H}{ii} regions and their radial distributions. Among the
analyzed properties we include: (1) the equivalent width of H$\alpha$, (2) the
amount of dust attenuation (A$_V$), (3) the most widely used diagnostic line
ratios ($\frac{[\ion{O}{iii}]}{H\beta}$ and $\frac{[\ion{N}{ii}]}{H\alpha}$), (4) the
ionizing parameter (U), (5) the oxygen abundance (O/H), (6) the
{\em effective} Str\"omgren radius and (7) the electron density.

We derive the following results: 
\begin{enumerate}
\item There is statistically significant change in the
ionization conditions across the field of view. The fraction of \ion{H}{ii}
regions affected either by AGNs, shocks or post-AGB stars is much larger 
for the central regions ($r< 0.25 r_e$), than in the outer ones. 
\item The average abundances for the central regions are comparable with the
mass-metallicity relation derived by \cite{tremonti04}. This is particularly
important considering that both the technique to derive the stellar masses and
the central metallicities (including the adopted calibrator) are different. 
The fully independent confirmation of the proposed relation through integral 
field spectroscopy is particularly important as aperture biases can be 
virtually eliminated. 
\item The oxygen abundance and the equivalent width of
H$\alpha$ present a radial gradient that, statistically, has the same slope
for all the galaxies in our sample, when normalized to the effective
radius. The derived slopes for each galaxy are compatible with a
Gaussian random distribution and are independent of 
the morphology of the analyzed galaxies (barred/non-barred,
grand-design/flocculent). 
\item Other properties show no gradient but seem to vary across each
galaxy, and galaxy by galaxy (like the electron density), without a clear
characteristic value, or they are well described by the average value either
galaxy by galaxy or among the different galaxies (like the dust attenuation).
\end{enumerate} 

The \ion{H}{ii} regions catalogues of this analysis will be made publicly available
for a better use on the astronomical community (see Appendix \ref{ape} and 
\noindent {\url{ftp://ftp.caha.es/CALIFA/early_studies/HII/tables/} }).

\begin{acknowledgements}

We thank the referee for the detailed and helpful comments that have improved considerably the quality of this article.

We thank the director of CEFCA, Dr. M. Moles, for his sincere support to this project.

We thank the {\it Viabilidad , Dise\~no , Acceso y Mejora } funding program,
ICTS-2009-10, for funding the data acquisition of this project.

S.F.S., F.F.R.O. and D. Mast thank the {\it Plan Nacional de Investigaci\'on y Desarrollo} funding programs, AYA2010-22111-C03-03 and AYA2010-10904E, of the Spanish {\it Ministerio de   Ciencia e Innovacion}, for the support given to this project. S.F.S. wants to note that the project continues despite of the lack of support of the same program during the 2012 period.

F.F.R.O. acknowledges the Mexican National Council for Science and
Technology (CONACYT) for financial support under the programme
Estancias Posdoctorales y Sabáticas al Extranjero para la
Consolidación de Grupos de Investigación, 2010-2011

J.M. and J.P. acknowledge financial support from the Spanish grant
AYA2010-15169 and Junta de Andaluc\'{\i}a TIC114 and Excellence Project P08-TIC-03531.

D. M. and A. M.-I. are supported by the Spanish Research Council within
the program JAE-Doc, Junta para la Ampliaci\'on de Estudios, co-funded by
the FSE.

R.A. Marino was also funded by the spanish programme of International Campus of Excellence Moncloa (CEI).

J. I.-P., J. M. V., A. M.-I. and C. K. have been partially funded by the projects AYA2010-21887 from the Spanish PNAYA,  CSD2006 - 00070  ``1st Science with
GTC''  from the CONSOLIDER 2010 programme of the Spanish MICINN, and TIC114 Galaxias y Cosmolog\'{\i}a of the Junta de Andaluc\'{\i}a (Spain). 

This paper makes use of the Sloan Digital Sky Survey data. Funding for the
SDSS and SDSS-II has been provided by the Alfred P. Sloan Foundation,  the
Participating Institutions,  the National Science Foundation,  the
U.S. Department of Energy,  the National Aeronautics and Space Administration, 
the Japanese Monbukagakusho,  the Max Planck Society,  and the Higher Education
Funding Council for England. The SDSS Web Site is http://www.sdss.org/.

The SDSS is managed by the Astrophysical Research Consortium for the
Participating Institutions. The Participating Institutions are the
American Museum of Natural History,  Astrophysical Institute Potsdam, 
University of Basel,  University of Cambridge,  Case Western Reserve
University,  University of Chicago,  Drexel University,  Fermilab,  the
Institute for Advanced Study,  the Japan Participation Group,  Johns Hopkins
University,  the Joint Institute for Nuclear Astrophysics,  the Kavli
Institute for Particle Astrophysics and Cosmology,  the Korean Scientist
Group,  the Chinese Academy of Sciences (LAMOST),  Los Alamos National
Laboratory,  the Max-Planck-Institute for Astronomy (MPIA),  the
Max-Planck-Institute for Astrophysics (MPA),  New Mexico State University, 
Ohio State University,  University of Pittsburgh,  University of Portsmouth, 
Princeton University,  the United States Naval Observatory,  and the
University of Washington.

\end{acknowledgements}

\bibliography{CALIFAI}
\bibliographystyle{aa}

\appendix

\section{Catalogues of the \ion{H}{ii} regions}
\label{ape}

\begin{figure*}
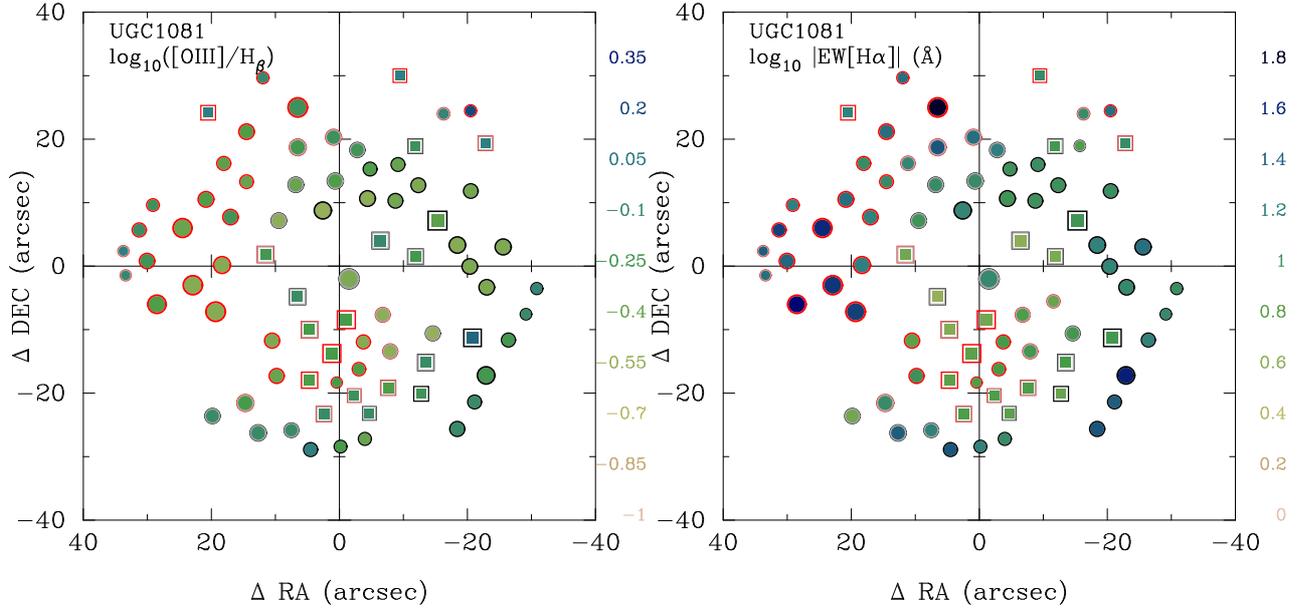

\centering
\includegraphics[width=8cm,angle=270]{figs/55UGC1081.ps}\includegraphics[width=8cm,angle=270]{figs/62UGC1081.ps}
\caption{{\it Left-panel:} Spatial distribution of the [\ion{O}{iii}]/H$\beta$ line
  ratio along the spatial extension of the UGC\,1081 galaxy, derived from the
  analysis of the \ion{H}{ii} regions. Each symbol corresponds to a \ion{H}{ii} region, its
  filling color corresponding to the shown parameter, scaled as displayed in the
  right-size color-table. Red symbols correspond to \ion{H}{ii} regions associated
  with arm 1, and black ones correspond to those associated with arm 2 (where
  the indexing of the arms was selected in arbitrary way). Grey symbols
  represent \ion{H}{ii} regions without a clear association with a particular arm,
  following the criteria described in Sect. \ref{spiral}. The circles
  represent those \ion{H}{ii} regions below the \cite{kauffmann03} demarcation line in
  the BPT diagram shown in Fig. \ref{fig:diag}, and the squares corresponds
  to those ones located in the intermediate region between that line and the
  \cite{kewley01} one. The size of the symbols are proportional to the
  H$\alpha$ intensity. {\it Right-panel:} Similar spatial distribution for the
  absolute value of the equivalent width of H$\alpha$ in logarithm
  scale. Symbols are similar to those described for the left
  panel. \label{fig:ape}}
\end{figure*}

The results of the overall analysis on the properties of the \ion{H}{ii}
regions is compiled in three different catalogues per galaxy, each of
them comprising different information. The nomenclature of each
catalogue is {\tt table.HII.TYPE.GALNAME.csv} (for the coma separated version)
or {\tt table.HII.TYPE.GALNAME.txt} (for the space separated ones), where {\tt
  TYPE} correspond to each of the following types: {\tt
  coords} (coordinates of the \ion{H}{ii} regions), {\tt flux} (fluxes of the stronger emission lines) 
or {\tt EW} (equivalent width of the stronger emission lines). {\tt GALNAME} corresponds
to the galaxy name as listed in Table \ref{table.gal.prop}.

The catalogues are stored in the CALIFA public ftp server:

\noindent {\url{ftp://ftp.caha.es/CALIFA/early_studies/HII/tables/} }
The content of each table is described in the following sections.

\subsection{{\tt coords} tables}

They include all the information regarding the location of each \ion{H}{ii} region
within the galaxy, and additional information regarding their relation with
the galaxy morphology, kinematics and the H$\alpha$ luminosity. The tables
have the following columns:

\begin{enumerate}

\item {\tt ID}, unique identifier of the \ion{H}{ii} region, including the name of the
  galaxy ({\tt NAME}) and a running index ({\tt NN}), in the form: {\tt NAME-NNN}.

\item {\tt RA}, the right ascension of the \ion{H}{ii} region. 

\item {\tt DEC}, the declination of the \ion{H}{ii} region. 

\item {\tt Xobs}, the relative distance in right ascension to the center of the galaxy, in arcsec. 

\item {\tt Yobs}, the relative distance in declination to the center of the galaxy, in arcsec.

\item {\tt Xres}, the deprojected and derotated distance in the X-axis from the center, in kpc. 

\item {\tt Yres}, the deprojected and derotated distance in the Y-axis from the center, in kpc.

\item {\tt R}, the deprojected and derotated distance to the center, in kpc.

\item {\tt Theta}, the deprojected position angle of the \ion{H}{ii} region, in degrees.

\item {\tt N$_{Arm}$}, the ID of the nearest spiral arm.

\item {\tt Inter$_{Arm}$}, a flag indicating if the \ion{H}{ii} region is most probably associated to a particular arm, 1, or most likely an inter-arm region, 0.

\item {\tt D$_{min-Arm}$}, the minimum distance in arcsec to the nearest spiral arm.

\item {\tt D$_{Arm}$}, the spiralcentric distance, i.e., the distance in arcsec along the nearest spiral arm.

\item {\tt Theta$_{Arm}$}, the angular distance in degree to the nearest spiral arm.

\item {\tt vel$_{H\alpha}$}, the H$\alpha$ rotational velocity, in km s$^{-1}$.

\item {\tt e\_vel$_{H\alpha}$}, error of the H$\alpha$ rotational velocity, in km s$^{-1}$.

\item {\tt log10(L$_{H\alpha}$)}, decimal logarithm of the dust corrected luminosity of H$\alpha$, in using of Erg s$^{-1}$.

\item {\tt e\_log10(L$_{H\alpha}$)}, error of the decimal logarithm of the dust corrected luminosity of H$\alpha$, in using of Erg s$^{-1}$.

\end{enumerate} 

\subsection{{\tt flux} tables}

They comprise the fluxes and ratios of the stronger emission lines detected in
each \ion{H}{ii} region, as derived from the analysis described in Sect.
\ref{decouple}. They include the following columns:

\begin{enumerate}

\item {\tt ID}, unique identifier of the \ion{H}{ii} region, described in the previous section.

\item {\tt Flux$_{H\beta}$}, observed flux of H$\beta$, in units of \Funits.

\item {\tt e\_Flux$_{H\beta}$}, error of the observed flux of H$\beta$, in units of \Funits.

\item {\tt  rat\_OII\_Hb }, [\ion{O}{ii}]~$\lambda$3727/H$\beta$ line ratio.

\item {\tt e\_rat\_OII\_Hb}, error of the [\ion{O}{ii}]~$\lambda$3727/H$\beta$ line ratio.

\item {\tt rat\_OIII\_Hb}, [\ion{O}{iii}]~$\lambda$5007/H$\beta$ line ratio.

\item {\tt e\_rat\_OIII\_Hb}, error of the [\ion{O}{iii}]~$\lambda$5007/H$\beta$ line ratio error.

\item {\tt rat\_OI\_Hb}, [\ion{O}{i}]~$\lambda$6300/H$\beta$ line ratio.

\item {\tt e\_rat\_OI\_Hb}, error of the [\ion{O}{i}]~$\lambda$6300/H$\beta$ line ratio.

\item {\tt rat\_Ha\_Hb}, H$\alpha$/H$\beta$ line ratio.

\item {\tt  e\_rat\_Ha\_Hb}, error of the H$\alpha$/H$\beta$ line ratio.

\item {\tt rat\_NII\_Hb}, [\ion{N}{ii}]~$\lambda$6583/H$\beta$ line ratio.

\item {\tt e\_rat\_NII\_Hb}, error of the [\ion{N}{ii}]~$\lambda$6583/H$\beta$ line ratio.

\item {\tt rat\_He\_Hb}, HeI\,6678/H$\beta$ line ratio.

\item {\tt e\_rat\_He\_Hb}, error of the HeI\,6678/H$\beta$ line ratio.

\item {\tt rat\_SII\,6717\_Hb}, [\ion{S}{ii}]~$\lambda$6717/H$\beta$ line ratio.

\item {\tt  e\_rat\_SII\,6717\_Hb}, error of the [\ion{S}{ii}]~$\lambda$6717/H$\beta$ line ratio.

\item {\tt  rat\_SII\,6731\_Hb}, [\ion{S}{ii}]~$\lambda$6731/H$\beta$ line ratio.

\item {\tt  e\_rat\_SII\,6731\_Hb}, error of the [\ion{S}{ii}]~$\lambda$6731/H$\beta$ line ratio.

\item {\tt BPT\_type}, flag indicating the location of the ionized gas region
  in the classical BPT [\ion{O}{iii}]/H$\beta$ vs. [\ion{N}{ii}]/H$\alpha$ diagnostic diagram
  shown in Fig. \ref{fig:diag}, with the following values: (0) location
  undetermined, due to the lack of any of the required line ratios with
  sufficient S/N for this analysis; (1) region located below the
  \cite{kauffmann03} demarcation line, i.e., corresponding to a classical
  star-forming \ion{H}{ii} region; (2) region located in the { intermediate} area within the
  \cite{kauffmann03} and the \cite{kewley01} demarcation lines, i.e.; (3)
  region located above the \cite{kewley01} demarcation line, i.e.,
  corresponding to the area expected from being ionized by an AGN and/or a
  shock.

\end{enumerate}

\subsection{{\tt EW} tables}

They comprise the equivalent width, in Angstroms, of the stronger emission
lines detected in each \ion{H}{ii} region, as derived from the analysis described in
Sect. \ref{decouple}. They include the following columns:

\begin{enumerate}

\item {\tt ID}, unique identifier of the \ion{H}{ii} region, described in the previous section.

\item {\tt EW\_OII}, equivalent width of [\ion{O}{ii}]~$\lambda$3727.

\item {\tt e\_EW\_OII}, error of the equivalent width of [\ion{O}{ii}]~$\lambda$3727.

\item {\tt EW\_Hbeta}, equivalent width of the H$\beta$ emission line.

\item {\tt e\_EW\_Hbeta}, error of the equivalent width of H$\beta$.

\item {\tt EW\_OIII}, equivalent width of [\ion{O}{iii}]~$\lambda$5007.

\item {\tt e\_EW\_OIII}, error of the equivalent width of [\ion{O}{iii}]~$\lambda$5007.

\item {\tt EW\_OI}, equivalent width of [\ion{O}{i}]~$\lambda$6300.

\item {\tt e\_EW\_OI}, error of the equivalent width of [\ion{O}{i}]~$\lambda$6300.

\item {\tt EW\_Halpha}, equivalent width of H$\alpha$.

\item {\tt e\_EW\_Halpha}, error of the equivalent width of H$\alpha$.

\item {\tt EW\_NII}, equivalent width of [\ion{N}{ii}]~$\lambda$6583.

\item {\tt e\_EW\_NII}, error of the equivalent width of [\ion{N}{ii}]~$\lambda$6583.

\item {\tt EW\_SII}, equivalent width of the [\ion{S}{ii}]~$\lambda\lambda$6717,6731 doublet.

\item {\tt e\_EW\_SII}, error of the equivalent width of the [\ion{S}{ii}]~$\lambda\lambda$6717,6731 doublet.

\end{enumerate}


An example of the use of the parameters listed in these tables is
shown in Fig. \ref{fig:vel}, and in the analysis performed in
Sect. \ref{struct}. Fig. \ref{fig:ape} shows an example of the
two dimensional distribution of the properties included in the described
catalogues, which makes use of the many different properties studied
for the \ion{H}{ii} regions.

\section{Empirical correction of the [\ion{N}{ii}] contamination in H$\alpha$ narrow band images}
\label{ape2}

\begin{figure*}
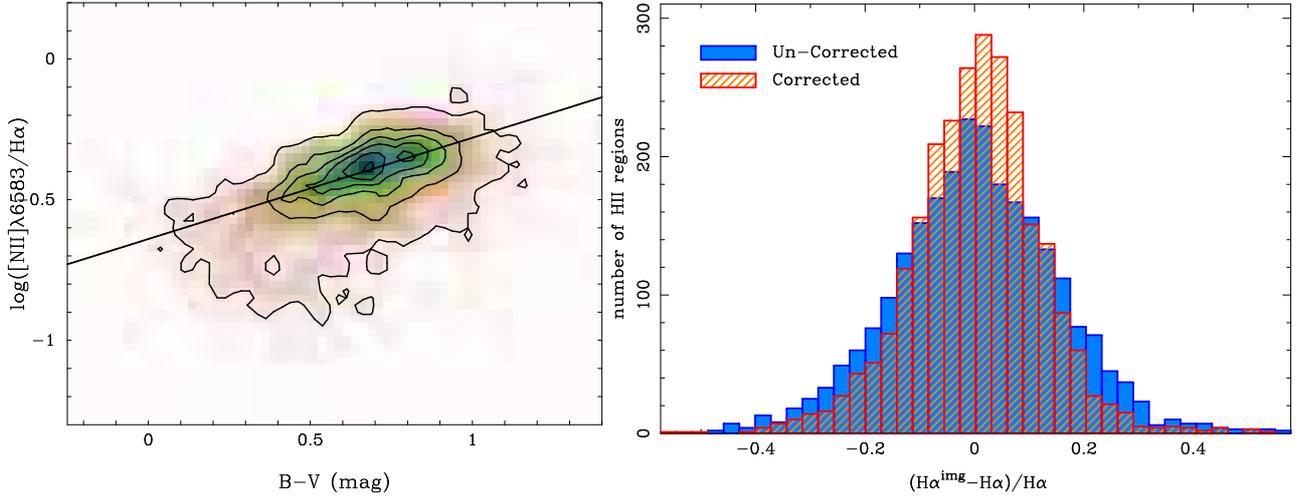

\centering
\includegraphics[width=6.5cm,angle=270]{figs/NIIHa_BV.ps}
\includegraphics[width=6.5cm,angle=270]{figs/Ha_cor.ps}
\caption{{\it left panel:} Distribution of the [\ion{N}{ii}]/H$\alpha$ line ratio
  along the $B-V$ color for all the detected \ion{H}{ii} regions. The image and
  contours show the density distribution in this space of parameters. The
  first contour is at the mean density, with a regular spacing of four times
  this value for each consecutive contour. The solid line shows the actual
  regression found between the two represented parameters. {\it right panel:}
  Histograms of the relative difference between the H$\alpha$ intensity
  derived from the narrow band images described in Sect. \ref{HIIanalyzer},
  and the ones derived from the fitting technique over the extracted spectra,
  described in Sect. \ref{decouple}, before (blue-solid histogram) and after
  (red-shaded histogram) applying the correction from the [\ion{N}{ii}] contamination
  based on the correlation shown in the left-panel. \label{fig:cor}}
\end{figure*}

The H$\alpha$ luminosity observed in spiral and irregular galaxies is
believed to be a direct tracer of the ionization of the inter-stellar
medium (ISM) by the ultraviolet (UV) radiation which is produced by
young high-mass OB stars. Since only high-mass, short-lived, stars contribute
significantly to the integrated ionizing flux, this luminosity is a direct
tracer of the star formation rate (SFR), independent of the previous star
formation history. This is why dust-corrected H$\alpha$ luminosity is one of
the most widely used observables to derive SFR in galaxies.

Among the several methods used to derive the H$\alpha$ luminosity and
its distribution across the optical extension of the galaxies,
narrow-band image is by far the {\it cheapest} one in terms of
telescope time and complexity. Thus, it is the most frequently used too
\citep[e.g.,][]{jame04,pere03}. The technique is rather simple: (i)
the galaxy is observed using both a narrow-band filter centered at the
wavelength of H$\alpha$ and a broader filter covering a wider range
around the same wavelength range; (ii) The broad-band image is used to
correct for the underlying continuum and provide a relative flux
calibration of the continuum subtracted narrow-band image; (iii) by
performing a flux calibration of the broad-band image it is possible
to have an accurate calibration of the decontaminated H$\alpha$ emission
map.

One of the major limitations of narrow-band H$\alpha$ imaging is
contamination by the [\ion{N}{ii}] line doublet, which can hardly be derived
from a single narrow-band imaging \citep[e.g.][]{jame05}. Most
frequently used narrow-band filters have a width of $\sim$50-80\AA, and in
most of the cases the emission from the [\ion{N}{ii}]~$\lambda\lambda$6548,6583
contaminate the derived H$\alpha$ emission map, and therefore, the
derived H$\alpha$ luminosity and SFR. Despite of the fact that the
[\ion{N}{ii}]/H$\alpha$ line ratio present a strong variation across the field
for star-forming galaxies \citep[e.g.][and Fig \ref{fig:diag}]{sanchez11}, 
it is generally assumed an average correction. 

The most commonly used corrections for entire galaxies are those
derived by \cite{ken83} and \cite{kenken83}. Spectrophotometric
[\ion{N}{ii}]/H$\alpha$ ratios of individual extragalactic \ion{H}{ii} regions from
the literature (see Kennicutt \& Kent 1983 and references therein)
were compiled from 14 spiral galaxies (mostly of type Sc) and 7
irregular galaxies. The average H$\alpha$/(H$\alpha$ + [\ion{N}{ii}]) ratio
was found to be fairly constant, spanning the ranges 0.75$\pm$0.12 for
the spirals, and 0.93$\pm$0.05 for the irregulars (Kennicutt 1983). In
terms of the ratio [\ion{N}{ii}]-total/H$\alpha$ this corresponds to a median
value of 0.33 for spirals and 0.08 for irregulars. These values were
calculated by finding the [\ion{N}{ii}]-total/H$\alpha$ ratio of the brightest
\ion{H}{ii} regions, averaging for each galaxy and then determining the mean
value for spiral and irregular types. This implicitly assumes that all
\ion{H}{ii} regions have the same proportion of [\ion{N}{ii}]-total to H$\alpha$
emission as those regions measured. We have already illustrated that
this may not be the case in general, e.g.,
Fig. \ref{fig:diag}. However, recent results show that the total
integrated spectra of galaxies may have a stronger [\ion{N}{ii}]
emission than the one reported before, with a large variation
line ratios for different galaxy types \cite{ken92}.

\cite{jansen00} already showed that there is a trend of the average
[\ion{N}{ii}]/H$\alpha$ ratio with galaxy luminosity. In their study, the lines ratios
for galaxies brighter than M$_{\rm B}$ = $-$19.5 are in agreement with the
values found by \cite{ken92}, but fainter than this a striking trend is seen
towards much lower values of this ratio. A similar trend is found by
\cite{gava04}.

To our knowledge the only attempt to make a correction across the
optical extension of each galaxy is the one introduced by
\cite{jame04}. In this study they used the spatial distribution of
the [\ion{N}{ii}]/H$\alpha$ line ratio, instead of an average correction for
all the entire galaxy. However, their estimations of the distribution
of the line ratio are based on physical principles rather than in
direct measurements.

Making use of our extensive catalogue of individual emission line
regions, we have explored possible empirical corrections that allow us
to perform spatially resolved corrections across the optical extension
of each individual galaxy. For doing so we correlated the measured
[\ion{N}{ii}]/H$\alpha$ line ratios for each individual region with different
parameters easy to address using broad-band photometry, like the
H$\alpha$ intensity, $B$ and $V$-band magnitudes, $B-V$ colors and
radial distance relative to the effective radius. 

Among the different explored linear relations, the one with stronger
correlation coefficient ($r\sim$0.5), and better defined zero-point
and slope, was the one with the $B-V$ color:

\begin{equation}
\log_{\rm 10} ([\ion{N}{ii}]/H\alpha) = -0.64_{\pm 0.20} + 0.36_{\pm 0.18} (B-V)
\end{equation}

Fig. \ref{fig:cor}, left panel, shows the distribution of
[\ion{N}{ii}]/H$\alpha$ line ratios along the $B-V$ color for all the detected
\ion{H}{ii} regions. The image and contours show the density distribution in
this space of parameters, together with a solid line showing the best
fitted linear regression described before. This relation can be used as a
simple proxy of the [\ion{N}{ii}]/H$\alpha$ line ratio, and used to decontaminate the
described H$\alpha$ narrow-band images.

We applied the proposed correction to the H$\alpha$ narrow-band images
described in Sect. \ref{HIIanalyzer}, to demonstrate its improvement
over a {\it classical} single correction over the entire galaxy. We derive for
each detected \ion{H}{ii} region the corresponding flux in the H$\alpha$ narrow-band
images, using the segmentation maps provided by {\sc HIIexplorer}
(H$\alpha^{\rm img}$). Then, we obtain the relative difference between this
{\it contaminated} flux and the real H$\alpha$ flux measured using the fitting
procedure describe in Sect. \ref{decouple}. The mean value of this
difference is $\sim$0.28$\pm$0.19, which is consistent with the typical
average correction found in the literature (e.g., Kennicutt 1983). Fig.
\ref{fig:cor}, right panel, shows a solid-blue histogram of the
relative difference between the H$\alpha$ flux derived from the
narrow-band images, and the ones derived using the emission line
fitting procedure after applying this average correction.

The derived correction is then applied on the same data using a
iterative processes. In each iteration the $B-V$ color is used to
guess the [\ion{N}{ii}]/H$\alpha$ line ratio. This ratio, together with the
decontaminated H$\alpha$ flux derived from in the previous iteration,
is used to derive the [\ion{N}{ii}] intensity. Finally, this intensity is
subtracted to the original contaminated H$\alpha^{\rm img}$ intensity
to derive a new decontaminated flux. For the first iteration it is
assumed that the intensity decontaminated by the average correction is
a good starting estimation of the H$\alpha$ flux. After three
iterations the decontaminated H$\alpha$ flux converge with a few
percent. Fig. \ref{fig:cor}, right panel, shows a hashed-red
histogram of the relative difference between the the H$\alpha$ flux
derived from the narrow-band images, and the ones derived using the
emission line fitting procedure after applying the proposed
correction. It is clear that the new histogram has a lower dispersion
that the previous one, with a standard deviation of $\sim$0.15. The
proposed correction improves the accuracy of the derived H$\alpha$
intensity by a a $\sim$60\%, in average.

The effect of this correction is stronger when analyzing the spatial
distribution of the H$\alpha$ emission and/or the relative strength
of the SFR. As we already indicated the [\ion{N}{ii}]/H$\alpha$ ratio tends
to decrease with the radius. Therefore, an average correction would
overestimate the H$\alpha$ emission and the SFR in the inner regions,
and underestimate it in the outer ones. Indeed, some authors have corrected the image-based H$\alpha$ fluxes for [\ion{N}{ii}] contamination using available spectroscopical data of selected \ion{H}{ii} regions within the galaxy \cite[e.g.][]{angel08}.

So far, we did pay no attention to the possible physical connections
between the explored parameters. The described relation may indicate a
physical connection between the conditions of the ionized gas and the
ionizing stellar population, or between both of them and a third
parameter not considered here. A possible origin of this connection
could be a co-evolution of both components of the \ion{H}{ii} regions. The
study of this connection, that we will address in future works, is { beyond
 the scope } of the current analysis.

\end{document}